%% file: ANA-HDBS-2020-16-PAPER.tex
\documentclass[cernpreprint, atlasdraft=false, texlive=2023, UKenglish, texmf, orcidlogo]{atlasdoc}  
\PreprintIdNumber{CERN-EP-2022-176}
 
\usepackage{atlaspackage}
\usepackage{atlasbiblatex}
 
\usepackage{atlasphysics}

\usepackage{amsmath, bm}

\graphicspath{{logos/}{figures/}}
 
\usepackage{ANA-HDBS-2020-16-PAPER-defs}
\usepackage{booktabs}

\AtlasTitle{Search for a new scalar resonance in flavour-changing neutral-current
top-quark decays $t \rightarrow qX$ ($q=u,c$),
with $X \rightarrow b\bar{b}$,
in proton-proton collisions at $\sqrt{s}=13$ TeV with the ATLAS detector}
 
\AtlasVersion{3.7}
 
\AtlasAbstract{
A search for flavour-changing neutral-current decays of a top quark into an up-type quark
(either up or charm) and
a light scalar particle $X$ decaying into a bottom anti-bottom quark pair is presented.
The search focuses on top-quark pair production where one top quark decays to $qX$, with $X \rightarrow b\bar{b}$,
and the other top quark decays according to the Standard Model,
with the $W$ boson decaying leptonically.
The final state is thus characterised by an isolated electron or muon and at least four jets.
Events are categorised according to the multiplicity of jets and
jets tagged as originating from $b$-quarks,
and a neural network is used to discriminate between signal and background processes.
The data analysed correspond to 139~fb$^{-1}$ of proton--proton collisions at a centre-of-mass energy of 13 TeV,
recorded with the ATLAS detector at the LHC.
The 95\% confidence-level upper limits
between 0.019\% and 0.062\% are derived for the branching fraction $\mathcal{B}$($t \rightarrow uX$)
and between 0.018\% and 0.078\% for the branching fraction $\mathcal{B}$($t \rightarrow cX$),
for masses of the scalar particle $X$ between 20 and 160 GeV.
}

\AtlasRefCode{HDBS-2020-16}  
 
\PreprintIdNumber{CERN-EP-2022-176}

\arXivId{2301.03902}
 
\HepDataRecord{https://www.hepdata.net/record/ins2621899}
 
\AtlasJournalRef{JHEP 07 (2023) 199}
\AtlasDOI{10.1007/JHEP07(2023)199}

\hypersetup{pdftitle={ATLAS document},pdfauthor={The ATLAS Collaboration}}

\begin{document}
 
\maketitle
 
\tableofcontents 
 
\section{Introduction}
\label{sec:intro}

Flavour-changing neutral-current (FCNC) interactions do not exist at tree level in the Standard Model (SM) and are strongly suppressed at higher orders due to the Glashow-Iliopoulos-Maiani (GIM) mechanism~\cite{Glashow:1970gm}.
Several extensions of the SM,
like the quark-singlet model~\cite{AguilarSaavedra:2002kr},
the two-Higgs doublet model with or without flavour-conservation~\cite{Bejar:2000ub, Guasch:1999jp,Cao:2007dk,Cao:2014udj}, SUSY with R-parity violation~\cite{Eilam:2001dh},
or composite Higgs models with partial compositeness~\cite{Azatov:2014lha},
predict the presence of FCNC contributions already at tree level,
and would enhance the top-quark FCNC decay branching fractions by more than ten orders of magnitude.
Other extensions of the SM predict the existence of new particles,
either a neutral scalar, a pseudoscalar, or an axion-like particle (ALP),
which are strongly coupled with third generation quarks and include FCNC couplings with first and second generation quarks.
One of the simplest extensions to the SM is the Froggatt-Nielsen mechanism~\cite{Froggatt:1979, Banerjee_2018},
which introduces a non-SM Higgs field $X$ with flavour charge,
the so-called \textit{flavon},
and a flavour dependent extra symmetry $U(1)_F$ that is broken so that the hierarchy between the couplings depends on the vacuum expectation value of the flavon.
In this model,
and for masses of a neutral scalar particle $X$ below 200~GeV,
the leading decay mode is \Xbb.
 
\begin{figure}[htb]
\centering
\includegraphics[width = 0.5\textwidth]{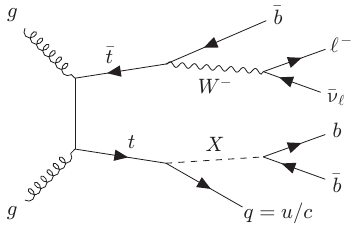}
\caption{
Leading-order Feynman diagram for the production of a scalar particle $X$ in association with a top quark.
}
\label{fig:Feynman}
\end{figure}

The ATLAS and CMS collaborations have searched for FCNC \tqH\ decays,
where $H$ is the SM Higgs boson and $q$ either an up- or charm-quark,
in $pp$ collisions
at $\sqrt{s}=13$~TeV~\cite{TOPQ-2017-07, TOPQ-2019-17, CMS-TOP-17-003, CMS:2021gfa, CMS-TOP-20-007}.
However,
searches for similar signatures involving a FCNC decay of the top-quark into a beyond-the-SM (BSM)
particle lighter than the top quark are uncovered in the literature.
This paper presents a generic search for top-quark pair production where one of the top quarks decays
to a light scalar particle $X$ (with mass $m_{X} < m_{\text{top}}$),
with \Xbb,
and an up-type quark (either $u$ or $c$),
while the other top quark decays to $Wb$ according to the SM
with the $W$ boson decaying leptonically,
as shown in Figure~\ref{fig:Feynman}.
This search uses the full Run 2 dataset of $pp$ collisions recorded at $\sqrt{s}=13$~TeV.
Events with one charged lepton ($l=e,\mu$) and jets in the final state are considered,
and separate regions are defined according to the overall number of jets and the number of jets tagged as containing a $b$-hadron.
In order to distinguish signals from SM backgrounds,
a neural network (NN) is employed,
with basic information from the jets and the lepton
as well as invariant masses and angular separation between pairs of jets.
Limits on the \tqX\ branching fraction are set by means of a simultaneous fit to the NN output distributions in the different analysis regions.
 
\section{ATLAS detector}
\label{sec:detector}

ATLAS~\cite{PERF-2007-01, ATLAS-TDR-2010-19, PIX-2018-001} is a multipurpose detector
with a forward--backward symmetric cylindrical geometry and a near \(4\pi\) coverage in solid angle.\footnote{\AtlasCoordFootnote}
It consists of an inner tracking detector (ID) surrounded by a thin superconducting solenoid providing a \SI{2}{\tesla} axial magnetic field,
electromagnetic and hadron calorimeters,
and a muon spectrometer (MS).
The ID covers the pseudorapidity range \(|\eta| < 2.5\).
It consists of silicon pixel, silicon microstrip, and transition radiation tracking detectors.
Lead/liquid-argon (LAr) sampling calorimeters provide electromagnetic (EM) energy measurements with high granularity.
A steel/scintillator-tile hadron calorimeter covers the central pseudorapidity range (\(|\eta| < 1.7\)).
The endcap and forward regions are instrumented with LAr calorimeters
for EM and hadronic energy measurements up to \(|\eta| = 4.9\).
The MS surrounds the calorimeters and is based on
three large air-core toroidal superconducting magnets with eight coils each.
The field integral of the toroids ranges between \num{2.0} and \SI{6.0}{\tesla\metre}
across most of the detector.
The MS includes a system of precision tracking chambers and fast detectors for triggering.
The first-level trigger is implemented in hardware and uses a subset of the detector
information to accept events at a rate below 100 kHz~\cite{TRIG-2016-01}.
This is followed by a software-based trigger that
reduces the accepted event rate to 1 kHz on average depending on the data-taking conditions.
An extensive software suite~\cite{ATL-SOFT-PUB-2021-001} is used for real and simulated data reconstruction and analysis,
for operation and in the trigger and data acquisition systems of the experiment.


\section{Object definition and event selection}
\label{sec:objectsEventSel}

Data were recorded from $pp$ collisions at $\sqrt{s}=13$~TeV with the ATLAS detector between 2015 and 2018.
Only data consistent with the beam collision region
and for which all relevant detector components were functional are used~\cite{DAPR-2018-01}.
The total integrated luminosity is 139~\ifb~\cite{ATLAS-CONF-2019-021, Avoni:2018iuv}.
Events were recorded with a single-electron or a single-muon trigger,
with minimum thresholds on the transverse momentum (\pT) varying from 20 to 26 GeV depending on the lepton flavour and peak instantaneous luminosity during data taking.
The triggers with the lowest \pT thresholds include isolation
requirements based on the ID or EM calorimeter
measurements ~\cite{TRIG-2018-01, TRIG-2018-05, TRIG-2019-04, TRIG-2019-03}.
In each event, the primary vertex is defined as the reconstructed vertex consistent with originating from the beam collision region in the $x-y$ plane having the highest scalar sum of the
squared \pT of associated tracks with \pT~$\ge$~0.5~GeV~\cite{ATL-PHYS-PUB-2015-026}.
 
Electrons are reconstructed from energy clusters in the EM calorimeter matched to tracks reconstructed in the ID~\cite{EGAM-2018-01},
and are required to have $p_\text{T}>27$ GeV and $| \eta |<2.47$.
Candidates in the calorimeter barrel–endcap transition region ($1.37<| \eta |<1.52$) are excluded.
Electrons must satisfy a ``Tight'' identification criterion based on a likelihood discriminant
described in Ref.~\cite{ATLAS-CONF-2016-024},
the longitudinal impact parameter must be smaller than 0.5~mm,
and the transverse impact parameter significance smaller than 5,
both defined with respect to the beam line.
An isolation criterion is applied to the selected electrons so that the scalar sum of the transverse energy clusters of the calorimeter in a cone of $\Delta R=0.2$ around the electron is less than 6\% of the electron \pT,
excluding clusters originating from the electron itself.
In addition,
the scalar sum of the \pT\ of the tracks above 1 GeV within $\Delta R = 0.2$ around the electron
is required to be smaller than 6\% of the electron \pT.
 
Muons are reconstructed from MS tracks matched to tracks in the ID,
and are required to have $p_\text{T}>27$~GeV and $|\eta|<2.5$,
the longitudinal impact parameter smaller than 0.5 mm,
and the transverse impact parameter significance smaller than 3,
both defined with respect to the beam line.
Muons are identified based on ``Medium'' requirements defined in Ref.~\cite{MUON-2018-03}.
The isolation criterion for the selected muons is defined as the scalar sum of the \pT of tracks within a cone around the muon (excluding the associated track) to be smaller than 6\% of the muon \pT,
with the track isolation cone radius equal to min(10 GeV/$p_{\mathrm{T}}^{\mu}$, 0.3) for $\pT < 50$~GeV
and 0.2 for \pT~>~50~GeV.
 
Jets are reconstructed from topological energy clusters in the calorimeter
and tracks from the ID using the particle-flow method~\cite{PERF-2015-09},
based on the anti-$k_t$ clustering algorithm~\cite{Cacciari:2008gp}
implemented in the FastJet package~\cite{Fastjet}
with a radius parameter of 0.4.
The jet energy is calibrated at particle level~\cite{PERF-2016-04} and jets are required to have $|\eta|<2.5$ and a minimum \pT
of 25~GeV.
Quality criteria are imposed to identify jets arising from non-collision sources or detector noise,
and any event containing such a jet is removed~\cite{ATLAS-CONF-2015-029}.
For jets with $|\eta|<2.4$ and $p_\text{T}<60$~GeV,
the contribution of jets from additional $pp$ collisions in the same and neighbouring bunch crossings (pile-up) is suppressed by the use of the “jet-vertex-tagger” (JVT)~\cite{PERF-2014-03}.
 
Jets containing $b$-hadrons are identified with the DL1r $b$-tagging algorithm,
based on a deep feed-forward NN~\cite{FTAG-2018-01,ATL-PHYS-PUB-2015-039, FlavourTaggingRun2}.
A jet is $b$-tagged if the DL1r score is above a certain threshold,
referred to as operating point (OP).
Four OPs are defined with average expected efficiencies for
jets containing $b$-hadrons of 60\%, 70\%, 77\% and 85\%,
which are each progressively looser,
as determined in simulated \ttbar\ events.
The DL1r $b$-tagging score is divided into five exclusive bins according to the OPs,
and the distribution obtained by ordering these five bins from higher to lower $b$-jet efficiency is referred to as “pseudo-continuous” $b$-tagging score.
Jets fulfilling either of the two tightest bins,
the 60\% OP ($b$-jets) or the 70\% and not the 60\% OPs ($bl$-jets, for looser $b$-tagging score),
are used in the analysis.
 
The missing transverse momentum (\MET) in the event is computed as the magnitude of the negative vector sum of
the \pT\ of all selected and calibrated physics objects in the event.
Low-momentum tracks from the primary vertex that cannot be associated with any of the reconstructed physics objects
described above are also included in the \MET calculation,
as a soft term~\cite{ATL-PHYS-PUB-2015-027,PERF-2014-04}.
 
A sequential overlap removal procedure is applied to ensure that the same calorimeter energy deposit or the
same track is not associated with two or more reconstructed objects,
following the prescription described in Ref.~\cite{TOPQ-2018-01}.
 
Events are required to have exactly one selected electron or muon that matches the lepton that fired the trigger,
and at least four jets.
At least two of the jets are identified as $b$-jets,
and an additional one as a $bl$-jet.
Further requirements on \MET\
and on the transverse mass of the lepton and \MET\
($m_{\text{T}}^W$~\footnote{$m_{\text{T}}^W = \sqrt{ (\pT^l)^2 + (\MET)^2 }$,
where $\pT^l$ is the transverse momentum of the lepton.})
are $\MET\geq20$~GeV and $\MET+m_{\text{T}}^W\geq60$~GeV,
to further reject multi-jet background.


\section{Monte Carlo samples}
\label{sec:MonteCarlo}

Monte Carlo (MC) samples are used to model signal and background processes,
as well as to derive related modelling uncertainties.
The generated events are processed through either a simulation~\cite{SOFT-2010-01}
of the ATLAS detector geometry and response using Geant4~\cite{Agostinelli:2002hh}
or through a faster simulation where the full Geant4 simulation of the calorimeter response is replaced by
a detailed parameterisation of the shower shapes~\cite{ATL-PHYS-PUB-2010-013}.
The effect of pile-up is modelled by overlaying the simulated hard-scatter event with inelastic $pp$ events
generated with \textsc{Pythia}~8.186~\cite{Sjostrand:2007gs} using the NNPDF2.3$_{\text{LO}}$~\cite{Ball:2014uwa} set of parton distribution functions (PDF) and the A3 set of tuned parameters (tune)~\cite{ATL-PHYS-PUB-2016-017}.
 
The \tqX\ signal process is modelled with the \textsc{Powheg-Box} v2~\cite{Nason:2004rx,Frixione:2007vw,Alioli:2010xd}
generator at next-to-leading order (NLO) with the NNPDF3.0$_{\text{NLO}}$~\cite{Ball:2014uwa} PDF set
and the $h_{\textrm{damp}}$ parameter\footnote{The $h_{\textrm{damp}}$ parameter is a resummation damping factor and one of the parameters that controls the matching of \textsc{Powheg} matrix elements to the parton shower and thus effectively regulates the high-\pT\ radiation against which the \ttbar\ system recoils.} set to 1.5 times the mass of the top quark.
The events are interfaced to \textsc{Pythia} 8.244~\cite{Sjostrand:2014zea} with the A14 tune~\cite{ATL-PHYS-PUB-2014-021}
to model the parton shower,
hadronisation,
and underlying event.
The top-quark decays are modelled with \textsc{MadSpin}~\cite{Frixione:2007zp,Artoisenet:2012st}
for both the SM, \tWb,
and BSM, \tqX,
decays assuming a neutral scalar particle $X$~\cite{Degrande_2015}
generated based on the leading order (LO) NNPDF2.3$_{\text{LO}}$ model
with \BR(\Xbb)=100\% and the SM Higgs-boson decay width (0.004~GeV).
 
A total of 52 samples are generated with the \ttbar\ cross-section,
(see below)
and 13 different $X$ masses, \mX,
and four different \tqX\ and \tbarqbarX\ decays: \tuX, \tbarubarX, \tcX\ and \tbarcbarX.\footnote{For simplicity,
all signal processes are denoted as \tqX,
with the charge-conjugate \tbarqbarX\ implied, unless stated otherwise.}
The $m_X$ hypotheses range from 20 to 160~GeV in steps of 10~GeV,
with the exception of the 110 and 130~GeV mass points.
The contribution of the $qg \rightarrow tX$ process has been neglected.
The acceptance times efficiency of the selection presented
in the previous section for the \tuX\ (\tcX) process ranges
from 0.16\% (0.22\%) for the 20~GeV mass sample to
1.65\% (1.70\%) for the 140~GeV mass sample,
decreasing slightly to 1.52\% (1.51\%) for the 160~GeV mass sample.
As the mass of the scalar approaches the mass of the top quark,
the efficiency to reconstruct the $u$- or $c$-jet decreases.
Four additional samples, \tuH, \tbarubarH, \tcH\ and \tbarcbarH,
where $H$ has a mass of 125~GeV and decays according to the SM Higgs boson,
are generated using the same setup as for the \tqX\ signal samples.
 
The main background for this search originates from \ttbar\ production in association with jets,
followed by smaller contributions from single-top-quark,
$Z$ and $W$ bosons plus jets (referred to as $V$+jets),
$t\bar t V$,
$t\bar t H$
and diboson,
as well as rare processes involving the production of a top quark.
The background due to non-prompt leptons is expected to be negligible based on studies of data using
multiple lepton isolation criteria~\cite{HIGG-2017-04} and on the analysis of low \MET\ events.
 
The production of \ttbar~+~jets events is modelled using the \textsc{Powheg-Box}~v2 NLO generator
with the NNPDF3.0$_{\text{NLO}}$ PDF set and the $h_{\textrm{damp}}$ parameter set to 1.5 times the top-quark mass.
The parton shower and hadronisation are modelled by \textsc{Pythia}~8.230 with the appropriate A14 tune.
The $b$-quark production from the matrix element is modelled using the
five-flavour scheme (5FS)~\cite{Maltoni:2012pa, Maltoni_2012_erratum}.
The sample is normalised to the Top++2.0~\cite{Czakon:2011xx} theoretical cross-section of $832^{+46}_{-51}$~pb,
calculated at next-to-next-to-leading order (NNLO) in QCD that includes resummation of next-to-next-to-leading logarithmic (NNLL) soft-gluon terms~\cite{Cacciari:2011hy,Baernreuther:2012ws,Czakon:2012zr,Czakon:2012pz,Czakon:2013goa}.
Theoretical uncertainties result from variations of the factorisation and renormalisation scales,
as well as from uncertainties in the PDF and $\alpha_{S}$,
which represent the largest contribution to the overall theoretical uncertainty in the cross-section and are calculated using the PDF4LHC~\cite{Butterworth:2015oua} prescription.
 
Analogously to previous similar searches performed in ATLAS~\cite{HIGG-2017-04,HDBS-2018-51},
the simulated \ttbar~+~jets events are categorised depending on the flavour content of
additional jets not originating from the decay of the \ttbar\ system.
Events that have at least one jet originating from a $b$-hadron
are labelled as \ttbin;
those with no jets originating from $b$-hadrons
but at least one jet originating from a $c$-hadron are labelled as \ttcin;
finally,
events containing no additional heavy-flavour jets,
aside those from top-quark or $W$-boson decays,
are labelled as \ttlight.
 
Additional samples are generated to estimate \ttbar~+~jets systematic uncertainties.
The impact of using different parton shower and hadronisation models is evaluated by comparing
the nominal sample with another sample produced with the same \textsc{Powheg-Box}~v2 generator and
the NNPDF3.0$_{\text{NLO}}$ PDF set,
but interfaced to \textsc{Herwig}~7.04~\cite{Bellm:2015jjp,Bahr:2008pv}.
To assess the uncertainty in the NLO generator,
the \textsc{Powheg-Box}~v2 sample is compared to a
sample of events generated with \amcatnlo~2.6.0
and the NNPDF3.0$_{\text{NLO}}$ PDF set,
interfaced to \textsc{Pythia}~8.230.
Radiation systematic uncertainties are derived by multiplying the factorisation
and renormalisation scales values in the nominal sample by factors of 0.5 and 2 and
using the \textit{Var3c} variation of the A14 parton shower tune.
 
To evaluate systematic uncertainties on the modelling of the \ttbin\ process,
samples with the \textsc{Powheg-Box~Res}~\cite{Jezo:2018yaf} generator and \textsc{OpenLoops}~\cite{Buccioni:2019sur,Cascioli:2011va,Denner:2016kdg} are produced,
using a pre-release of the implementation of this process in \textsc{Powheg-Box~Res} provided
by the authors~\cite{ttbbPowheg},
with the NNPDF3.0$_{\textsc{NLO}}$ PDF set.
The four-flavour (4FS) scheme is used with the $b$-quark mass set to 4.95~GeV.
They are interfaced to \textsc{Pythia} 8.240 with the A14 tune.
The factorisation scale is set to
$0.5\times\sum_{i=t,\bar{t},b,\bar{b},j}m_{\text{T},i}$,\footnote{$m_{\text{T},i}=\sqrt{m_i^2+p_{\text{T},i}^2}$}
the renormalisation scale to $\prod_{i=t,\bar{t},b,\bar{b}} m_{\text{T},i}^{1/4}$,
and the $h_{\textrm{damp}}$ parameter to $0.5\times\sum_{i=t,\bar{t},b,\bar{b}}m_{\text{T},i}$.
 
All generated \ttbar\ samples assume a diagonal Cabibbo-Kobayashi-Maskawa matrix,
thus the \wcb\ contribution is not included (\BR\ = $5.72\times 10^{-4}$).
Additional \ttbar~+~jets events are produced
with one of the $W$s decaying leptonically and the other to $cb$,
using the SM with non-zero Wolfenstein coefficients and 5FS.
As for the nominal \ttbar,
\textsc{Powheg+Herwig}~7.1.6 and \amcatnlo\ samples are produced to estimate systematic uncertainties related to the
parton shower and hadronisation and to the MC generator, respectively.
 
The associated production of top quarks with $W$ bosons ($tW$),
single-top-quark $t$-channel and single-top-quark $s$-channel productions are modelled with the
\textsc{Powheg-Box}~v2 generator at NLO accuracy in QCD using the 5FS scheme
and the NNPDF3.0$_{\text{NLO}}$ set of PDFs.
The events are interfaced to \textsc{Pythia}~8.230.
The uncertainty due to the parton shower and hadronisation model is evaluated by comparing the nominal samples with
samples where events generated with \textsc{Powheg-Box}~v2 are interfaced to \textsc{Herwig}~7.04.
To assess the uncertainty in the NLO generator,
the nominal samples are compared to samples generated with
\amcatnlo~2.6.2 at NLO in QCD using the 5FS scheme and the NNPDF2.3$_{\text{LO}}$ PDF set.
 
The $tW$ process is modelled using the diagram removal scheme~\cite{Frixione:2008yi}
to handle interference and overlap with \ttbar\ production.
A related uncertainty is estimated by comparing it with an alternative sample generated using the diagram subtraction scheme~\cite{Frederix:2012dh}.
The $tW$ single-top-quark process inclusive cross-section is corrected to the theory
prediction calculated at NLO in QCD with NNLL soft-gluon corrections~\cite{Kidonakis:2010ux,Kidonakis:2013zqa},
whereas the inclusive cross-section for $t$-channel and $s$-channel single-top-quark production is calculated at NLO in QCD with \textsc{Hathor}~2.1~\cite{Aliev:2010zk,Kant:2014oha}.
 
Samples of $W/Z$+jets events are generated with the \textsc{Sherpa}~v2.2.1~\cite{Bothmann_2019}
generator.
The matrix-element calculation is performed up to two additional partons at NLO and
up to four additional partons at LO with the Comix~\cite{Gleisberg:2008fv} and \textsc{OpenLoops} libraries.
The matrix-element calculation is merged with the \textsc{Sherpa} parton shower~\cite{Schumann:2007mg} using the ME+PS@NLO prescription~\cite{Hoeche:2011fd,Hoeche:2012yf,Catani:2001cc,Hoeche:2009rj}.
The PDF set used for the matrix-element calculation is NNPDF3.0$_{\text{NNLO}}$.
Both the $W$+jets and $Z$+jets samples are normalised to their respective inclusive NNLO theory cross-sections calculated with \textsc{FEWZ}~\cite{Anastasiou:2003ds}.
 
Diboson processes ($WW$, $WZ$, and $ZZ$, collectively denoted as $VV$) are simulated with the \textsc{Sherpa}~2.2.1 or~2.2.2 generator depending on the process,
and include off-shell effects and Higgs boson contributions when appropriate.
Fully leptonic and semi-leptonic final states are generated using matrix elements
at NLO in QCD for up to one additional parton,
and at LO for up to three additional parton emissions.
The matrix-element calculations are matched and merged with the \textsc{Sherpa} parton shower using the MEPS@NLO prescription.
Virtual QCD corrections are provided by the \textsc{OpenLoops} library,
and the NNPDF3.0$_{\text{NNLO}}$ set of PDFs is used along with the dedicated set of tuned parton-shower parameters developed by the \textsc{Sherpa} authors.
 
The production of \ttH\ events is modelled with \textsc{Powheg+Pythia}~8.230,
generated at NLO with the NNPDF3.0$_{\text{NLO}}$ PDF set.
The uncertainty due to the parton shower and hadronisation model is evaluated by comparison with samples modelled with \textsc{Herwig}~7.0.4.
To assess the uncertainty in the generator choice,
the nominal samples are compared to samples generated with \amcatnlo~2.6.2 at NLO in QCD using the 5FS scheme and the NNPDF2.3$_{\text{NLO}}$ PDF set.
The \ttH\ cross-section is calculated at NLO QCD and NLO electroweak accuracies using \amcatnlo,
as reported in Ref.~\cite{deFlorian:2016spz}.
The production of $t\bar{t}V$,
$tHq$ and $tZq$ events is modelled with the \amcatnlo~2.3.3 generator at NLO and the NNPDF3.0$_{\text{NLO}}$ PDF,
and interfaced to \textsc{Pythia}~8.210.
The $tZq$ total cross-section is calculated at NLO using \amcatnlo~2.3.3 with the NNPDF3.0$_{\text{NLO}}$ PDF set.
 
In all samples the top-quark mass is set to 172.5 GeV,
and the decays of $b$- and $c$-hadrons are performed
by \textsc{EvtGen} v1.6.0~\cite{Lange:2001uf},
except in samples simulated by the \textsc{Sherpa} event generator.
All samples and their basic generation
parameters are summarised in Table~\ref{tab:MCsummary}.
 
\begin{table}[htb]
\caption{Nominal simulated signal and background event samples.
The ME generator,
PS generator and calculation accuracy of the
cross-section in QCD used for normalisation are shown.
The \tqH\ samples are generated using the same setup as for the \tqX\ signal samples.
Either \textsc{Sherpa}~2.2.1 or \textsc{Sherpa}~2.2.2 was used for different diboson contributions.
The rightmost column shows whether fast or full simulation was used to produce the samples.}
\small
\begin{center}
\begin{tabular}{llllc}
\hline
Physics process      &  ME generator              & PS generator                & Normalisation  & Simulation \\
\hline
\hline
\tqX                 &  \textsc{Powheg-Box}~v2    &  \textsc{Pythia}~8.244      &   NLO           &  Fast \\
\hline
$t\bar t$ + jets     &  \textsc{Powheg-Box}~v2    & \textsc{Pythia}~8.244      &   NNLO+NNLL     &  Fast  \\
\hline
Single-top $tW$      &  \textsc{Powheg-Box}~v2    &  \textsc{Pythia}~8.230     &   NNLO+NNLL     &  Full \\
Single-top $t$-chan  &  \textsc{Powheg-Box}~v2    &  \textsc{Pythia}~8.230     &   NNLO+NNLL     &  Full  \\
Single-top $s$-chan  &  \textsc{Powheg-Box}~v2    &  \textsc{Pythia}~8.230     &   NNLO+NNLL     &  Full \\
\hline
$V$ + jets           &  \textsc{Sherpa}~2.2.1     &  \textsc{Sherpa}~2.2.1     &   NNLO          &  Full  \\
\hline
Diboson              &  \textsc{Sherpa}~2.2       &  \textsc{Sherpa}~2.2       &   NLO           &  Full  \\
\hline
$t\bar t H$          &  \textsc{Powheg-Box}~v2    &  \textsc{Pythia}~8.230     &   NLO           &  Full  \\
$t\bar t V$          &  \amcatnlo~2.3.3           &  \textsc{Pythia}~8.210     &   NLO           &  Full \\
$tHq$                &  \amcatnlo~2.3.3           &  \textsc{Pythia}~8.210     &   NLO           &  Full  \\
$tZq$                &  \amcatnlo~2.3.3           &  \textsc{Pythia}~8.210     &   NLO           &  Full  \\
\hline
\end{tabular}
\end{center}
\label{tab:MCsummary}
\end{table}


\section{Analysis strategy}
\label{sec:analysis}

In order to optimise the sensitivity of the search,
events are categorised into separate regions according to the number of reconstructed jets (j) and $b$-jets (b) in the event. The analysis uses three regions enriched in signal,
4j 3b, 5j 3b and 6j 3b,
and three control regions with small expected signal yields,
4j 4b, 5j $\ge$4b and 6j $\ge$4b.
In all cases,
the $b$-jets are defined at the 60\% OP.
A discriminating variable based on a NN is defined in each signal region for each scalar mass hypothesis.
The binned output of this variable is used in a combined profile likelihood fit to separate the scalar signal
from the SM backgrounds,
whereas the control regions are also included in the fit
to constrain background normalisation factors.
Thus, the fit simultaneously determines both the signal and background yields,
while constraining the overall background model within the assigned systematic uncertainties.
 
The main background for this search originates from \ttbar\ production in association with jets.
It was observed that the simulation does not provide a fully satisfactory description of the data
jet multiplicity and the transverse energy distributions.
To improve the agreement between simulation and data,
three additional regions requiring two $b$-jets and one additional $bl$-jet,
named 2b+1bl in the following,
are used to extract weights to correct the \ttbar and \wcb\ simulated distributions,
similarly to the method developed in recent ATLAS searches~\cite{HDBS-2018-51, TOPQ-2020-10}.
As the signal samples are modelled as \ttbar\ events using the same MC generator,
the correction factors, which have small impact, are also applied to the signal.
 
Data and MC predictions are compared in the 2b+1bl regions separately for events with four,
five and six jets to extract reweighting factors.
Since the mismodelling is assumed to be mainly due to the additional radiation in the parton shower,
which is independent of the flavour of the associated jet,
the correction factors are expected to be appropriate for the 3b and $\geq$4b regions as well,
to the point that the remaining discrepancies would be covered by the \ttbar\ systematic uncertainty model.
The \ttbar\ corrections are derived for each jet multiplicity and as a function of \HTall,
defined as the scalar sum of the transverse momenta of all selected objects in the event including \MET.
The reweighting factors for each jet multiplicity can be expressed as:
 
\begin{equation}
R(H_\mathrm{T}^\mathrm{all}) =
\frac{N_{\text{Data}}(H_\mathrm{T}^\mathrm{all}) - N_{\text{MC}}^\text{non-\ttbar}(H_\mathrm{T}^\mathrm{all})}{N_{\text{MC}}^{\ttbar}(H_\mathrm{T}^\mathrm{all})}.
\label{eq:rew}
\end{equation}
 
In all jet multiplicities,
the reweighting factors are close to unity for $\HTall>800$~GeV and increase rapidly up to a factor of 2--3 towards lower values of \HTall.
Among several functions,
a hyperbola ($R(\HTall) = A+\frac{B}{(H_{\text{T}}^{\text{all}})^C}$) was found to be the best fit
to the weight functions to obtain the final corrections for each jet multiplicity,
as shown in Figure~\ref{fig:reweighting}.
Figure~\ref{fig:rw_jet0pt} shows the effect of the reweighting in the leading jet \pT distribution in the 3b signal regions.
 
\begin{figure}[htb]
\begin{center}
\subfloat[4j 2b+1bl]{\includegraphics[width=0.45\textwidth]{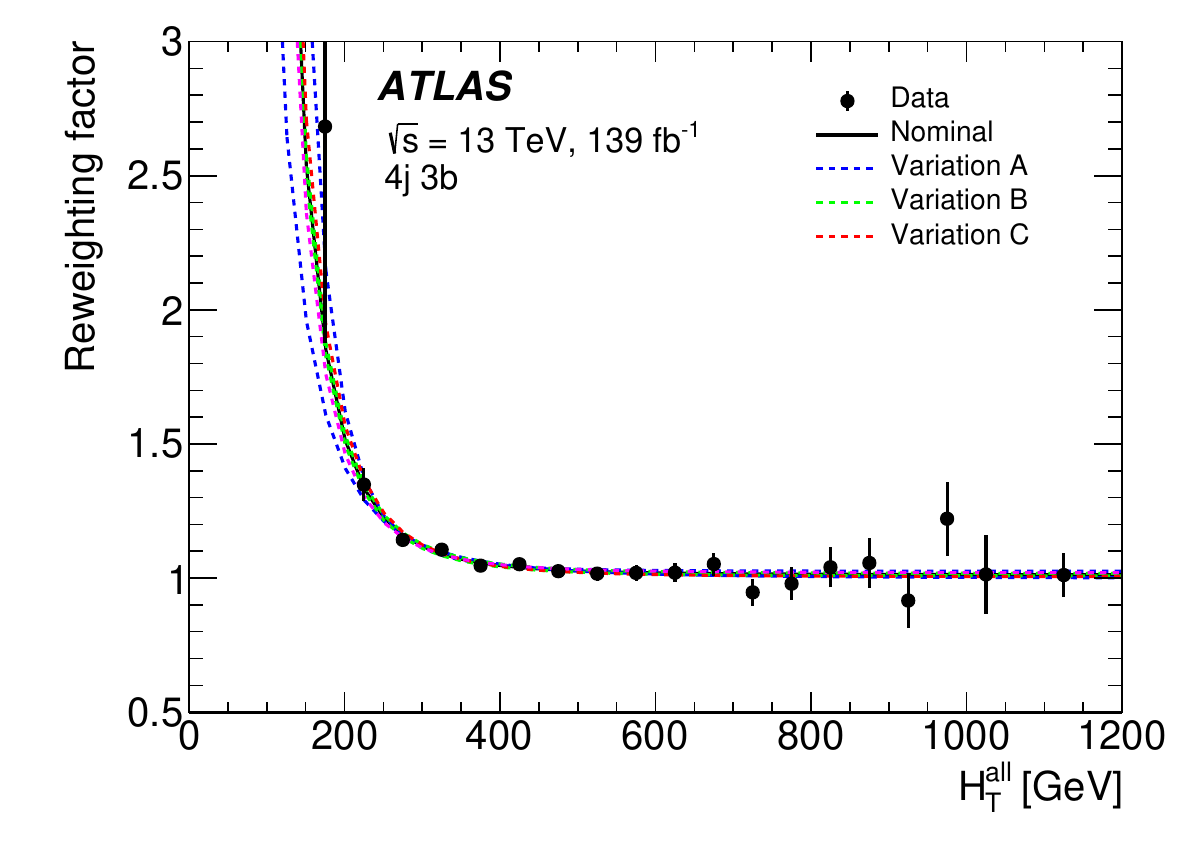}}
\subfloat[5j 2b+1bl]{\includegraphics[width=0.45\textwidth]{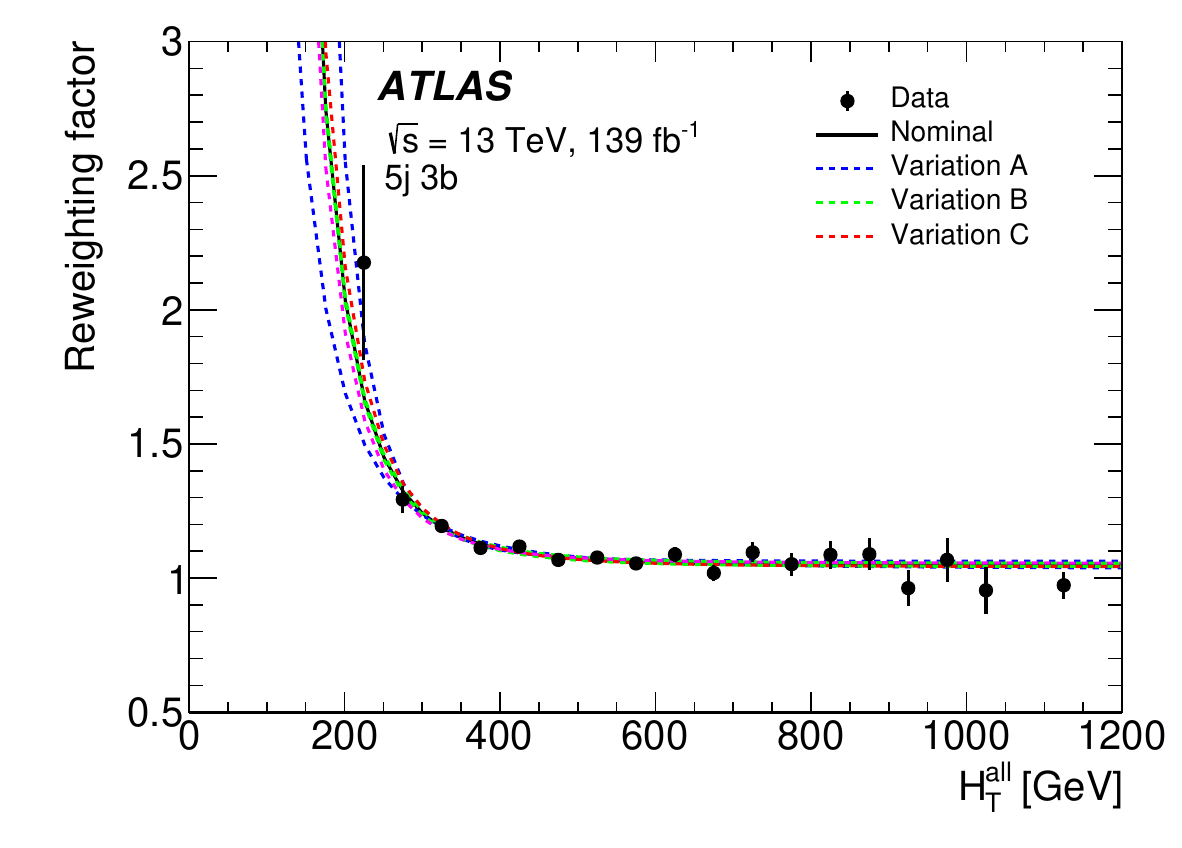}}\\
\subfloat[6j 2b+1bl]{\includegraphics[width=0.45\textwidth]{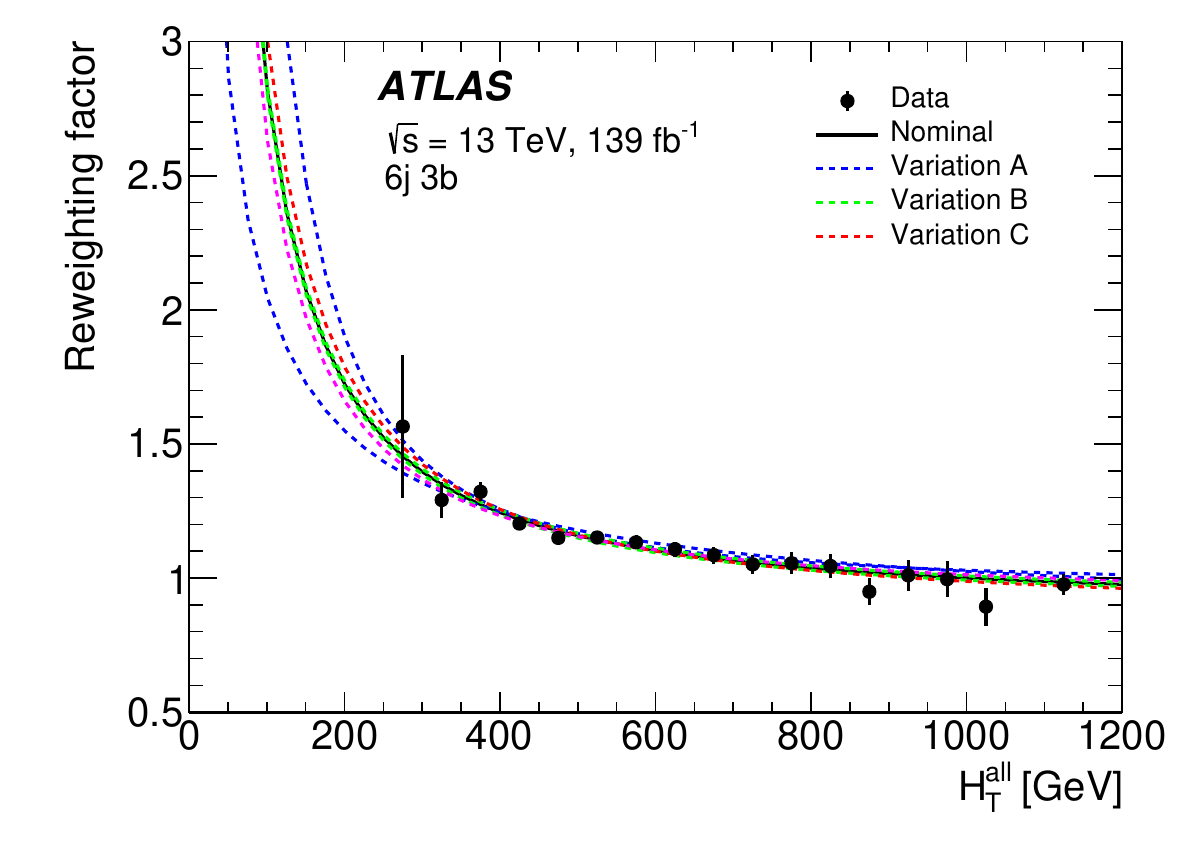}}
\caption{Weight functions obtained from the comparison between data and simulation of \HTall\
for the 2b+1bl regions and three different jet multiplicities,
with the uncertainty bands associated to the variations of the eigenvalues of the matrix error of the fit function,
namely $A$, $B$ and $C$.
The errors in the data points include the statistical uncertainties in data and MC predictions.
}
\label{fig:reweighting}
\end{center}
\end{figure}

\begin{figure}[htb]
\centering
\subfloat[4j 3b, unweighted.]{\includegraphics[width = 0.33\textwidth]{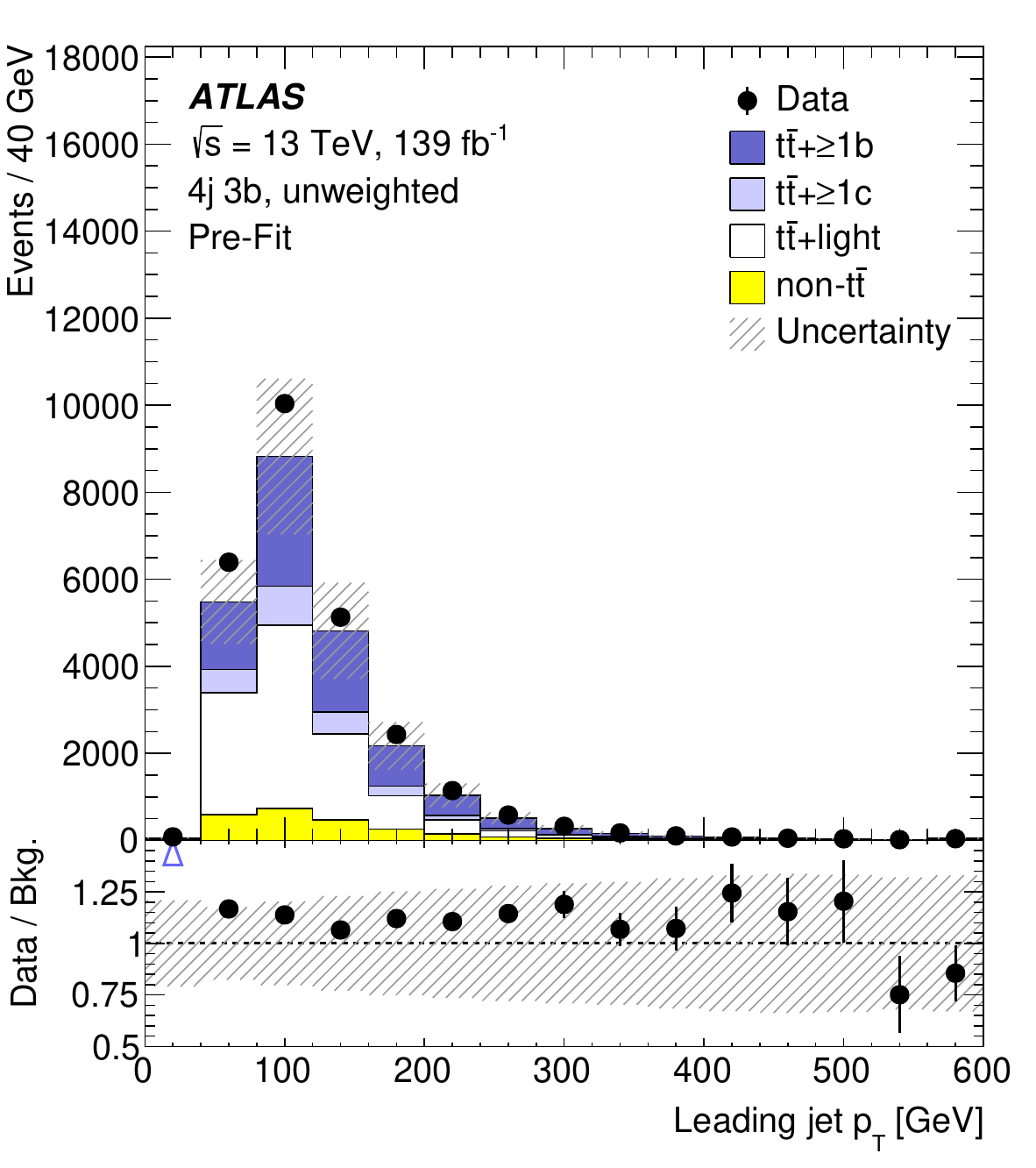}}
\subfloat[5j 3b, unweighted.]{\includegraphics[width = 0.33\textwidth]{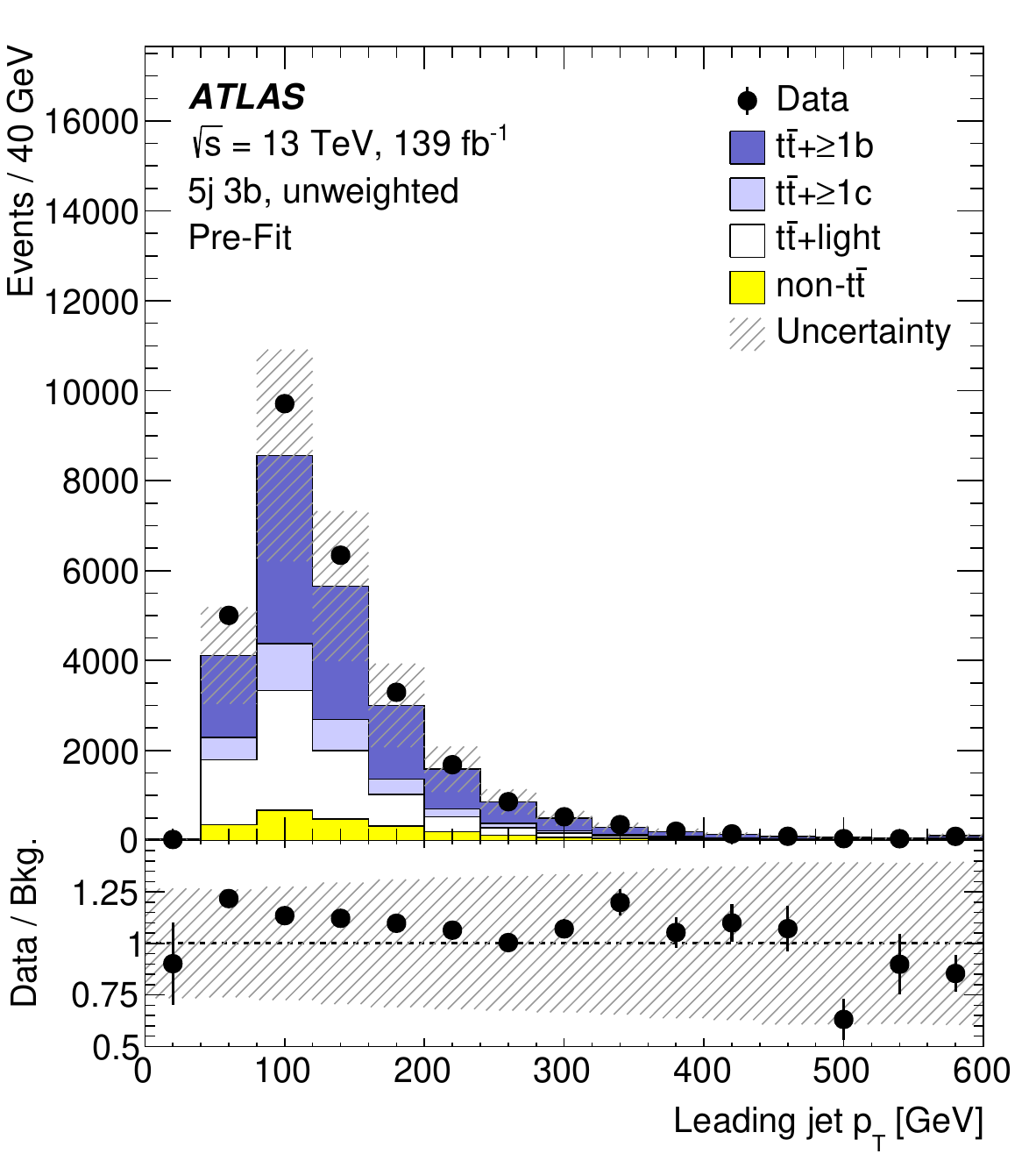}}
\subfloat[6j 3b, unweighted.]{\includegraphics[width = 0.33\textwidth]{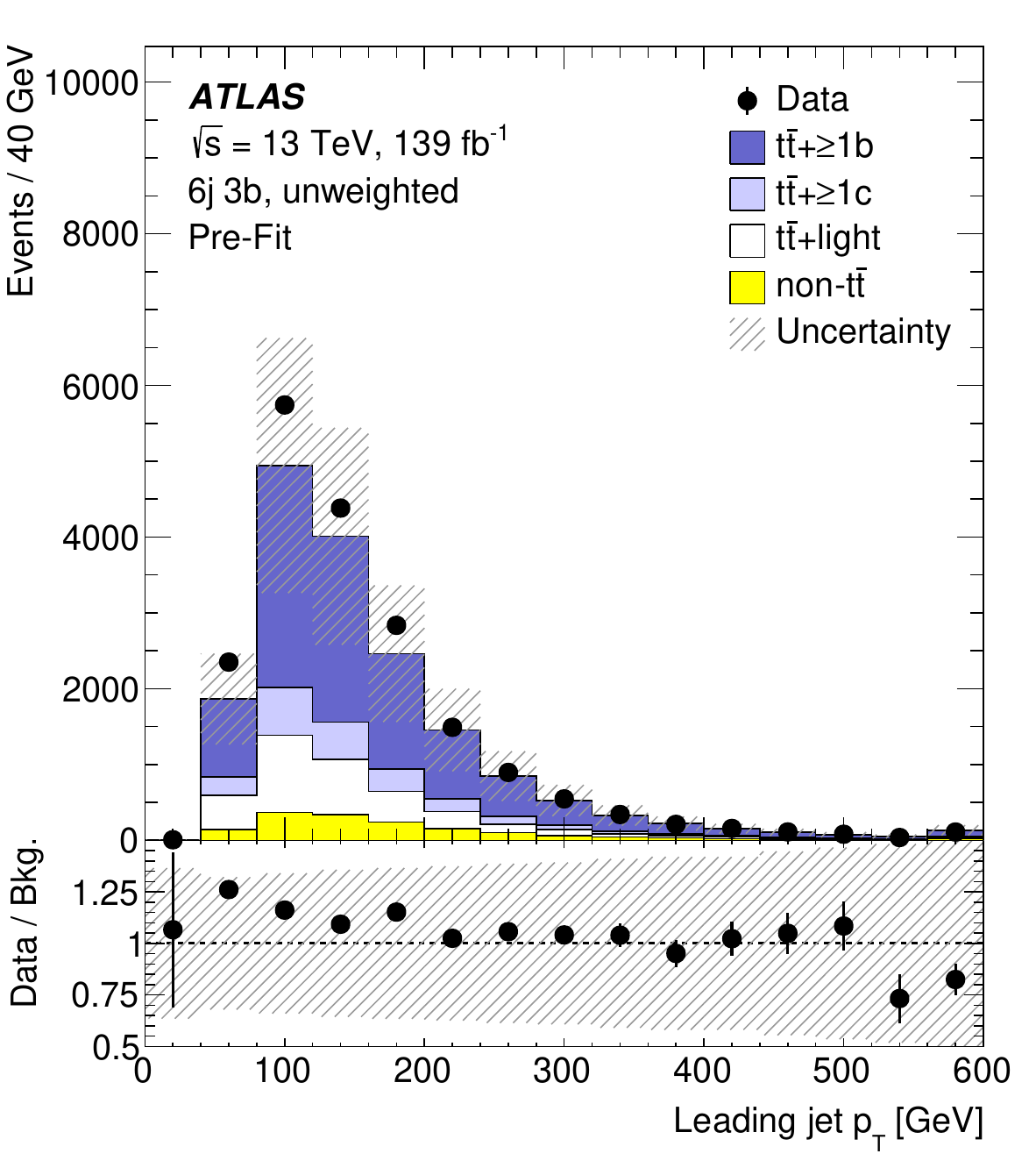}} \\
\subfloat[4j 3b, reweighted.]{\includegraphics[width = 0.33\textwidth]{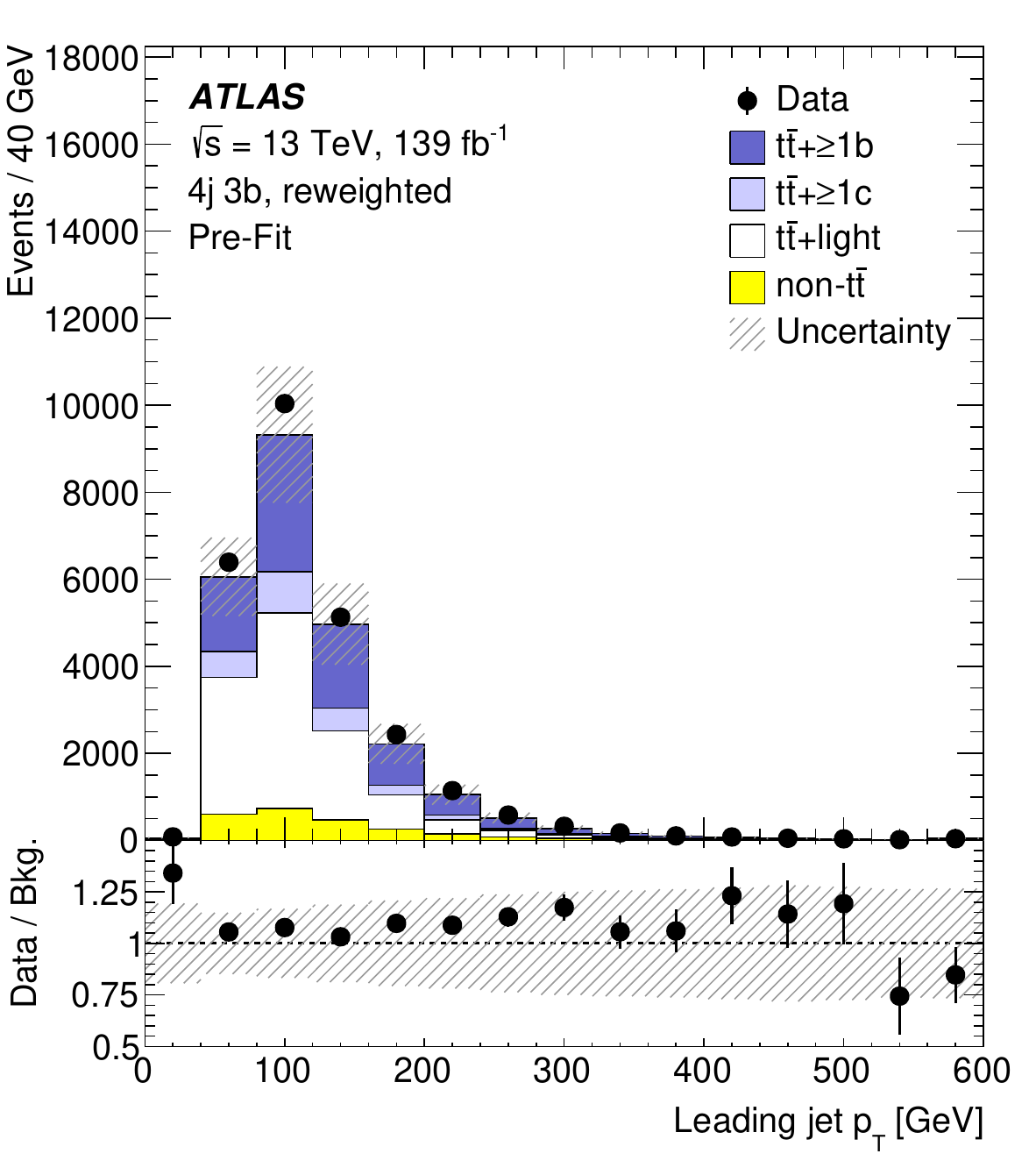}}
\subfloat[5j 3b, reweighted.]{\includegraphics[width = 0.33\textwidth]{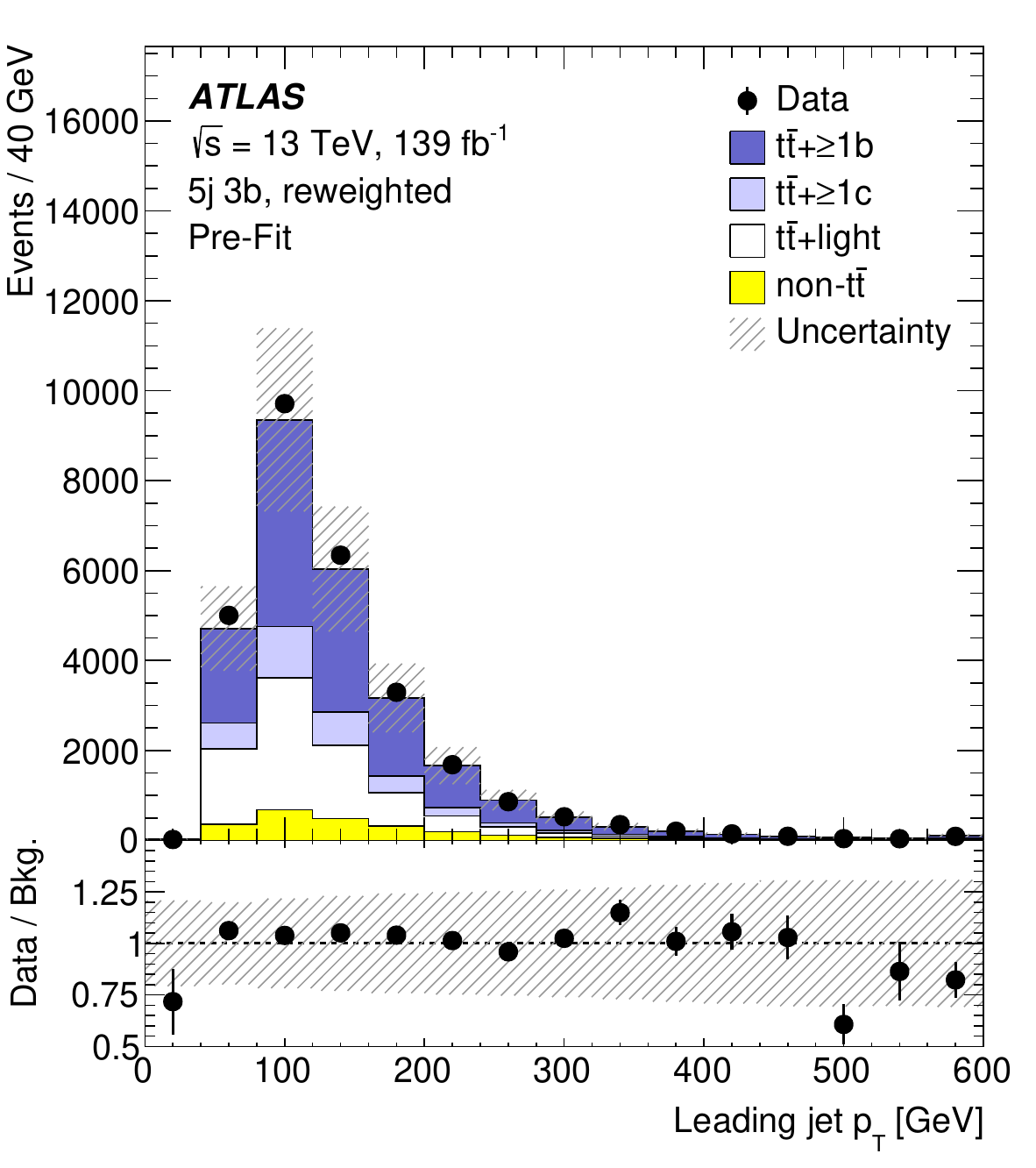}}
\subfloat[6j 3b, reweighted.]{\includegraphics[width = 0.33\textwidth]{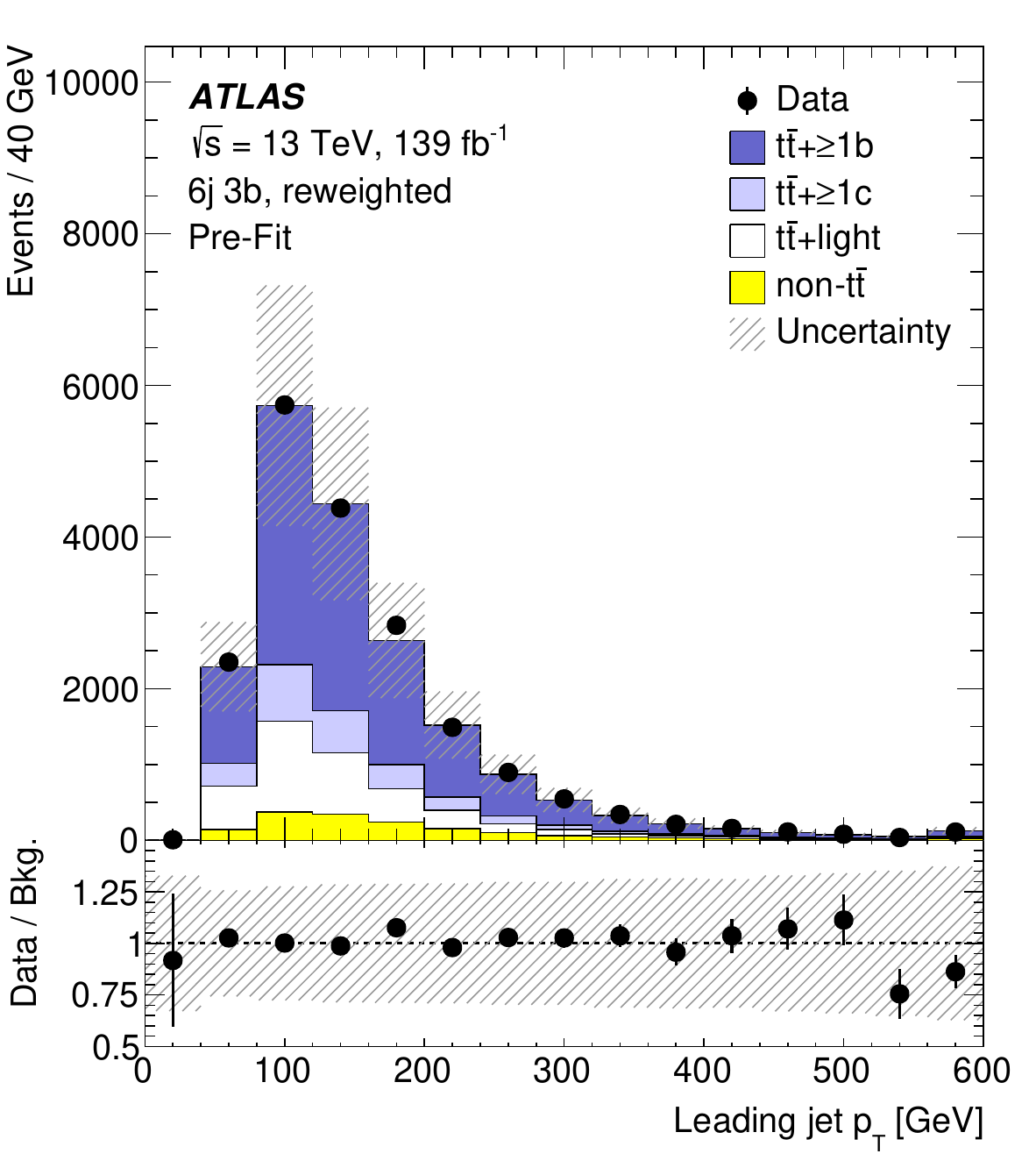}}
\caption{Distributions of the leading jet \pT\ before the fit to the data in the different 3b analysis regions before (from (a) to (c)) and after (from (d) to (f)) applying the \HTall-based reweighting.
The last bin includes the overflow.
The uncertainty bands include the correlated systematic uncertainties in the prediction and the statistical uncertainties uncorrelated across bins.
In cases (d) to (f),
the uncertainty bands are computed after the reweighting and include the associated uncertainties
(see Section~\ref{sec:systematic}).
}
\label{fig:rw_jet0pt}
\end{figure}
 
To enhance the separation between signal and background,
a NN with five hidden layers of 250 nodes each,
3000 batch size,
$10^{-0.75}$ learning rate,
rectified linear unit activation function,
and binary cross entropy loss function,
has been used and is implemented with the deep learning library,
Keras 2.4.3~\cite{chollet2015keras}.
Batch normalisation~\cite{ioffe2015batch} is performed to speed up the learning process,
dropout~\cite{hinton2012improving} is applied at a 25\% rate,
and the Adam algorithm~\cite{kingma2014adam} is used to optimise the parameters.
 
The variables used in the NN include information on the kinematics of the various reconstructed objects
in addition to observables meant to reconstruct the $X\rightarrow b \bar b$ system.
In the following,
jets are ordered by looser $b$-tagging OP,
and by \pT\ within a given bin.
With this convention,
the list of NN input variables includes:
 
\begin{itemize}
\item \pT, $\eta$ and $\phi$ of the first six leading jets.
\item \MET\ magnitude and $\phi$.
\item \pT and $\eta$ of the lepton.\footnote{The origin of the $\phi$ coordinate is chosen to be that of the lepton,
and the $\phi$s of the rest of the reconstructed objects are redefined relative to it.}
\item Bin of pseudo-continuous $b$-tagging distribution for the fourth, fifth and sixth jets.
\item Three invariant masses and three $\Delta R$ of two $b$-jets from the three leading jets combined in pairs.
These variables aim to reconstruct the decay of the scalar particle,
although the width of the reconstructed mass is dominated by experimental resolution effects.
\end{itemize}
 
A 5-fold training~\cite{Kohavi95} is performed using the events in the [4-6]j3b and [4-6]j$\geq$4b regions of the \tuX\ and \tcX\ signal processes separately.
For each training,
all background samples and all signal samples with $m_X\geq30$~GeV for the corresponding process are used.
The various backgrounds are normalised according to their cross-sections,
while the different signals are normalised to the total background.
The training also includes the value of the \mX\ parameter,
which for signal events is defined to be the true mass of the signal sample,
while for background events a random value of the $X$ mass,
taken from the fraction of signal masses in the input dataset,
is assigned to each event~\cite{Baldi:2016fzo}.
In addition to increasing the size of the training sample,
the use of a mass-parameterised NN allows the different signals to be differentiated.
Figures~\ref{fig:NNcomparisonuX} and~\ref{fig:NNcomparisoncX} compare the distributions of the NN output in the signal regions between either the \tuX\ or \tcX\ process and the background for three representative values of \mX: 30, 80, and 120~GeV.
For high values of \mX,
the invariant masses and angular distances of the $b$-jet pairs peak at similar values
in both signal and background events,
thus reducing the NN discriminating power.
The NN output distributions are used in a fit to extract the amount of signal in data.

\begin{figure}[htb]
\centering 
\subfloat[\mX = 30 GeV, 4j 3b]{\includegraphics[width = 0.3\textwidth]{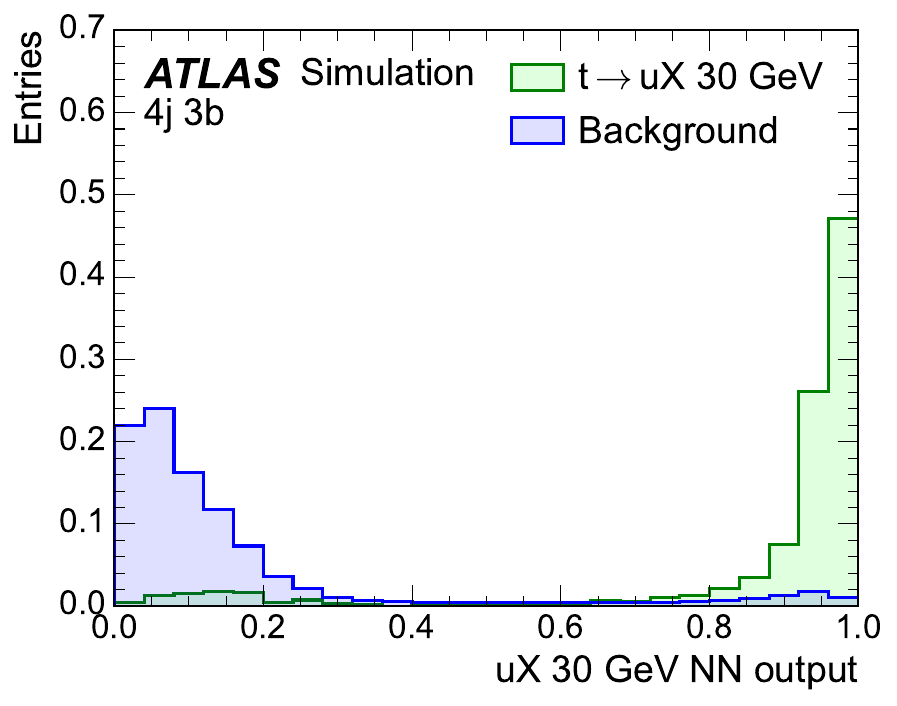}}
\subfloat[\mX = 30 GeV, 5j 3b]{\includegraphics[width = 0.3\textwidth]{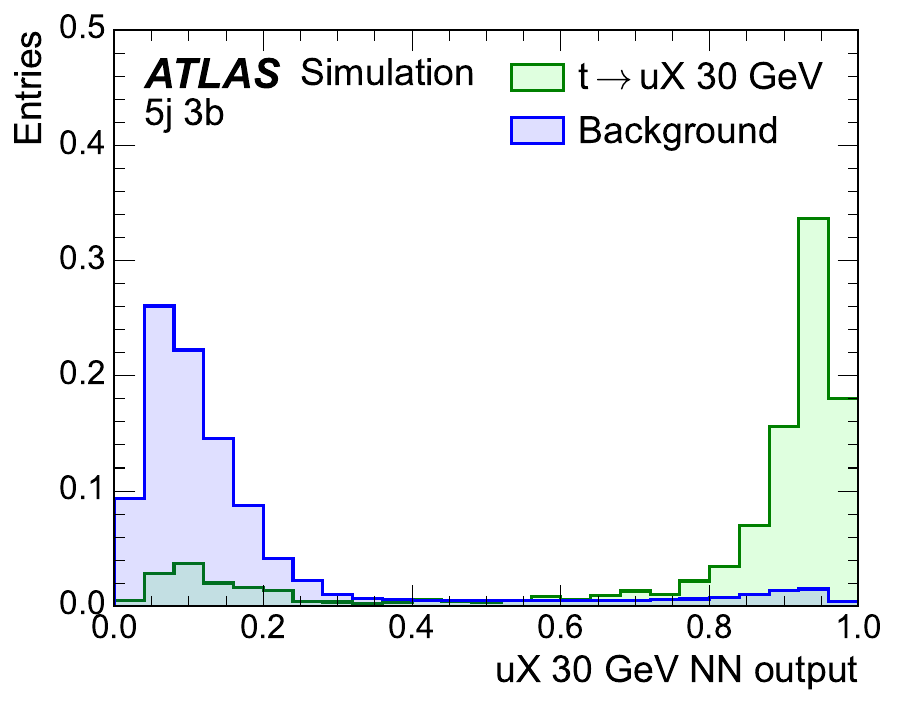}}
\subfloat[\mX = 30 GeV, 6j 3b]{\includegraphics[width = 0.3\textwidth]{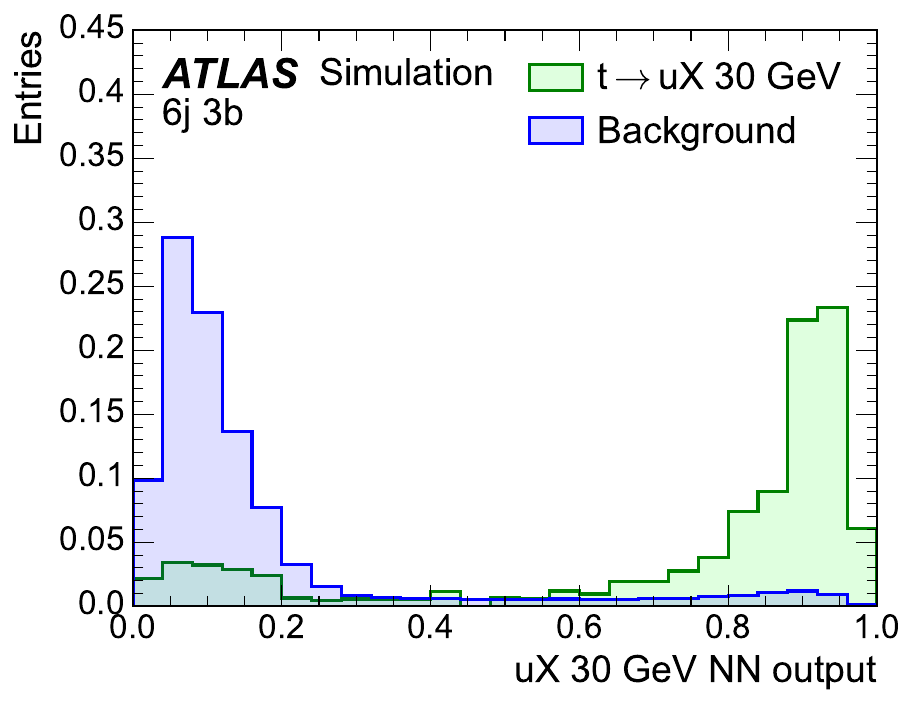}}  \\
\subfloat[\mX = 80 GeV, 4j 3b]{\includegraphics[width = 0.3\textwidth]{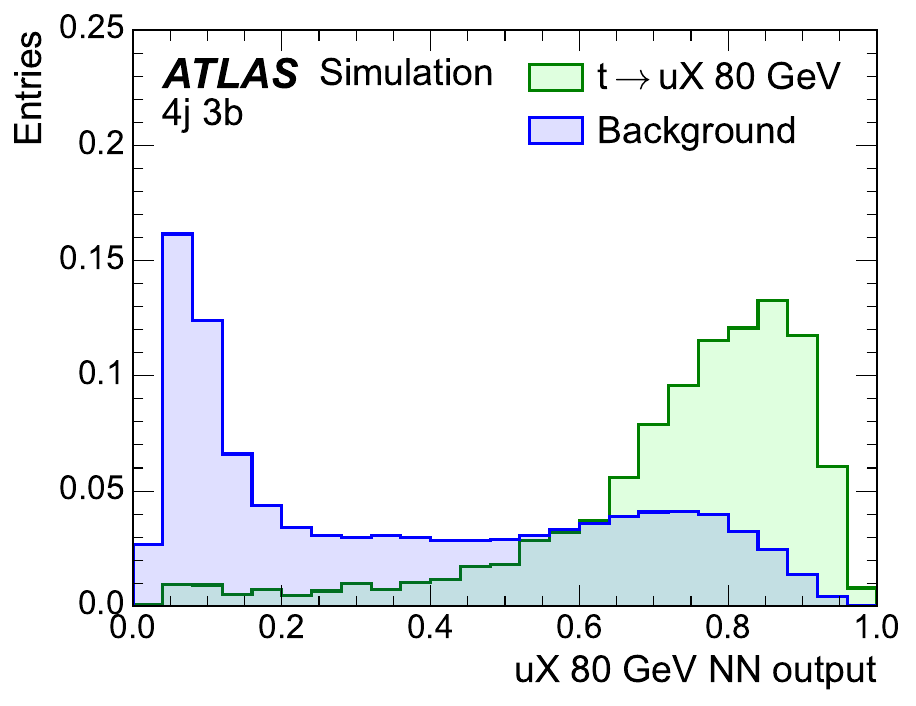}}
\subfloat[\mX = 80 GeV, 5j 3b]{\includegraphics[width = 0.3\textwidth]{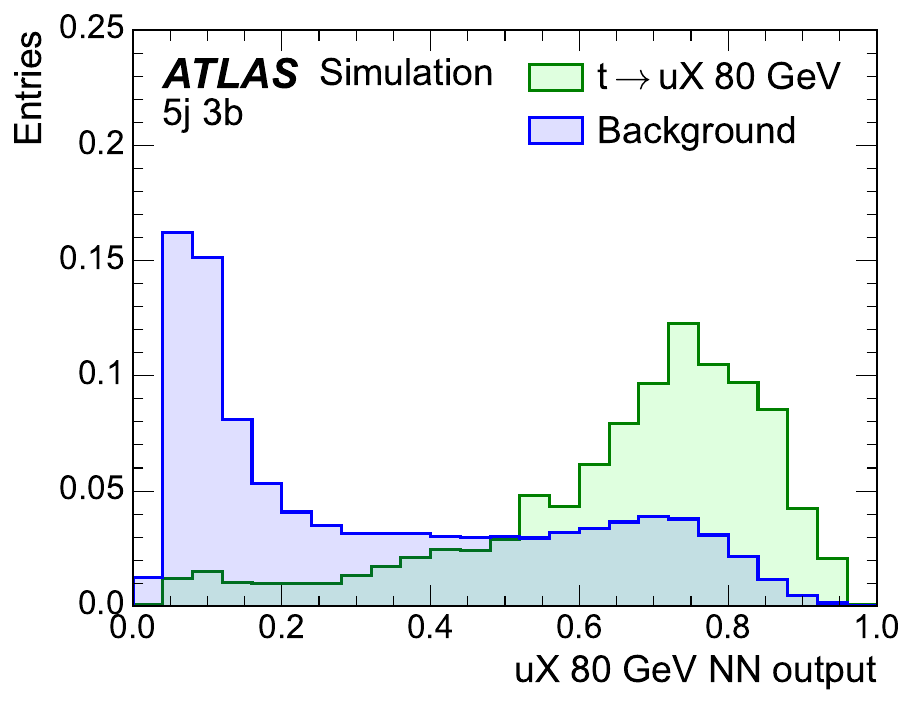}}
\subfloat[\mX = 80 GeV, 6j 3b]{\includegraphics[width = 0.3\textwidth]{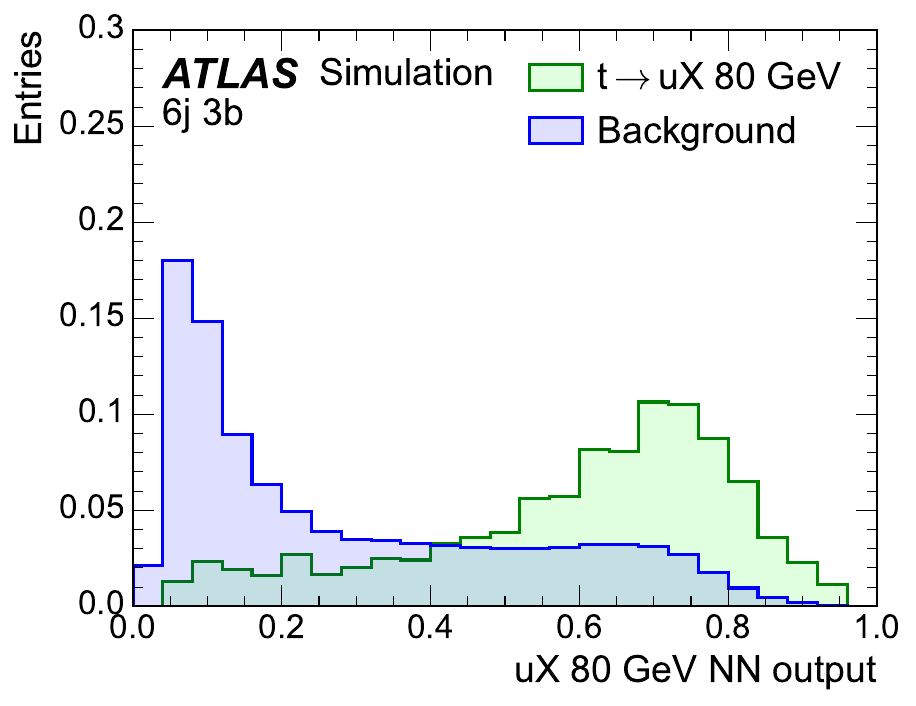}}  \\
\subfloat[\mX = 120 GeV, 4j 3b]{\includegraphics[width = 0.3\textwidth]{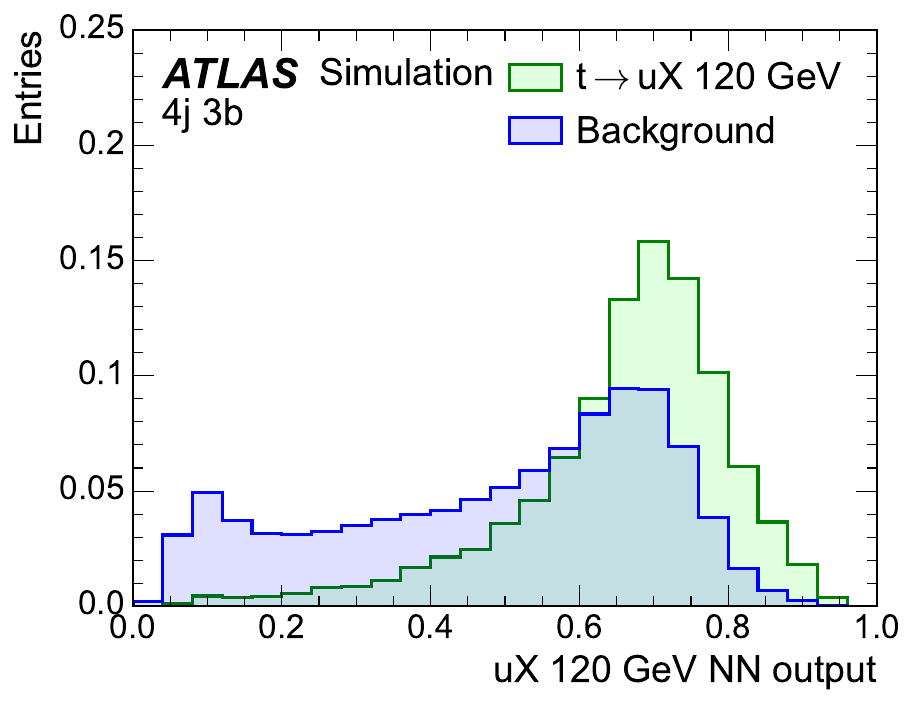}}
\subfloat[\mX = 120 GeV, 5j 3b]{\includegraphics[width = 0.3\textwidth]{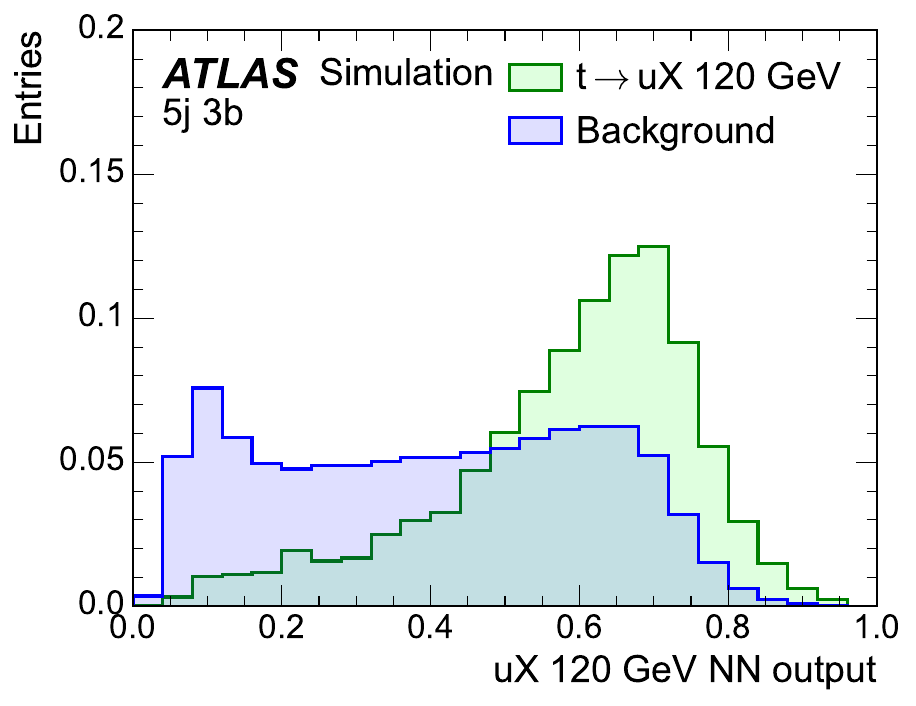}}
\subfloat[\mX = 120 GeV, 6j 3b]{\includegraphics[width = 0.3\textwidth]{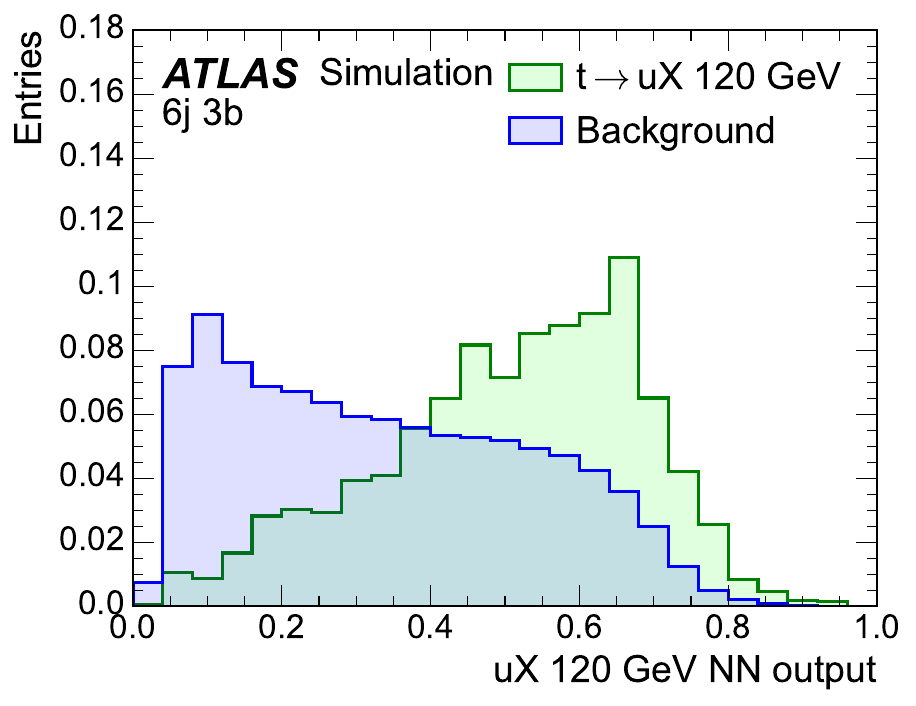}}  \\
\caption{ NN output distributions in the three signal regions for top-quark decays to $uX$ under the 30, 80 and 120 GeV $X$ mass hypotheses.
Background samples are normalised according to their cross-sections.
Signal and background distributions are finally normalised to the same area.}
\label{fig:NNcomparisonuX}
\end{figure}
 
\begin{figure}[htb]
\centering 
\subfloat[\mX = 30 GeV, 4j 3b]{\includegraphics[width = 0.3\textwidth]{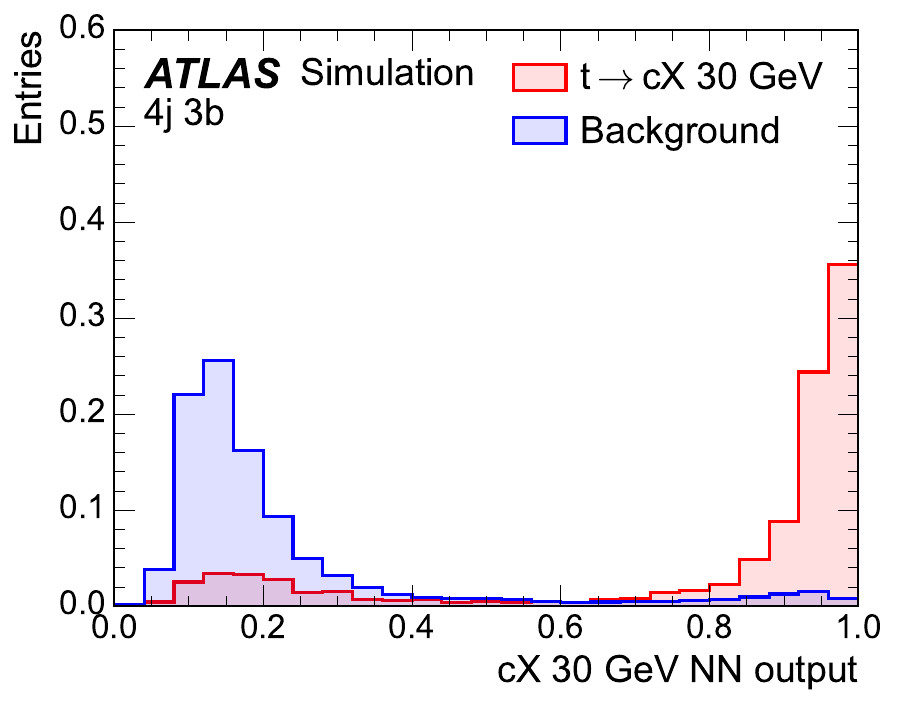}}
\subfloat[\mX = 30 GeV, 5j 3b]{\includegraphics[width = 0.3\textwidth]{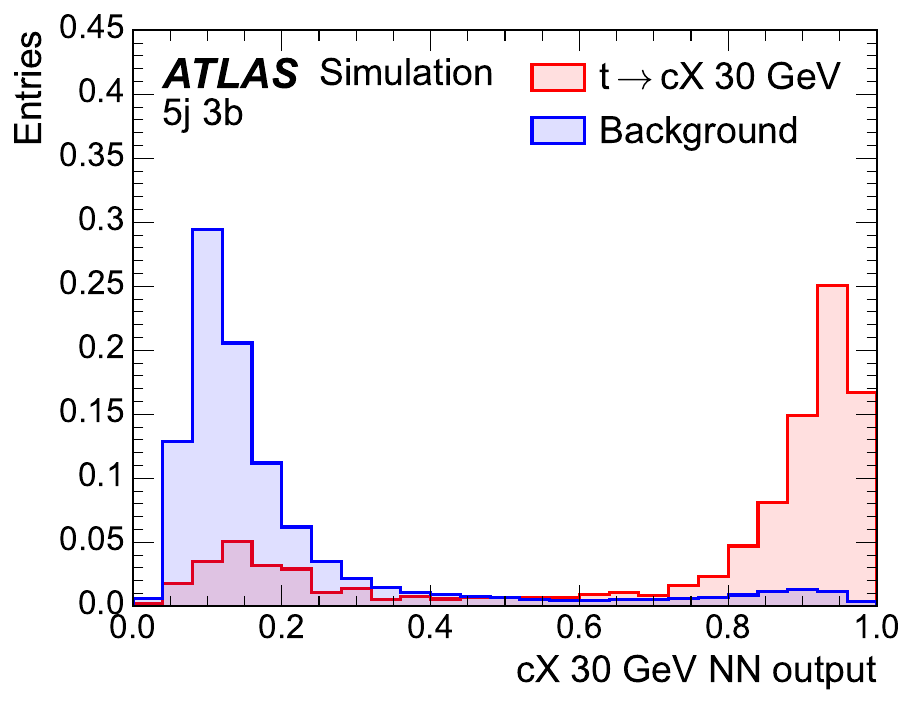}}
\subfloat[\mX = 30 GeV, 6j 3b]{\includegraphics[width = 0.3\textwidth]{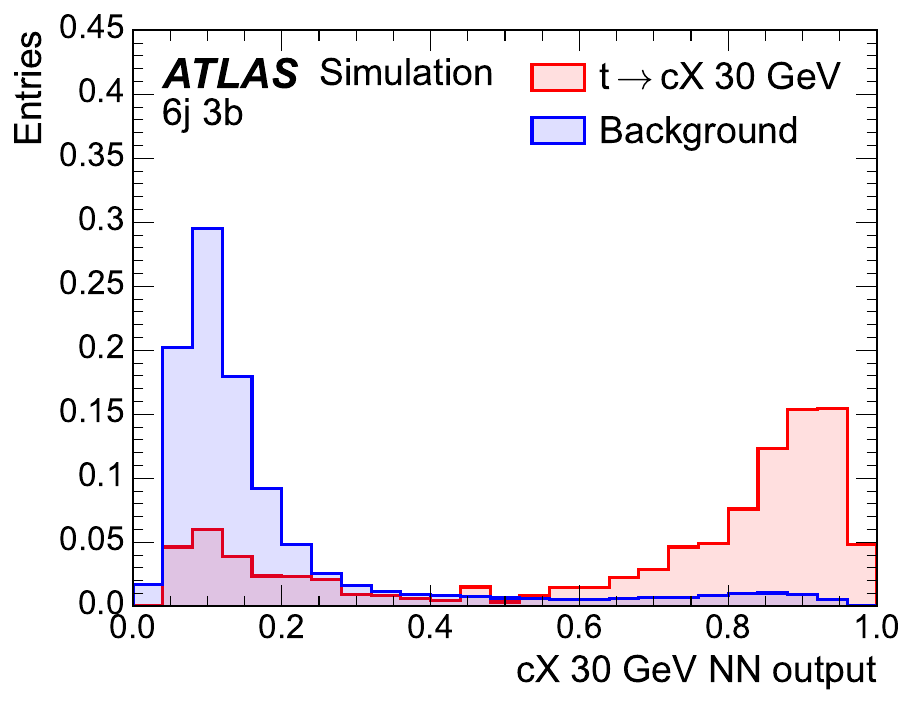}}  \\
\subfloat[\mX = 80 GeV, 4j 3b]{\includegraphics[width = 0.3\textwidth]{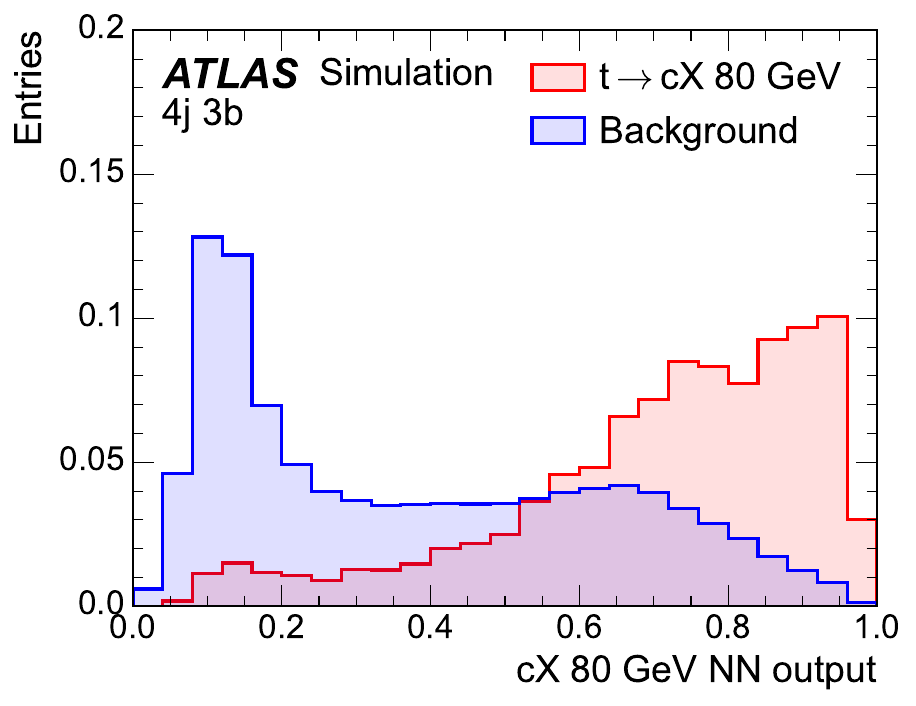}}
\subfloat[\mX = 80 GeV, 5j 3b]{\includegraphics[width = 0.3\textwidth]{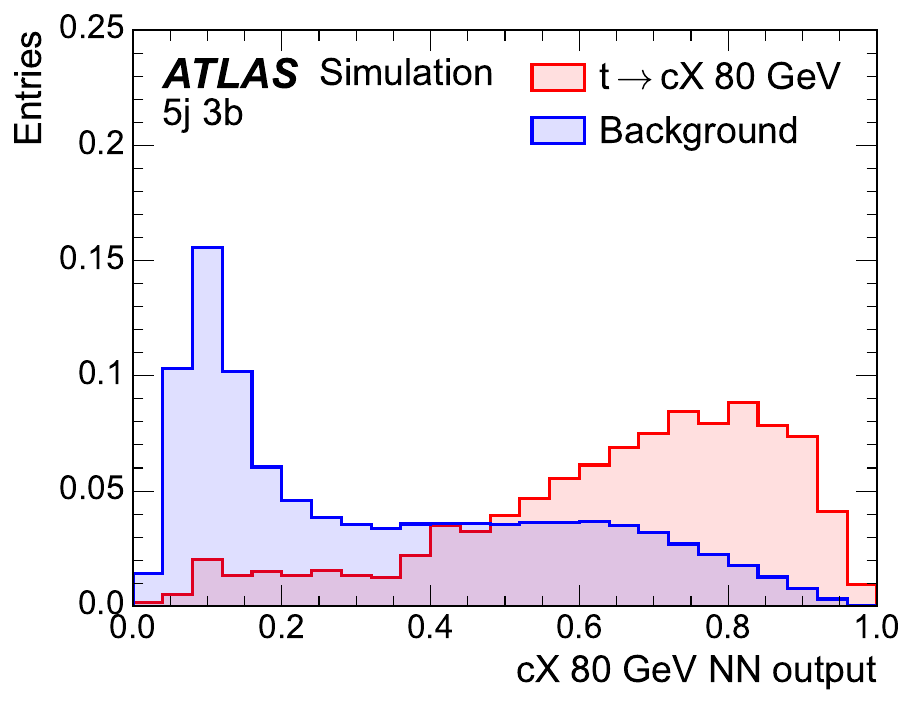}}
\subfloat[\mX = 80 GeV, 6j 3b]{\includegraphics[width = 0.3\textwidth]{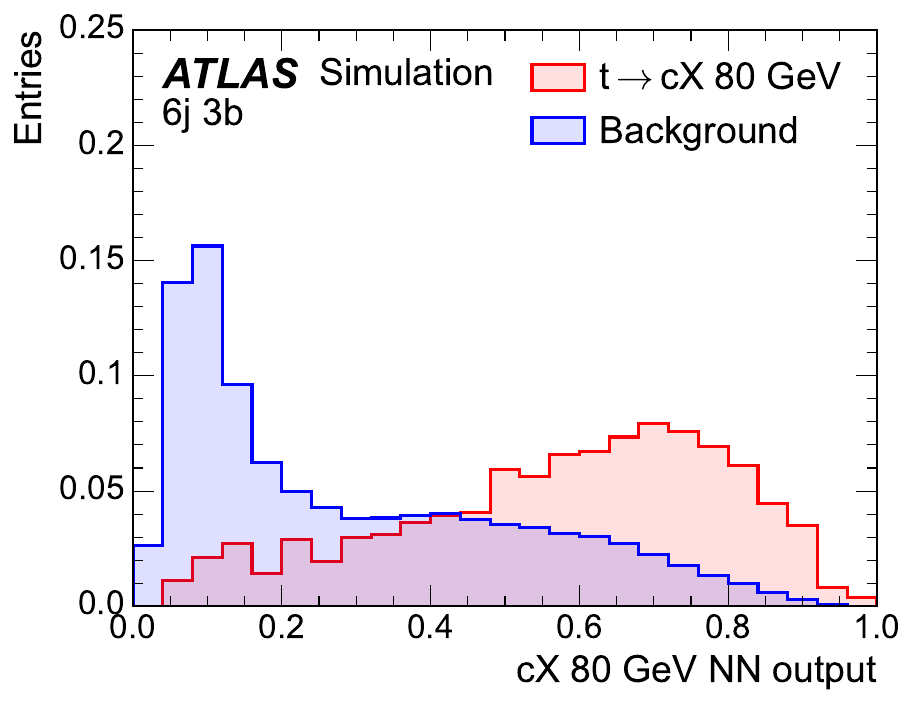}}  \\
\subfloat[\mX = 120 GeV, 4j 3b]{\includegraphics[width = 0.3\textwidth]{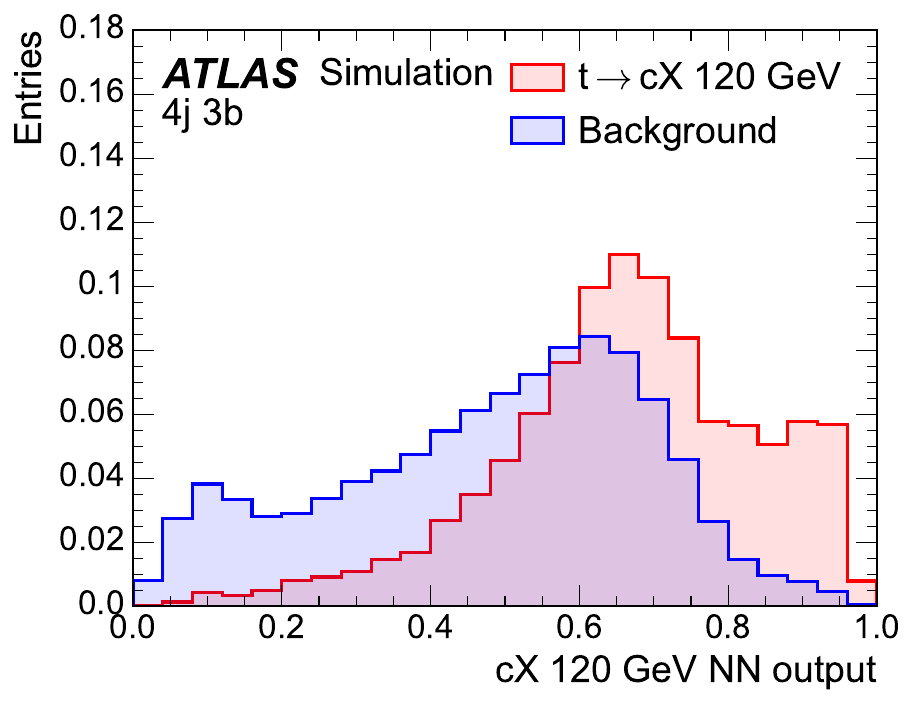}}
\subfloat[\mX = 120 GeV, 5j 3b]{\includegraphics[width = 0.3\textwidth]{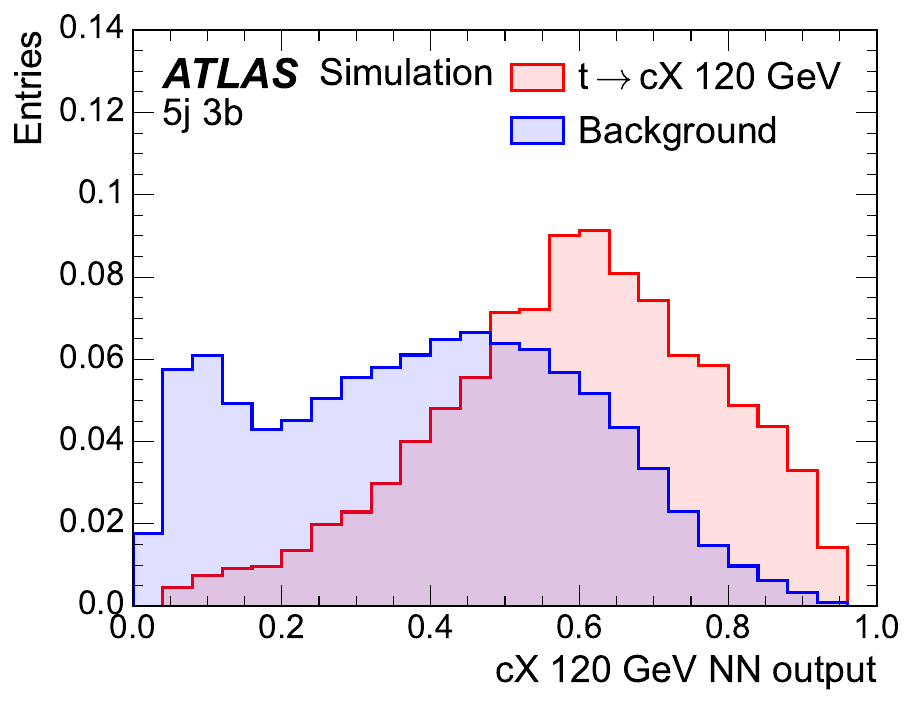}}
\subfloat[\mX = 120 GeV, 6j 3b]{\includegraphics[width = 0.3\textwidth]{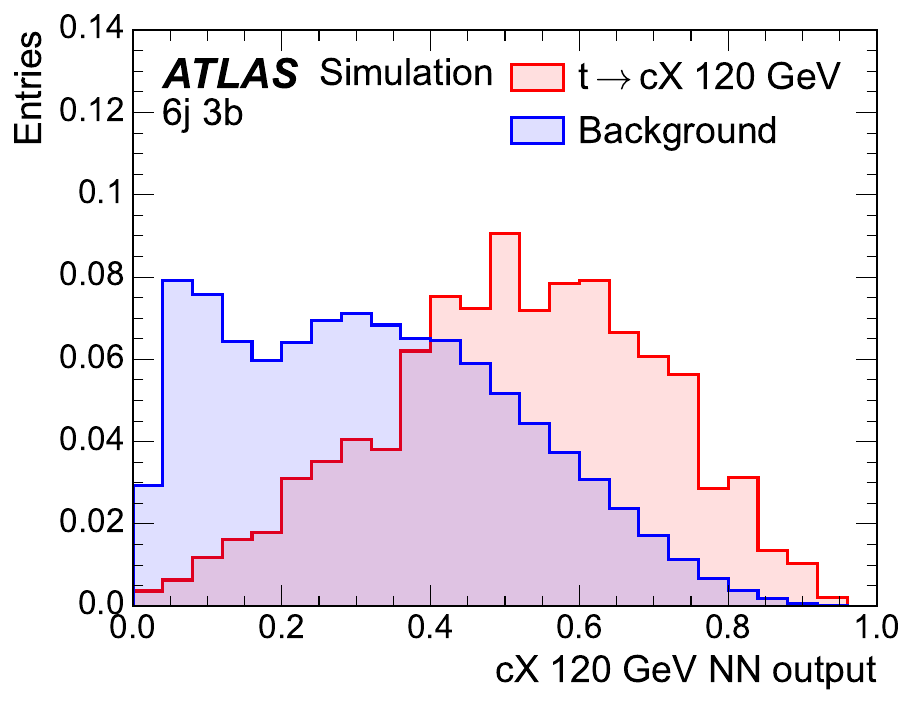}}  \\
\caption{NN output distributions in the three signal regions for top-quark decays to $cX$ under the 30, 80 and 120 GeV $X$ mass hypotheses.
Background samples are normalised according to their cross-sections.
Signal and background distributions are finally normalised to the same area.}
\label{fig:NNcomparisoncX}
\end{figure}
 
\clearpage


\section{Systematic uncertainties}
\label{sec:systematic}


Several sources of uncertainties,
which may affect the normalisation of the signal and backgrounds,
as well as the shape of their corresponding NN outputs,
are considered in this analysis.
Correlations of a given uncertainty are set across processes and event categories as appropriate.
 
Different sources of systematic uncertainties covering potential mismodelling of the \ttbar\ background depending on
jet and $b$-jet multiplicities have been considered.
The uncertainties associated to the \ttbin, \ttcin\ and \ttlight\  processes are treated as uncorrelated,
unless stated otherwise.
Uncertainties associated with the choice of matrix-element generator and parton shower and hadronisation models
are obtained by comparing the nominal \ttbar sample with alternative samples.
These uncertainties are evaluated by deriving data-based corrections for these alternative samples
in a procedure analogous to that used to correct the nominal \ttbar sample (see Section~\ref{sec:analysis}).
The \ttbar modelling uncertainties are thus evaluated by comparing the alternative \ttbar samples to the nominal ones after the corresponding data-based corrections are applied to both sets.
These uncertainties are further decorrelated between different jet multiplicity regions.
The uncertainty due to initial- and final-state radiation (ISR/FSR) is estimated by varying the parameters
of the A14 parton shower tune~\cite{ATL-PHYS-PUB-2017-007} as described in Ref.~\cite{TOPQ-2020-10}.
An uncertainty accounting for missing higher-order QCD corrections in the matrix-element calculation is estimated
by varying the renormalisation ($\mu_R$) and factorisation ($\mu_F$) scales in \textsc{Powheg Box} v2 by factors of 2 and 0.5 relative to the nominal scales.
Uncertainties due to high-order QCD corrections,
ISR and FSR modelling are treated as correlated between different jet multiplicity regions.
The \ttbin\ modelling uncertainty is evaluated by comparing the 5FS nominal sample to the alternative 4FS.
A normalisation uncertainty of 50\% is assumed separately for \ttbin\ and \ttcin.
For \ttbin\ the choice is motivated by the level of agreement between data and prediction
in the control regions for this background before the fit~\cite{TOPQ-2017-12}.
The \ttbin\ normalisation uncertainty is constrained from the fit,
while the \ttcin\ normalisation uncertainty is not.
 
The background originating from \ttbar\ events with a $W$ boson decaying into a charm and a bottom quark
is modelled with dedicated samples of simulated events.
Modelling uncertainties from the choice of the NLO generator as well as parton shower and hadronisation models for this
sub-set of \ttbar\ events are assigned by comparing the nominal prediction with alternative events generated
with \textsc{MG5\_aMC+Pythia8} and \textsc{Powheg+Herwig},
respectively,
as discussed in Section~\ref{sec:MonteCarlo}.
An additional cross-section uncertainty for this process is assigned by combining in quadrature
the 6\% uncertainty in the inclusive \ttbar\ production cross-section with a 3\% uncertainty in the $V_{cb}$  measurements~\cite{ParticleDataGroup:2020ssz}.
 
Systematic uncertainties in the data-based \ttbar\ corrections (Section~\ref{sec:analysis})
arise from the statistical uncertainty in the parametrisation of the correction factors and subtraction
of the non-\ttbar\ backgrounds.
These uncertainties are uncorrelated in each jet multiplicity,
but correlated across \ttbin, \ttcin\ and \ttlight\ background components,
as they cover the modelling of the parton shower.
Table~\ref{tab:syssummaryttbar} summarises the systematic uncertainties affecting the
modelling of the \ttbar~+~jets background.
 
\begin{table}[htbp]
\caption{
Summary of the sources of systematic uncertainty for \ttbarjets\ modelling.
The last column of the table indicates the \ttbarjets\ components to which a systematic uncertainty is assigned and if split in jet multiplicity.
Each contribution of the radiation systematic uncertainty is varied individually.
The NLO generator,
parton shower and hadronisation,
as well as radiation and reweighting uncertainties are also applied to signal samples.
}
\centering
\begin{tabular}{lll}
\hline
\multicolumn{1}{c}{Systematic uncertainty}  &
\multicolumn{1}{c}{Description} &
\multicolumn{1}{c}{Process} \\
\hline
\hline
NLO generator                    & \textsc{Powheg+Pythia8} vs. \textsc{MG5\_aMC+Pythia8} & \ttbar 4j, 5j, 6j \\
Parton shower \& hadronisation   & \textsc{Powheg+Pythia8} vs. \textsc{Powheg+Herwig7} & \ttbar 4j, 5j, 6j \\
Radiation                        & $\mu_R$, $\mu_F$, $h_{\text{damp}}$, ISR and FSR & \ttbar 4j, 5j, 6j \\
\ttbin\ modelling                & 5FS vs 4FS &  \ttbin\ 4j, 5j, 6j \\
\hline
\ttbin\ cross-section            & $\pm$50\%  &  \ttbin  \\
\ttcin\ cross-section            & $\pm$50\%  &  \ttcin  \\
\ttbar\ reweighting          	   & Stat. unc. derived from matrix error &  All \ttbarjets \\
\hline
\end{tabular}
\label{tab:syssummaryttbar}
\end{table}
 
Uncertainties affecting the modelling of the single-top-quark background include a $+5\%/-4\%$ uncertainty in the total cross-section estimated as a weighted average of the theoretical uncertainties in $t$-,
$Wt-$ and $s$-channel productions~\cite{Kidonakis:2010ux,Kidonakis:2011wy,Kidonakis:2010tc}.
Uncertainties associated with the choice of the NLO generator,
parton shower and hadronisation models are evaluated by using alternative samples introduced in Section~\ref{sec:MonteCarlo}.
These uncertainties are treated as uncorrelated among single-top-quark production processes.
An additional systematic uncertainty in the $Wt$-channel process concerning the separation between \ttbar\ and $Wt$ at NLO is assessed by comparing samples generated with the diagram subtraction or the diagram removal schemes.
The uncertainty in the ISR and FSR modelling is estimated with the same procedure used to
evaluate the corresponding source from the \ttbar\ background.
 
Uncertainties affecting the normalisation of the $V$+jets background are estimated for the sum of $W$+jets and $Z$+jets.
The agreement between data and the total background predictions is found to be within approximately 40\%,
which is taken to be the total normalisation uncertainty correlated across all $V$+jets processes.
This uncertainty is justified by variations of the factorisation and renormalisation scales and of the matching parameters in the Sherpa samples,
and by the uncertainty in the extraction from data of the correction factor for the heavy-flavour component~\cite{Ball:2014uwa,Bothmann:2016nao}.
An additional 25\% uncertainty is added in quadrature to the inclusive 40\% uncertainty for each additional jet multiplicity beyond four,
resulting in 47\% and 52\% in regions with five and six jets,
respectively~\cite{Alwall:2007fs}.
 
Uncertainties in the diboson background normalisation include 5\% from the NLO
theory cross-sections~\cite{Campbell:1999ah},
as well as an additional 24\% normalisation uncertainty added in quadrature for each additional inclusive jet-multiplicity bin, based on a comparison among different algorithms for merging LO matrix elements and parton showers~\cite{Alwall:2007fs}.
Therefore, the total uncertainty is 34\%, 42\% and 48\% for events with four,
five, and six jets, respectively.
Recent comparisons between data and \textsc{Sherpa}~2.1.1 for
$WZ\rightarrow \ell'\nu\ell\ell+\geq$4~jets show agreement within the experimental uncertainty of approximately 40\%~\cite{STDM-2015-19}, which further motivates the uncertainties above.
Uncertainties in the $\ttbar V$, $tZ$, \ttH\ and $tH$ cross-sections are
60\%, 60\%, +9/-12\% and 50\%,
respectively,
arising from the uncertainties in their respective NLO theoretical cross-sections~\cite{deFlorian:2016spz,TOPQ-2016-11}.
 
Finally, several normalisation and shape uncertainties are taken into account for the \tqX\ signal.
Since no alternative signal samples are used in the analysis,
the uncertainties from the choice of NLO generator,
parton shower and hadronisation,
and reweighting of the \ttlight\ background are assigned to the signal.
These uncertainties are chosen to be correlated with the \ttlight\ background,
and thus are included in the \ttlight\ modelling category in Tables~\ref{tab:systematicsuX}
and~\ref{tab:systematicscX},
and uncorrelated across jet multiplicities.
 
The uncertainty in the measurement of the integrated luminosity of the Run 2 dataset is 1.7\%~\cite{ATLAS-CONF-2019-021} and affects the overall normalisation of all simulated processes.
The uncertainty is derived using the LUCID-2 detector for the baseline luminosity measurements~\cite{Avoni:2018iuv}
from a calibration of the luminosity scale using $x-y$ beam-separation scans.
An uncertainty is assigned to the modelling of pile-up in simulation to account for differences in predicted and measured inelastic cross-sections in a given fiducial volume~\cite{STDM-2015-05}.
 
Uncertainties associated with electrons and muons arise from the
trigger, reconstruction, identification and isolation efficiencies~\cite{MUON-2018-03,EGAM-2018-01,TRIG-2018-01,TRIG-2018-05}.
These efficiencies are slightly different between data and simulation,
and scale factors derived using tag-and-probe techniques in data enriched in
$Z\rightarrow \ell^+\ell^-$ ($\ell=e,\mu$) events and in
the corresponding MC samples are applied to the simulation to correct for the differences~\cite{PERF-2017-01,PERF-2015-10}.
Additional uncertainties in the lepton momentum scale and resolution are derived using data enriched in
$Z\rightarrow \ell^+\ell^-$ and $J/\psi\rightarrow\ell^+\ell^-$ events~\cite{EGAM-2018-01,PERF-2015-10}.
In total, four independent components are considered for electrons and ten for muons.
The combined effect of all these uncertainties results in an overall normalisation uncertainty in the signal and background of approximately 1\%.
 
Uncertainties in jet measurements arise from the jet energy scale (JES) and resolution,
and from the efficiency to pass the JVT requirements~\cite{PERF-2016-04,PERF-2016-05,PERF-2016-06}.
The largest contribution results from the JES,
the uncertainty of which is split into 29 uncorrelated components,
and depends on jet \pT and $\eta$,
jet flavour, pile-up treatment,
and simulation of the hadronic shower shape.
The JES is calibrated with a series of simulation-based corrections and measurements in data samples enriched in photon,
$Z$-boson, or multi-jet production.
 
The $b$-tagging efficiencies for $b$- and $c$-jets in the simulation are corrected
by \pT-dependent factors to match the efficiencies in data,
whereas in the case of light-jets the efficiency is scaled by \pT- and $\eta$-dependent factors.
The $b$-jet efficiencies and $c$-jet mis-tagging rates are measured in a data sample enriched in \ttbar events~\cite{FTAG-2018-01, ATLAS:2021cxe},
while the light-jet mis-tagging rates are measured in a multi-jet data sample enriched in light jets.
Uncertainties affecting $b$-, $c$-, and light-jet efficiencies or mis-tagging rates are decomposed
into 45, 15 and 20 uncorrelated components,
respectively~\cite{ATLAS-CONF-2018-006}.
 
Uncertainties associated with energy scales and resolutions of leptons and jets are propagated to the \MET reconstruction.
Additional uncertainties affecting the reconstruction of low energy particles present in the event,
not associated with any leptons or jets,
are measured using data enriched in $Z\rightarrow \ell^+\ell^-$ events by studying the recoil of the $Z$-boson~\cite{PERF-2016-07}.


\section{Results}
\label{sec:results}

The presence of a \tqX\ signal is tested by means of a binned maximum-likelihood fit to the data
performed simultaneously in all analysis regions.
Each mass hypothesis and process,
\tuX\ or \tcX,
is tested separately.
The inputs to the fit are the NN output distributions in the three signal regions and
the total yields in the three control regions,
and the likelihood is constructed as a product of Poisson probability terms over all bins considered in the search.
The parameter of interest is the branching fraction of \tqX,
set to a reference value of 0.1\%.
The likelihood function also depends on a set of nuisance parameters that encode the effect of
systematic uncertainties in the signal and background expectations.
All nuisance parameters are subject to Gaussian, log-normal or Poisson constraints in the likelihood.
 
For a given value of $\mu$,
the nuisance parameters $\bm{\theta}$ allow variations of the expected amount of signal and
background according to the corresponding systematic uncertainties,
and their fitted values result in deviations from the nominal expectations that provide the best fit to the data for each mass hypothesis and quark flavour.
This procedure allows a reduction of the impact of systematic uncertainties on the search sensitivity
by taking advantage of the highly populated background-dominated bins included in the likelihood fit.
Statistical uncertainties in each bin of the predicted NN output are taken into account by dedicated parameters in the fit.
The best-fit branching fraction is obtained by performing a binned likelihood fit to the data under the signal-plus-background hypothesis,
maximising the likelihood function $\cal L (\mu, \bm{\theta)}$ over $\mu$ and $\bm{\theta}$.
 
The test statistic $q_{\mu}$ is defined as the profile likelihood ratio,
$q_{\mu} = -2 \log ({\cal L} (\mu, \bm{\hat{\theta}} _{\mu})/ {\cal L}(\hat{\mu}, \bm{\hat{\theta}}) )$,
where $\hat{\mu}$ and $\bm{\hat{\theta}}$
are the values of the parameters that maximise the likelihood function
(subject to the constraint $0 \le \hat{\mu} \le \mu$),
and $\bm{\hat{\theta}_{\mu}}$ are the values of the nuisance parameters that maximise the likelihood function
for a given value of $\mu$.
A related test statistic is used to determine whether the observed data is compatible with
the background-only hypothesis by setting $\mu = 0$ in the profile likelihood ratio and leaving
$\hat{\mu}$ unconstrained:
$q_{0} = -2 \log ({\cal L} (0, \bm{\hat{\theta}_{0}})/ {\cal L}(\hat{\mu}, \bm{\hat{\theta}}))$.
The $p$-value (referred to as $p_0$)
representing the level of agreement between the data and the background-only hypothesis,
is estimated by integrating the distribution of $q_0$ based on the asymptotic formulae in Ref.~\cite{Cowan:2010js},
above the observed value of $q_0$ in the data.
Upper limits on $\mu$,
and thus on the branching fraction \BR,
are derived by using $q_\mu$ in the $\mathrm{CL_s}$  method~\cite{Junk:1999kv,Read:2002hq}.
For a given signal scenario,
values of \BR\ yielding $\mathrm{CL_s} < 0.05$,
where $\mathrm{CL_s}$ is computed using the asymptotic approximation,
are excluded at $\le 95$\% confidence level (CL).
 
There are about 200 nuisance parameters considered in the fit,
and this number varies slightly across the range of mass hypotheses.
The \ttbin\ normalisation, modelling, parton shower and hadronisation uncertainties are the most constrained by the fit.
A summary of the systematic uncertainties
with similar sources grouped together is given for $m_X$ values of 30, 80 and 120~\GeV\ in Tables~\ref{tab:systematicsuX} and~\ref{tab:systematicscX} for the \tuX\ and \tcX\ processes,
respectively.
Depending on the process and the $m_X$ hypothesis,
the total systematic uncertainty is dominated by \ttbar\ modelling,
jet energy scale and resolution,
or jet tagging uncertainties.
 
\begin{table}[htb]
\caption{
Summary of the statistical and systematic uncertainties in $\mu = \BR(\tuX)$
shown for an $X$ signal with a mass of 30, 80 or 120~\GeV,
extracted from the fit to the data.
A pre-fit value of $\mu = 0.1\%$ is assumed for all $X$ mass hypotheses.
Due to correlations between the different sources of uncertainty,
the total systematic uncertainty in $\mu$ can be different from the sum in quadrature of the individual sources.
}
\begin{center}
\begin{tabular}{l c c c}
\hline
Uncertainty source   & $\Delta\mu(uX_{30})$ & $\Delta\mu(uX_{80})$ & $\Delta\mu(uX_{120})$ \\
\hline \hline
\ttbin\ modelling	                       &	0.040	&	0.060	&	0.098	\\
\ttcin\ modelling	                       &	0.033	&	0.055	&	0.091	\\
\ttlight\ modelling         	           &	0.034	&	0.058	&	0.040	\\
 
\ttbin\ normalisation                  	   &	0.012	&	0.011	&	0.039	\\
\ttcin\ normalisation	                   &	0.017	&	0.036	&	0.087	\\
 
$W\rightarrow cb$ modelling	               &	0.001	&	0.010   &	0.017	\\
 
\ttbar\ reweighting                        &	0.005	&	0.013	&	0.017	\\
 
Other backgrounds                          &	0.008	&	0.026	&	0.023	\\
 
Luminosity, JVT, pile-up            	   &	0.002	&	0.006	&   0.012 	\\
 
Lepton trigger, identification, isolation  &	0.001	&	0.004	&	0.007	\\
Jet energy scale and resolution	           &	0.008	&	0.037	&	0.040	\\
 
$b$-tagging efficiency for $b$-jets  	   &	0.007	&	0.008	&	0.041	\\
$b$-tagging efficiency for $c$-jets        &	0.014	&	0.027	&	0.079	\\
$b$-tagging efficiency for light jets      &	0.007	&	0.008	&	0.010	\\
 
\MET	                                   &	0.002	&	0.010	&	0.011	\\
\hline
Total systematic uncertainty        	   &	0.077	&	0.125	&	0.220	\\
\hline
Signal statistical uncertainty	           &	0.014	&	0.009	&	0.007	\\
\hline
Total statistical uncertainty	           &	0.064	&	0.070	&	0.065	\\
\hline \hline
Total uncertainty           	           &	0.098	&	0.141	&	0.230	\\
\hline
\end{tabular}
\end{center}
\label{tab:systematicsuX}
\end{table}
 
\begin{table}[htb]
\caption{
Summary of the statistical and systematic uncertainties in $\mu = \BR(\tcX)$
shown for an $X$ signal with a mass of 30, 80 or 120~\GeV,
extracted from the fit to the data.
A value of $\mu = 0.1\%$ is assumed for all $X$ mass hypotheses.
Due to correlations between the different sources of uncertainty,
the total systematic uncertainty in $\mu$ can be different from the sum in quadrature of the individual sources.
}
\begin{center}
\begin{tabular}{l c c c}
\hline
Uncertainty source   & $\Delta\mu(cX_{30})$ & $\Delta\mu(cX_{80})$ & $\Delta\mu(cX_{120})$ \\
\hline \hline
 
\ttbin\ modelling	                       &	  0.034	&	0.074	&	0.079	\\
\ttcin\ modelling	                       &	  0.010	&	0.012	&	0.040	\\
\ttlight\ modelling	                       &	  0.008	&	0.049	&	0.038	\\
 
\ttbin\ normalisation       	           &	  0.026	&	0.038	&	0.001	\\
\ttcin\ normalisation	                   &	  0.019	&	0.048	&	0.013	\\
 
$W \rightarrow cb$ modelling               &	0.001	&	0.020	&	0.015	\\
 
\ttbar\ reweighting                        &  	0.005	&	0.013	&	0.019	\\
 
Other backgrounds            	           &	0.009	&	0.057	&	0.047	\\

Luminosity, JVT, pile-up	               &	0.005	&	0.005	&	0.003	\\
 
Lepton trigger, identification, isolation  &	0.001	&	0.004 	&	0.003  \\
 
Jet energy scale and resolution	           &	0.017	&	0.049	&	0.051	\\
$b$-tagging efficiency for $b$-jets        &	0.003	&	0.016	&	0.023	\\
$b$-tagging efficiency for $c$-jets	       &	0.010	&	0.038	&	0.091	\\
$b$-tagging efficiency for light jets      &	0.009	&	0.065	&	0.125	\\
 
\MET	                                   &	0.001	&	0.003	&	0.008	\\
 
\hline
Total systematic uncertainty	           &	0.056	&	0.150	&	0.208	\\
\hline
Signal statistical uncertainty             &	0.017	&	0.012	&	0.008	\\
\hline
Total statistical uncertainty	           &	0.064	&	0.067	&	0.058	\\
\hline \hline
Total uncertainty	                       &	0.079	&	0.162	&	0.217	\\
\hline
\end{tabular}
\end{center}
\label{tab:systematicscX}
\end{table}

Table~\ref{tab:postfitYields30} shows the event yields after the background-plus-signal fit under the \tuX\ and \tcX\ hypotheses with the $X$ scalar mass of 30~GeV.
Figures~\ref{fig:NNfit30} to~\ref{fig:NNfit120} show the post-fit distributions
of the NN output in the 3b regions and the yields in the $\geq$4b regions
for the 30, 80 and 120 GeV $m_X$ hypotheses and the two signal processes.
The binning of the NN distributions is optimised for each signal,
so as to maximize the signal over the background per bin,
while keeping a sufficient fraction of events in each bin.
After the fit,
good agreement between the data and simulation is found in the input variables of the NN.

\begin{table}[!htb]
\caption{
Event yields of the signal and SM background processes in the six analysis regions
after the fit to the data under the \tuX\ (top) and \tcX\ (bottom) hypotheses assuming $\mX = 30$~GeV.
Total includes signal and background.
The quoted uncertainties take into account correlations and constraints of the nuisance parameters
and include both the statistical and systematic uncertainties.
Negative correlations between the \ttlight, \ttbin\ and \ttcin\ modelling uncertainties
can make the uncertainty in the total yields smaller than in the individual components.
}
\begin{center}
\begin{tabular}{l *{6}{r@{}c@{}l}}
\hline
\multicolumn{19}{c}{ \tuX, $\mX = 30$~\GeV}  \\
\hline
\hline
&& \makebox[0pt]{4j 3b} &&& \makebox[0pt]{4j 4b} &&& \makebox[0pt]{5j 3b} &&& \makebox[0pt]{5j $\geq$4b} &&& \makebox[0pt]{6j 3b} &&& \makebox[0pt]{6j $\geq$4b}   \\
\hline
\ttlight   & 9300 &$\,\pm\, $&900 & 4.0 &$\,\pm\, $&2.4 & 6200 &$\,\pm\, $&900 & 7 &$\,\pm\, $&5 & 2700 &$\,\pm\, $&500 & 5 &$\,\pm\, $&4 \\
\ttbin   & 11200 &$\,\pm\, $&1000 & 319 &$\,\pm\, $&22 & 15400 &$\,\pm\, $&1200 & 980 &$\,\pm\, $&50 & 12000 &$\,\pm\, $&900 & 1250 &$\,\pm\, $&60 \\
\ttcin   & 3400 &$\,\pm\, $&1100 & 12 &$\,\pm\, $&7 & 4200 &$\,\pm\, $&1300 & 33 &$\,\pm\, $&11 & 2900 &$\,\pm\, $&900 & 29 &$\,\pm\, $&10 \\
$W\rightarrow cb$   & 380 &$\,\pm\, $&60 & 8.1 &$\,\pm\, $&1.2 & 270 &$\,\pm\, $&50 & 11.4 &$\,\pm\, $&1.9 & 132 &$\,\pm\, $&22 & 7.4 &$\,\pm\, $&2.4 \\
Single-$t$   & 1200 &$\,\pm\, $&400 & 19 &$\,\pm\, $&11 & 1100 &$\,\pm\, $&400 & 49 &$\,\pm\, $&22 & 640 &$\,\pm\, $&280 & 60 &$\,\pm\, $&40 \\
$\ttbar H$   & 106 &$\,\pm\, $&14 & 6.6 &$\,\pm\, $&1.0 & 273 &$\,\pm\, $&32 & 45 &$\,\pm\, $&7 & 309 &$\,\pm\, $&35 & 75 &$\,\pm\, $&10 \\
$\ttbar V$   & 120 &$\,\pm\, $&80 & 7 &$\,\pm\, $&5 & 190 &$\,\pm\, $&120 & 25 &$\,\pm\, $&15 & 190 &$\,\pm\, $&120 & 33 &$\,\pm\, $&21 \\
$VV$, $V$+jets   & 870 &$\,\pm\, $&290 & 16 &$\,\pm\, $&5 & 770 &$\,\pm\, $&40 & 28.9 &$\,\pm\, $&3.0 & 459 &$\,\pm\, $&32 & 27.5 &$\,\pm\, $&3.2 \\
\hline
Signal   & 10 &$\,\pm\, $&40 & 0.02 &$\,\pm\, $&0.08 & 8 &$\,\pm\, $&33 & 0.2 &$\,\pm\, $&0.8 & 4 &$\,\pm\, $&16 & 0.1 &$\,\pm\, $&0.6 \\
\hline
Total  & 26580 &$\,\pm\, $&170 & 392 &$\,\pm\, $&17 & 28410 &$\,\pm\, $&180 & 1176 &$\,\pm\, $&33 & 19300 &$\,\pm\, $&150 & 1490 &$\,\pm\, $&40 \\
\hline
\hline
Data                    & & \makebox[0pt]{26614} &&& \makebox[0pt]{374} &&& \makebox[0pt]{28394} &&& \makebox[0pt]{1179} &&& \makebox[0pt]{19302} &&& \makebox[0pt]{1492} \\
\hline
\noalign{\vskip 1cm}
\hline
\multicolumn{19}{c}{ \tcX, $\mX = 30$~\GeV}  \\
\hline
\hline
&& \makebox[0pt]{4j 3b} &&& \makebox[0pt]{4j 4b} &&& \makebox[0pt]{5j 3b} &&& \makebox[0pt]{5j $\geq$4b} &&& \makebox[0pt]{6j 3b} &&& \makebox[0pt]{6j $\geq$4b}   \\
\hline
\ttlight   & 10200 &$\,\pm\,$&1200 & 6.4 &$\,\pm\,$&3.3 & 6700 &$\,\pm\,$&1000 & 10 &$\,\pm\,$&6 & 3000 &$\,\pm\,$&600 & 7 &$\,\pm\,$&5 \\
\ttbin   & 9800 &$\,\pm\,$&1200 & 284 &$\,\pm\,$&25 & 14500 &$\,\pm\,$&1300 & 970 &$\,\pm\,$&50 & 11400 &$\,\pm\,$&1000 & 1250 &$\,\pm\,$&60 \\
\ttcin   & 3900 &$\,\pm\,$&1300 & 17 &$\,\pm\,$&10 & 4600 &$\,\pm\,$&1400 & 41 &$\,\pm\,$&14 & 3300 &$\,\pm\,$&1100 & 35 &$\,\pm\,$&12 \\
$W\rightarrow cb$   & 400 &$\,\pm\,$&60 & 8.7 &$\,\pm\,$&1.2 & 280 &$\,\pm\,$&50 & 12.3 &$\,\pm\,$&2.1 & 134 &$\,\pm\,$&23 & 7.9 &$\,\pm\,$&2.6 \\
Single-$t$   & 1200 &$\,\pm\,$&400 & 25 &$\,\pm\,$&15 & 1100 &$\,\pm\,$&400 & 43 &$\,\pm\,$&19 & 550 &$\,\pm\,$&230 & 51 &$\,\pm\,$&31 \\
$\ttbar H$   & 109 &$\,\pm\,$&14 & 6.9 &$\,\pm\,$&1.0 & 280 &$\,\pm\,$&33 & 46 &$\,\pm\,$&7 & 316 &$\,\pm\,$&35 & 78 &$\,\pm\,$&11 \\
$\ttbar V$   & 140 &$\,\pm\,$&80 & 8 &$\,\pm\,$&5 & 220 &$\,\pm\,$&120 & 28 &$\,\pm\,$&16 & 220 &$\,\pm\,$&120 & 38 &$\,\pm\,$&22 \\
$VV$, $V$+jets   & 810 &$\,\pm\,$&260 & 16 &$\,\pm\,$&5 & 730 &$\,\pm\,$&50 & 27.0 &$\,\pm\,$&3.0 & 425 &$\,\pm\,$&34 & 25.0 &$\,\pm\,$&3.0 \\
\hline
Signal   & 20 &$\,\pm\,$&40 & 0.5 &$\,\pm\,$&1.2 & 14 &$\,\pm\,$&31 & 0.5 &$\,\pm\,$&1.1 & 7 &$\,\pm\,$&15 & 0.4 &$\,\pm\,$&1.0 \\
\hline
Total  & 26600 &$\,\pm\, $&180 & 373 &$\,\pm\, $&18 & 28400 &$\,\pm\, $&190 & 1183 &$\,\pm\, $&34 & 19310 &$\,\pm\, $&150 & 1490 &$\,\pm\,$&40 \\
\hline
\hline
Data                    & & \makebox[0pt]{26614} &&& \makebox[0pt]{374} &&& \makebox[0pt]{28394} &&& \makebox[0pt]{1179} &&& \makebox[0pt]{19302} &&& \makebox[0pt]{1492} \\
\hline
\end{tabular}
\end{center}
\label{tab:postfitYields30}
\end{table}

\begin{figure}[htb]
\centering
\subfloat[\tuX, 4j 3b]{\includegraphics[width = 0.30\textwidth]{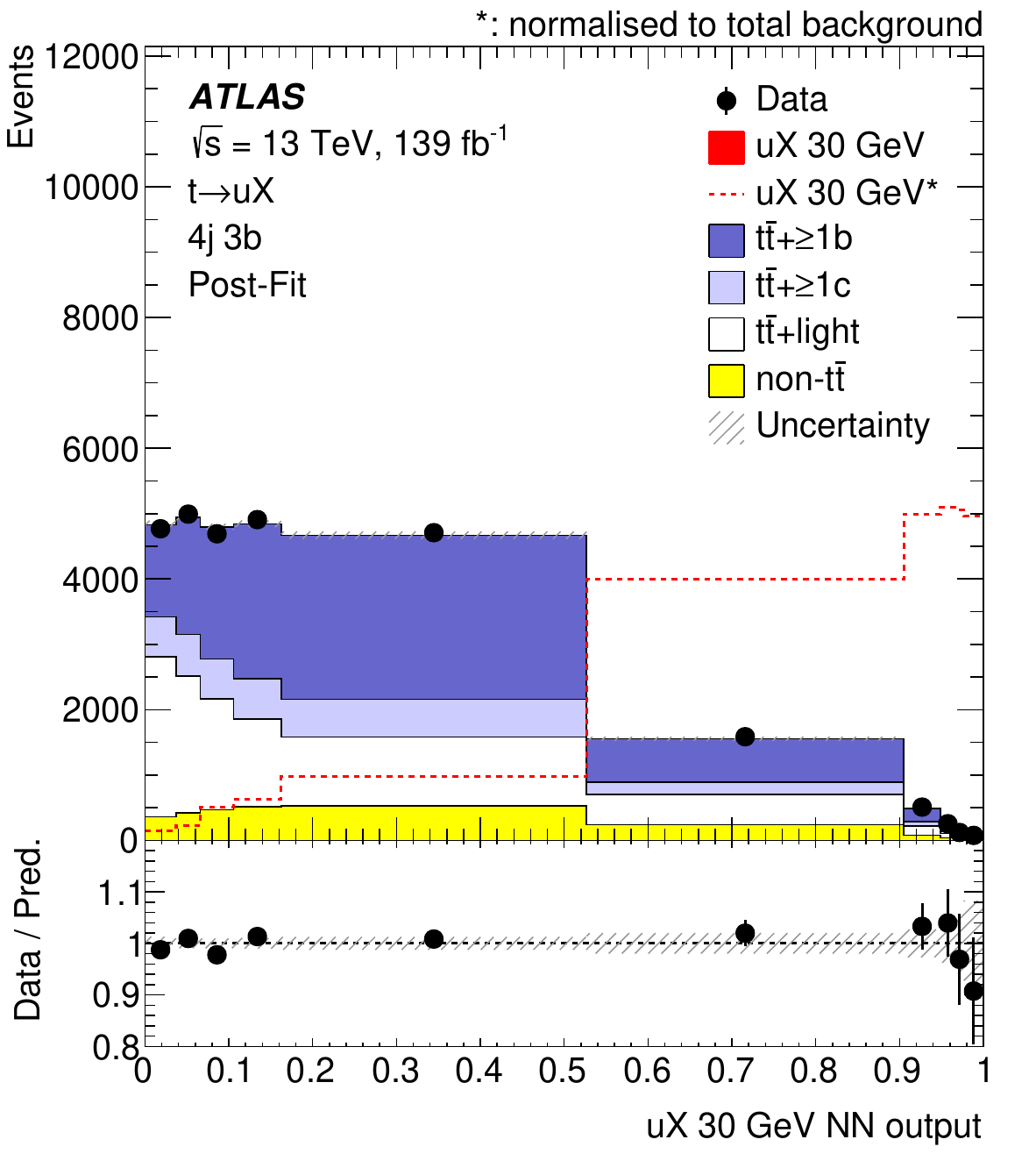}}
\subfloat[\tuX, 5j 3b]{\includegraphics[width = 0.30\textwidth]{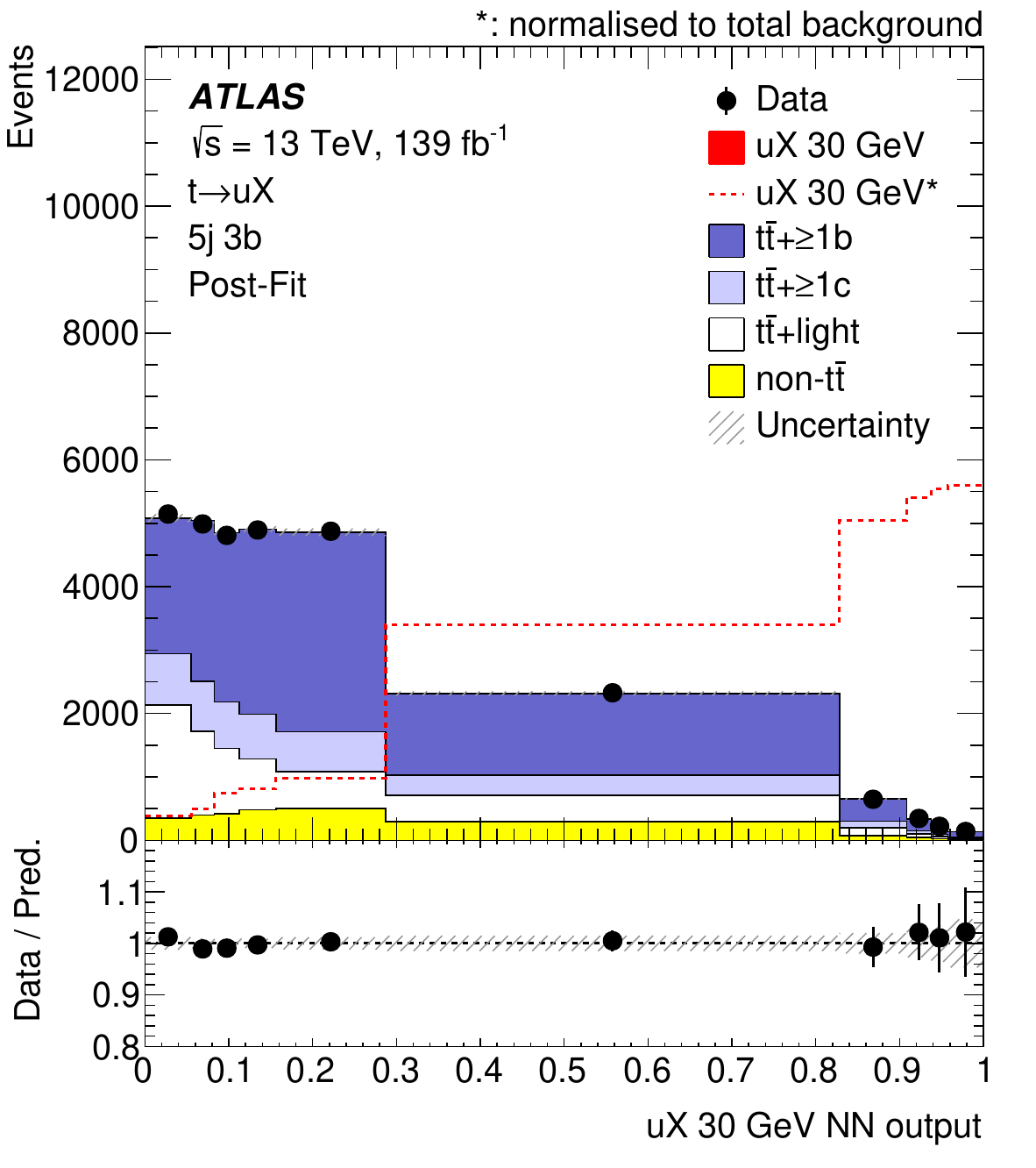}}
\subfloat[\tuX, 6j 3b]{\includegraphics[width = 0.30\textwidth]{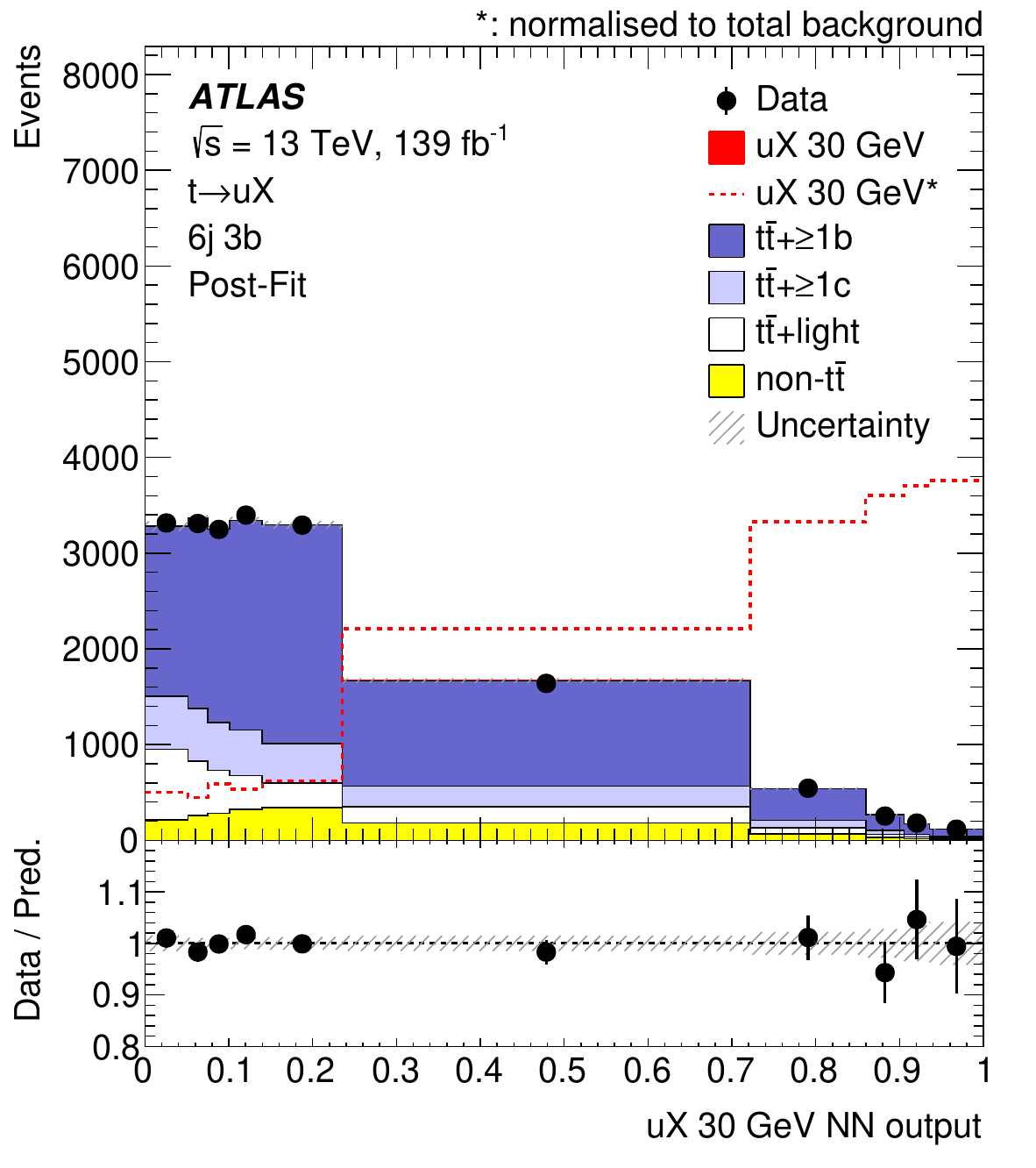}} \\
\subfloat[\tcX, 4j 3b]{\includegraphics[width = 0.30\textwidth]{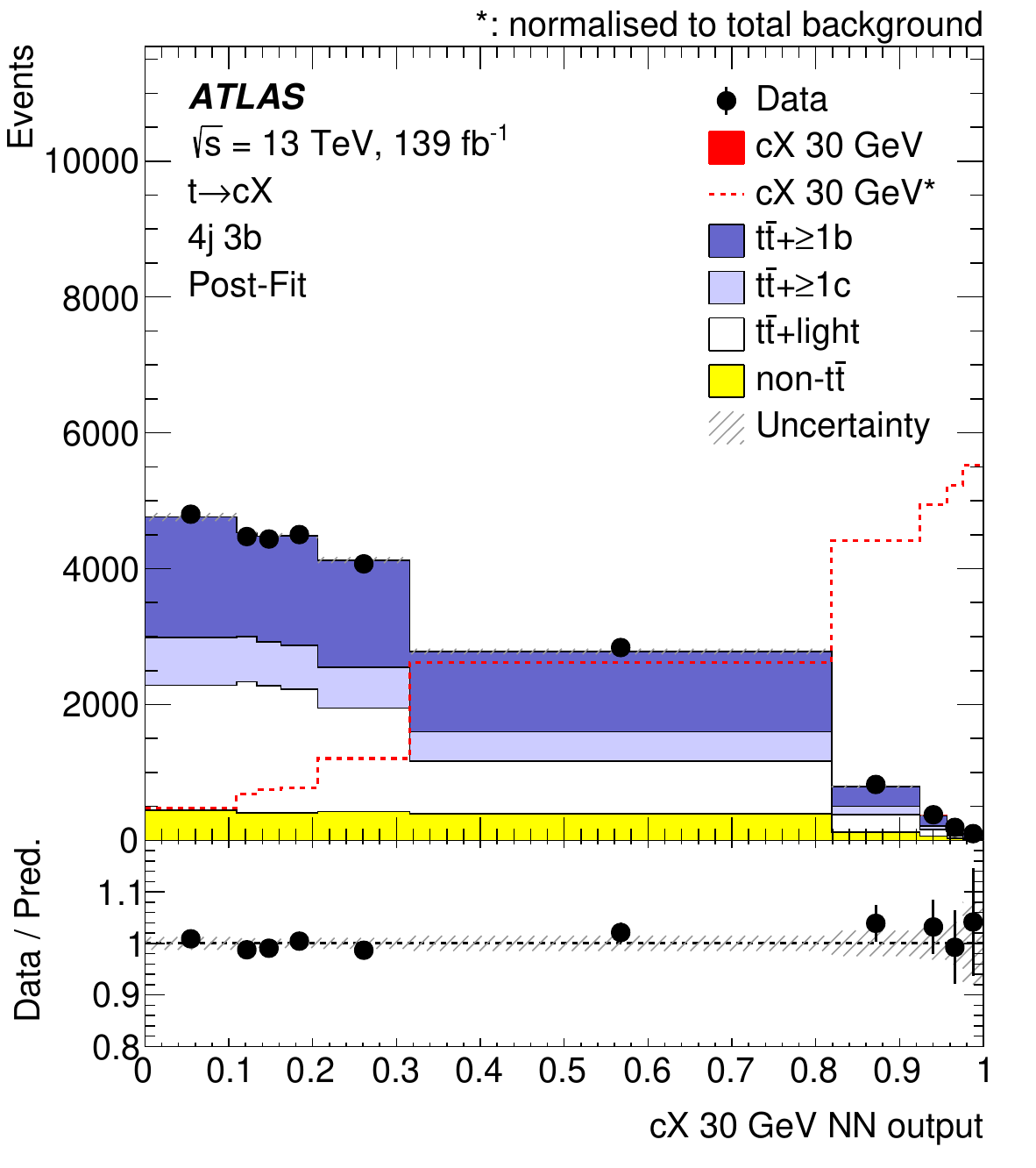}}
\subfloat[\tcX, 5j 3b]{\includegraphics[width = 0.30\textwidth]{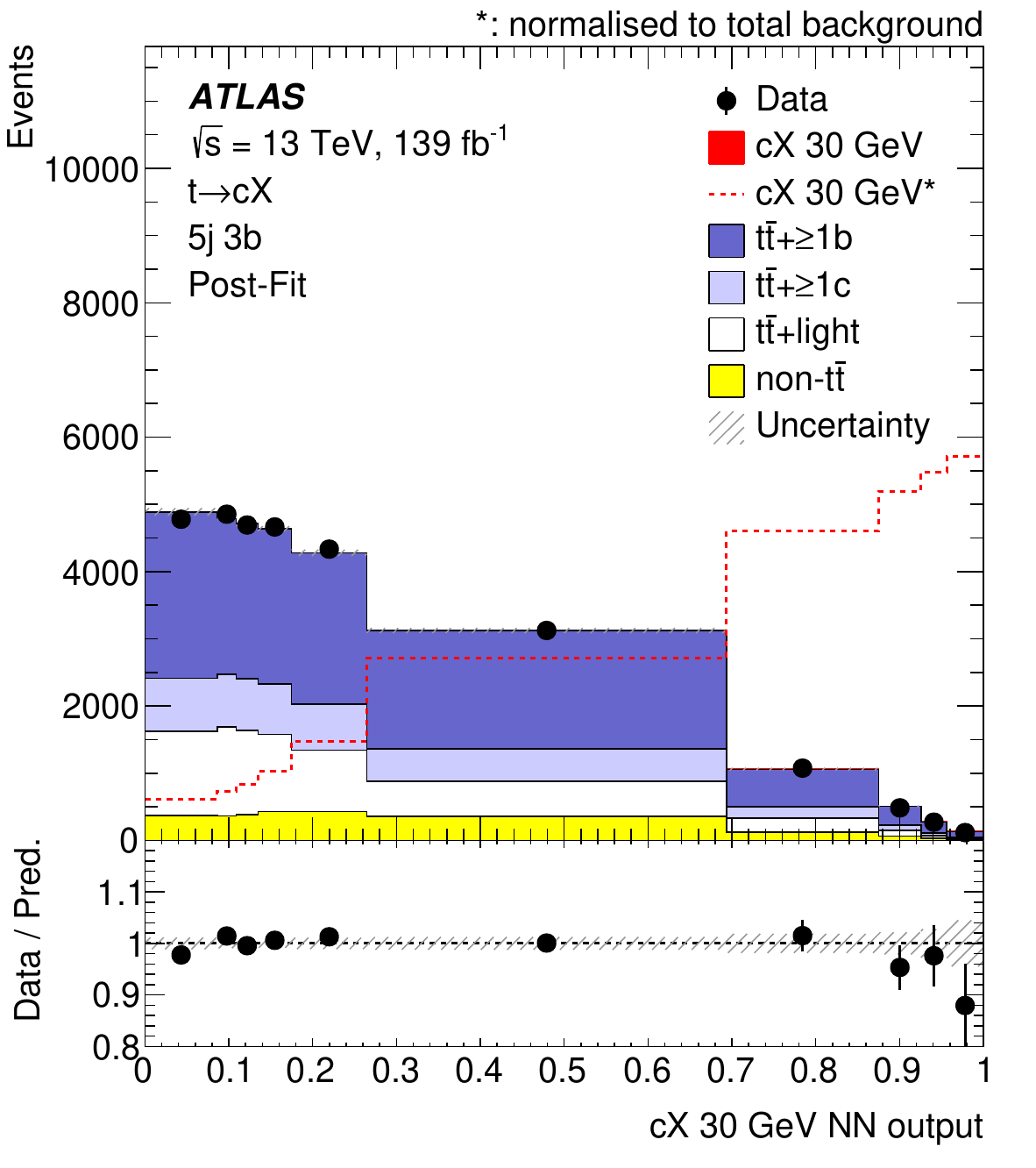}}
\subfloat[\tcX, 6j 3b]{\includegraphics[width = 0.30\textwidth]{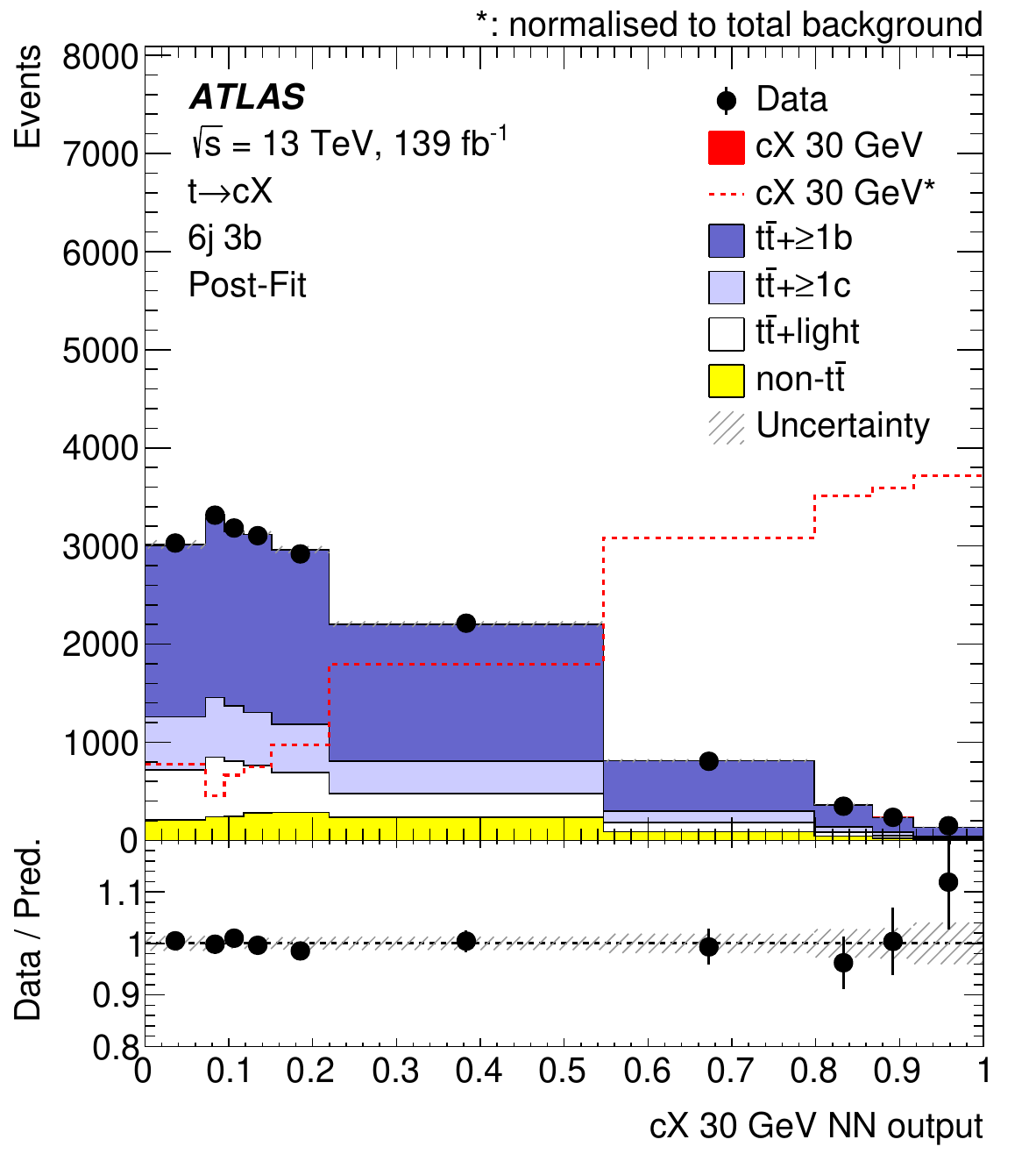}} \\
\subfloat[\tuX, $\ge 4$b]{\includegraphics[width = 0.30\textwidth]{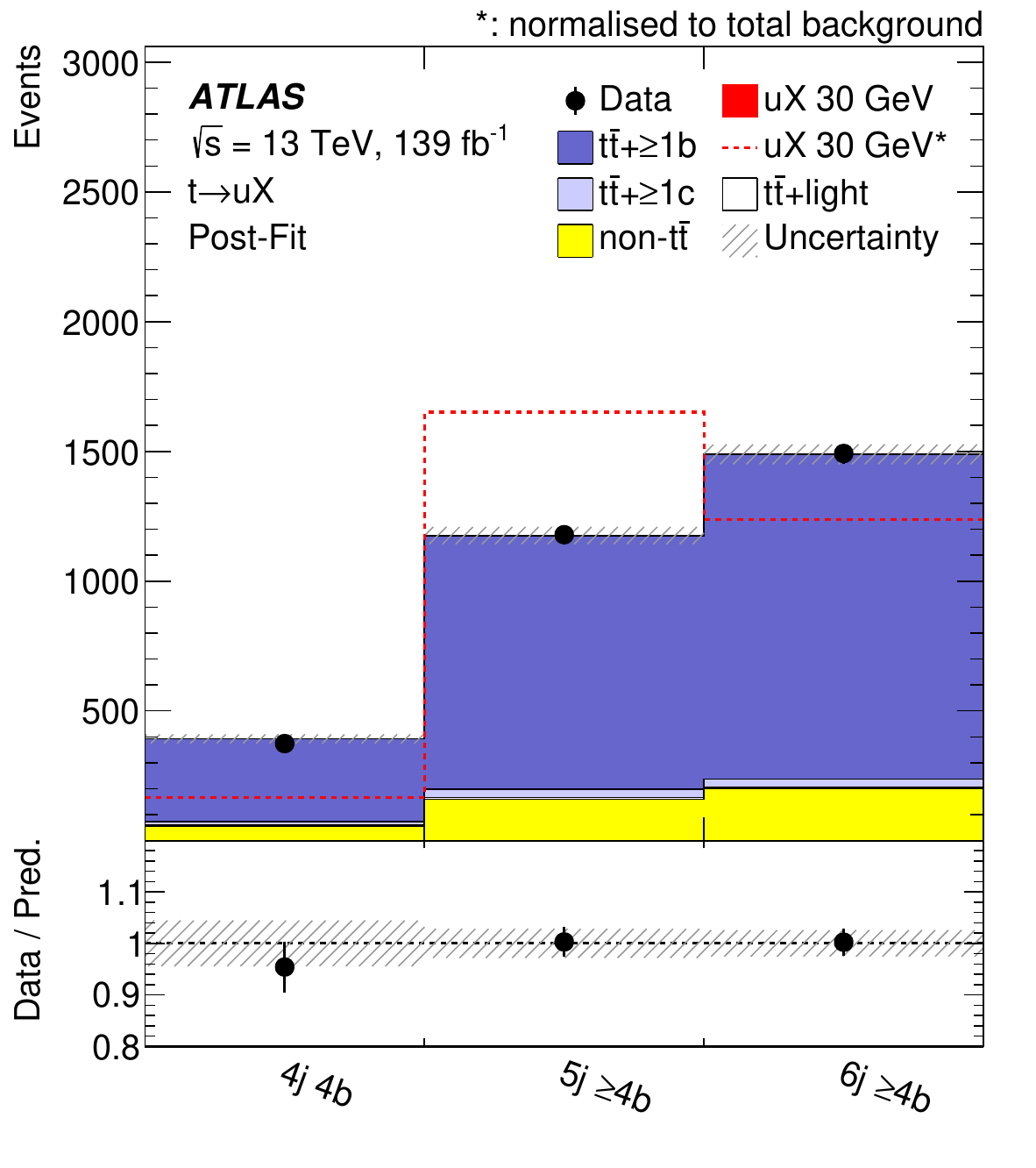}}
\subfloat[\tcX, $\ge 4$b]{\includegraphics[width = 0.30\textwidth]{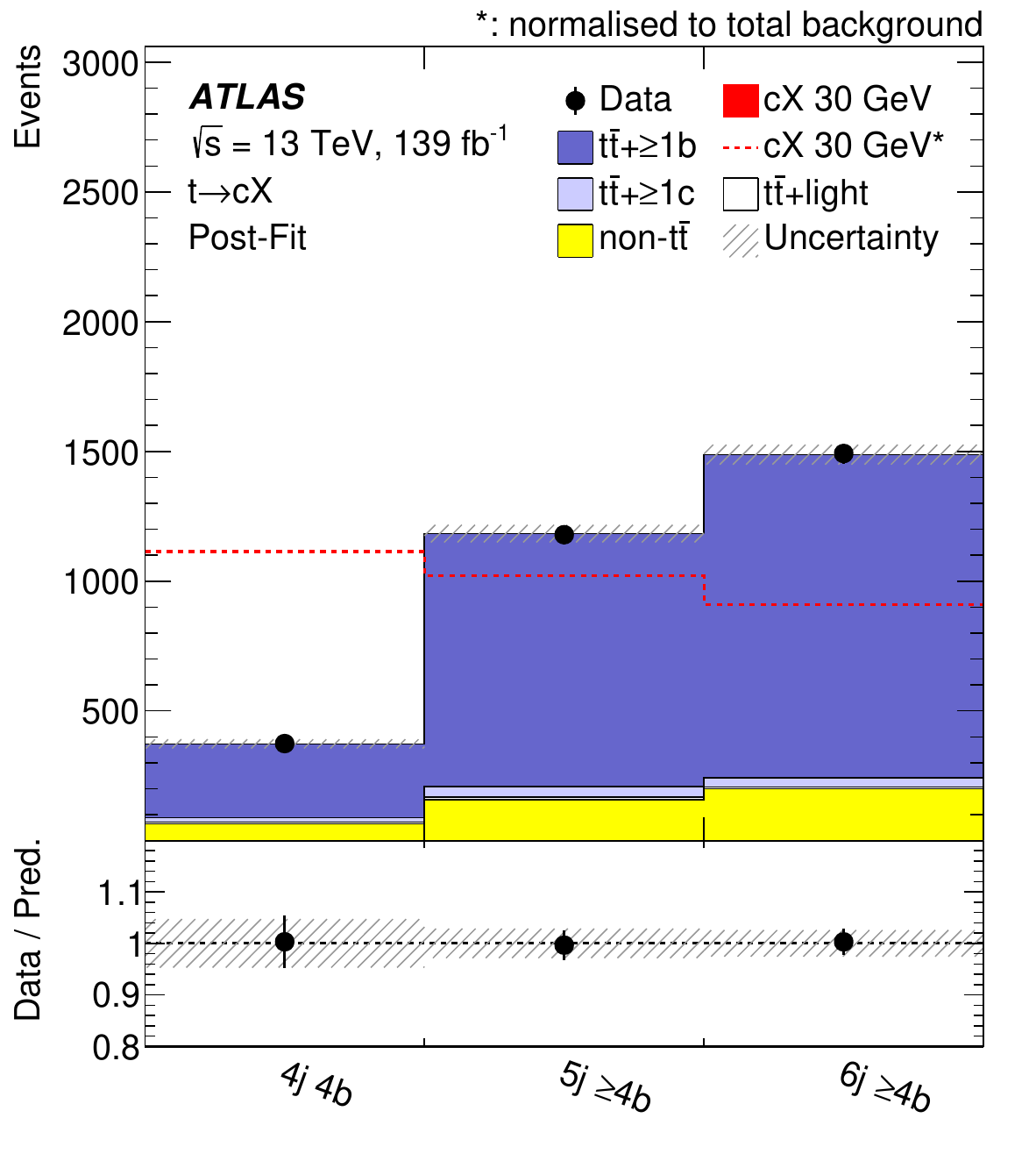}}
\caption{Comparison between the data and prediction for the NN output in the 3b regions for the \tuX\ ((a) to (c))
and the \tcX\ ((d) to (f)) processes,
and the yields in the $\ge 4$b regions for the \tuX\ (g) and the \tcX\ (h) processes
after the signal-plus-background fit to data for the 30 GeV $X$ scalar mass hypothesis.
The dashed red line corresponds to a $m_X = 30$~GeV signal normalised to the total background.
The uncertainty bands show the total uncertainty after the fit.}
\label{fig:NNfit30}
\end{figure}
 
\begin{figure}[htb]
\centering
\subfloat[\tuX, 4j 3b]{\includegraphics[width = 0.30\textwidth]{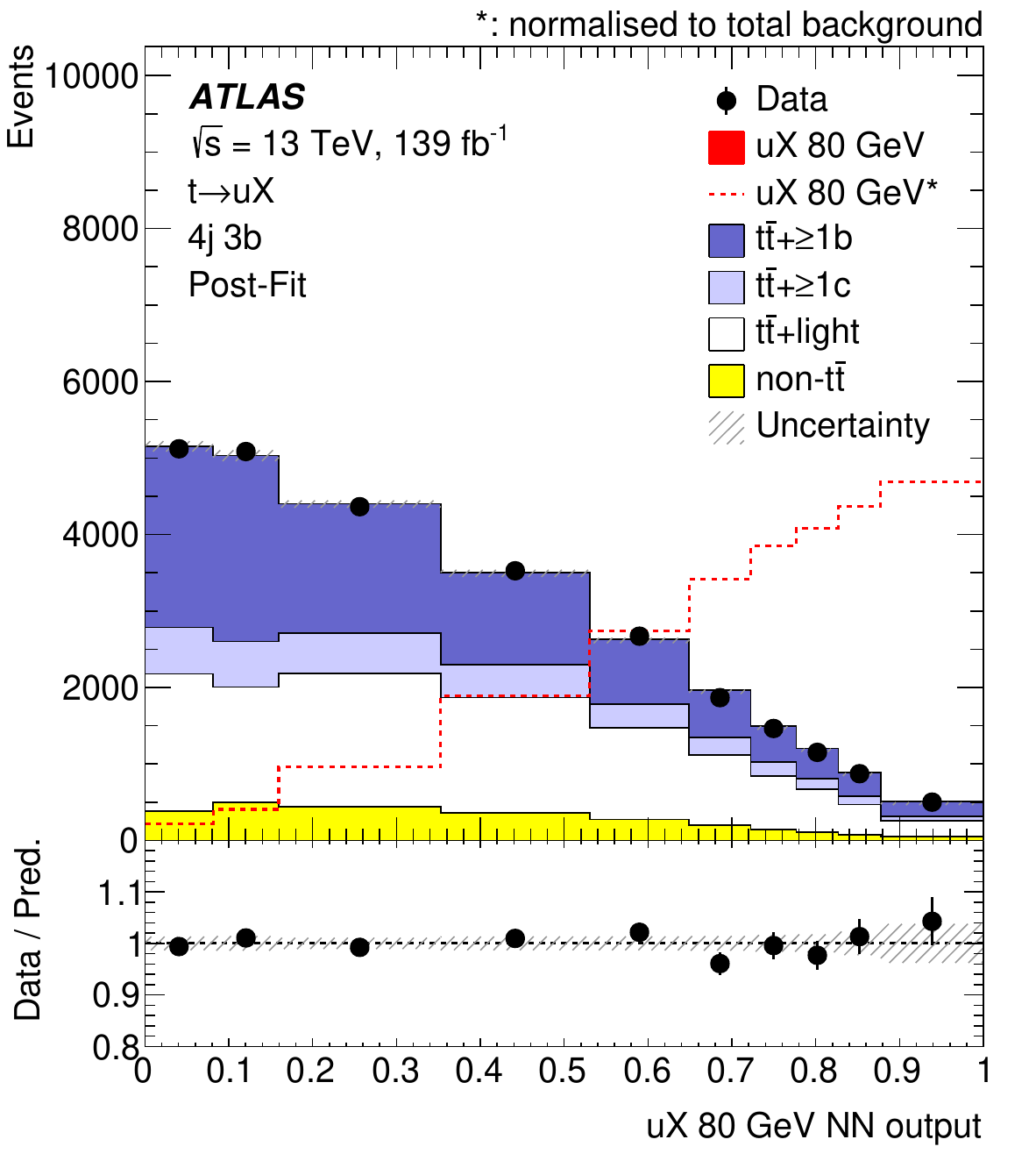}}
\subfloat[\tuX, 5j 3b]{\includegraphics[width = 0.30\textwidth]{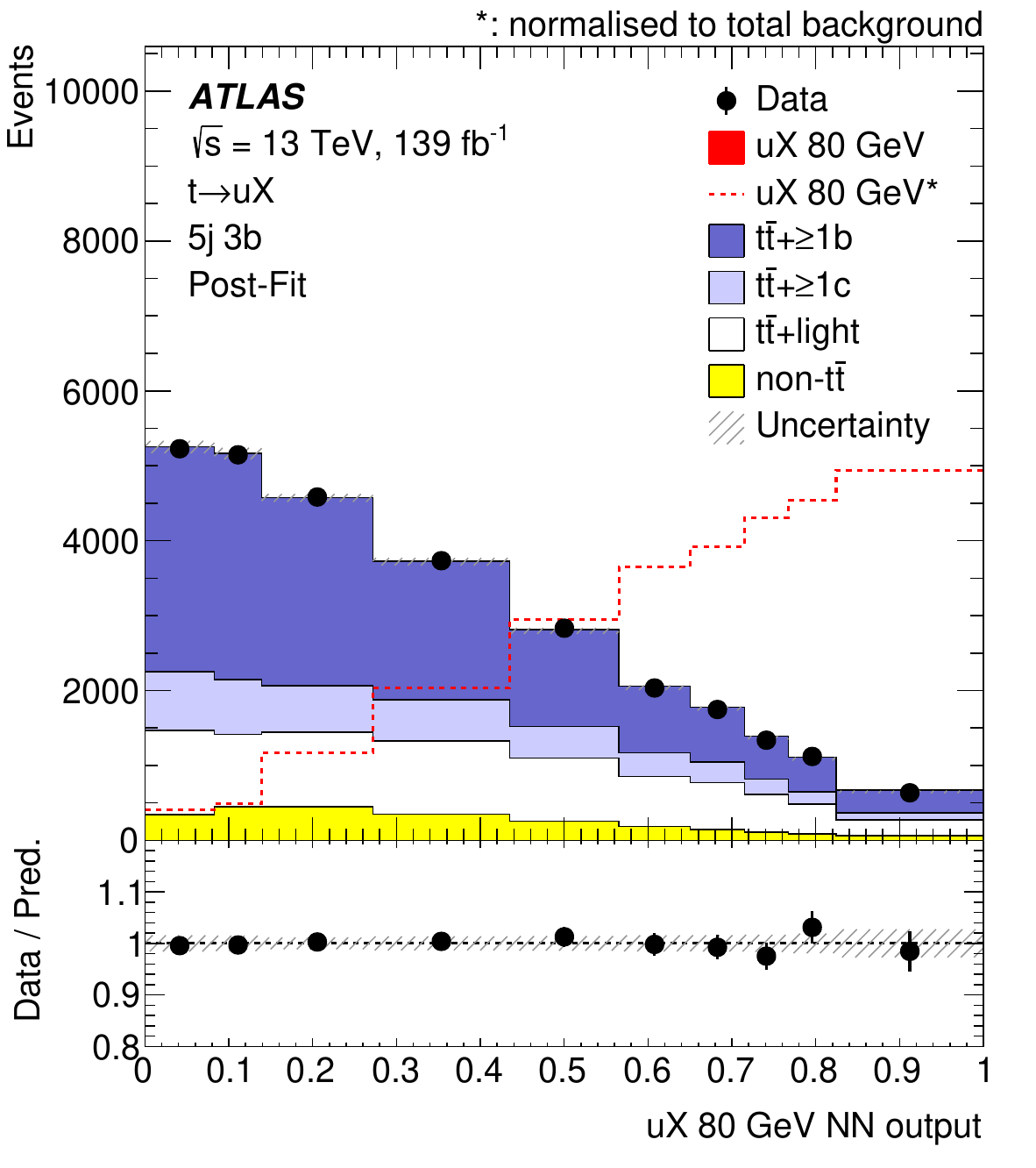}}
\subfloat[\tuX, 6j 3b]{\includegraphics[width = 0.30\textwidth]{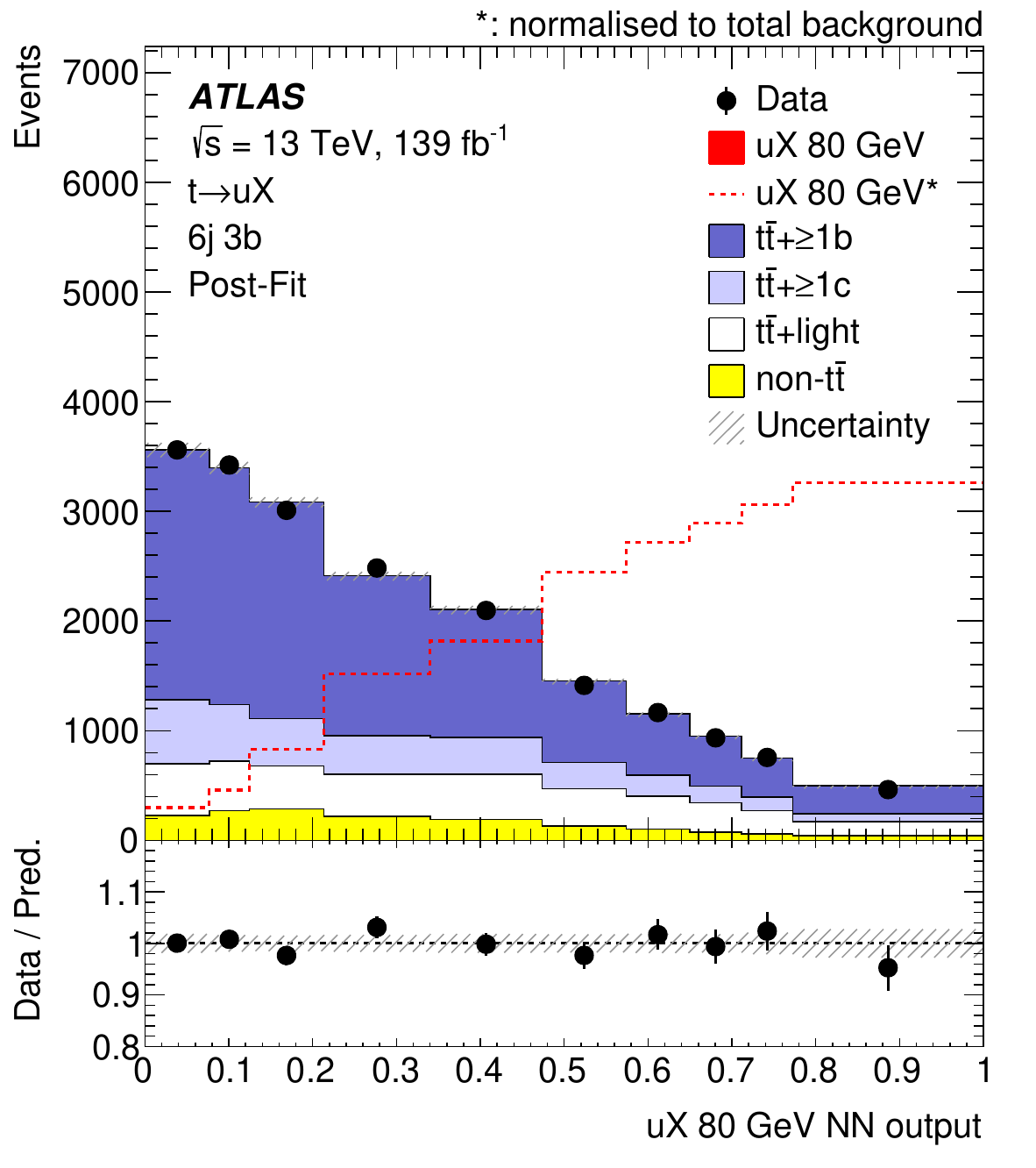}} \\
\subfloat[\tcX, 4j 3b]{\includegraphics[width = 0.30\textwidth]{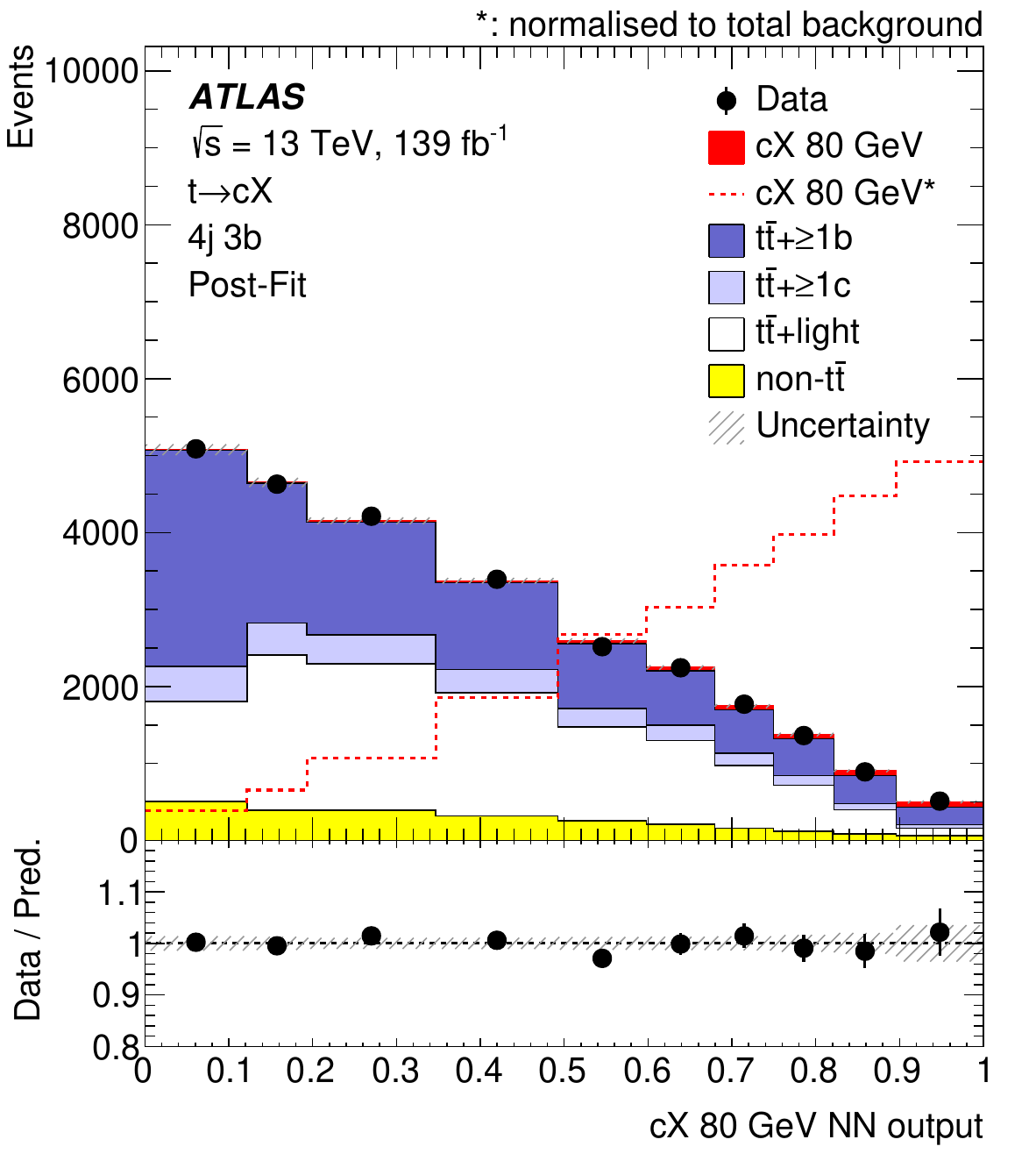}}
\subfloat[\tcX, 5j 3b]{\includegraphics[width = 0.30\textwidth]{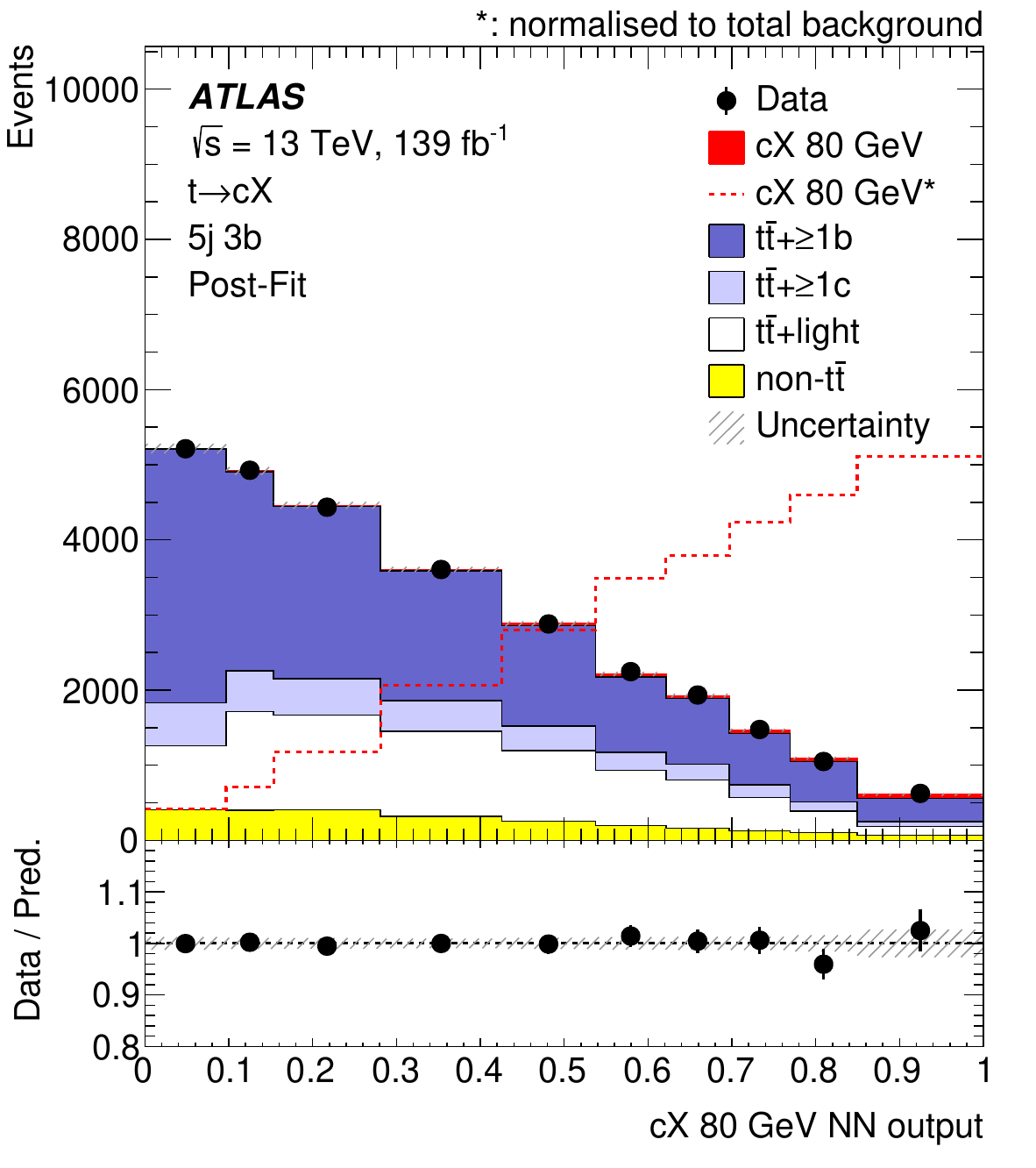}}
\subfloat[\tcX, 6j 3b]{\includegraphics[width = 0.30\textwidth]{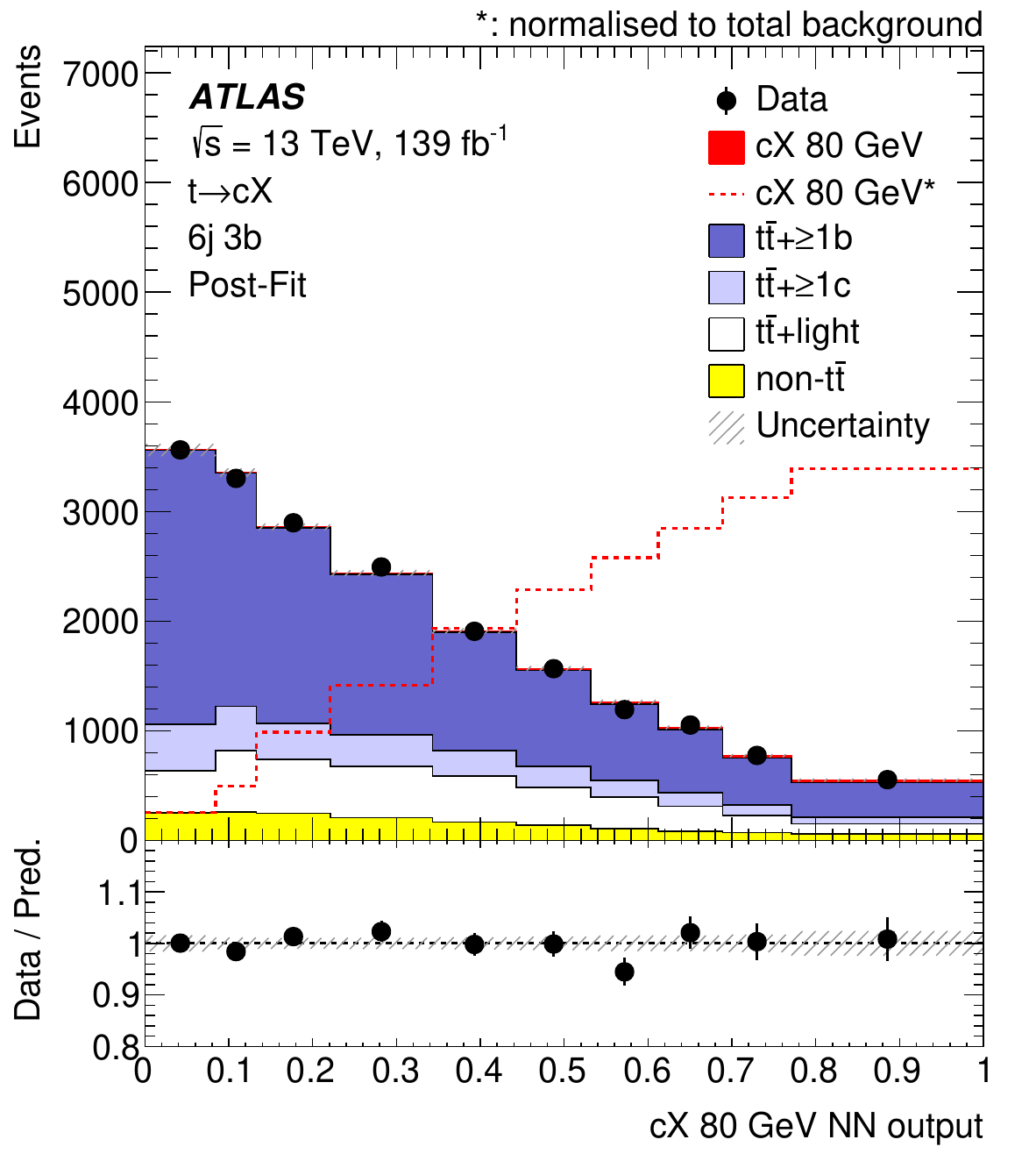}} \\
\subfloat[\tuX, $\ge 4$b]{\includegraphics[width = 0.30\textwidth]{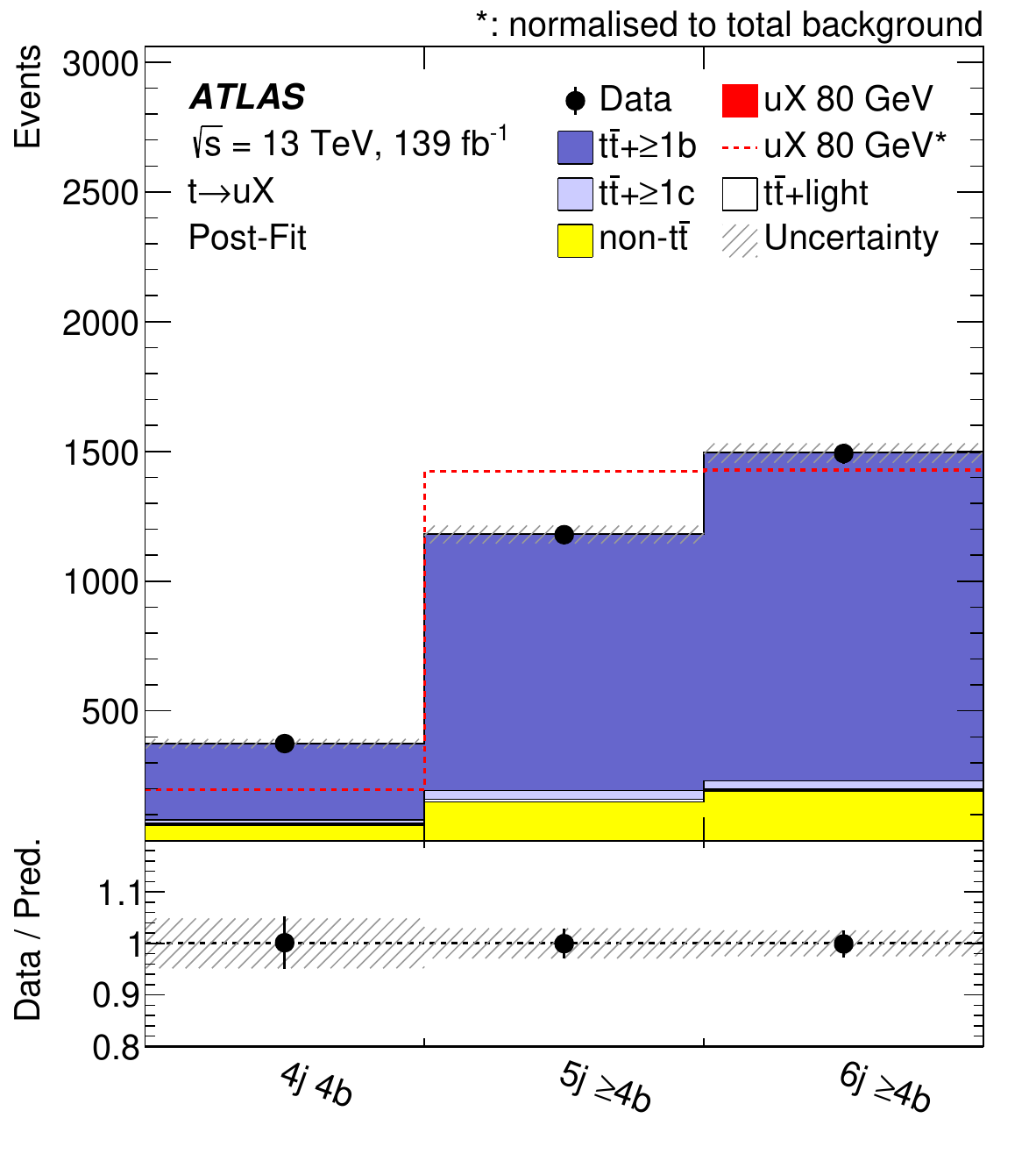}}
\subfloat[\tcX, $\ge 4$b]{\includegraphics[width = 0.30\textwidth]{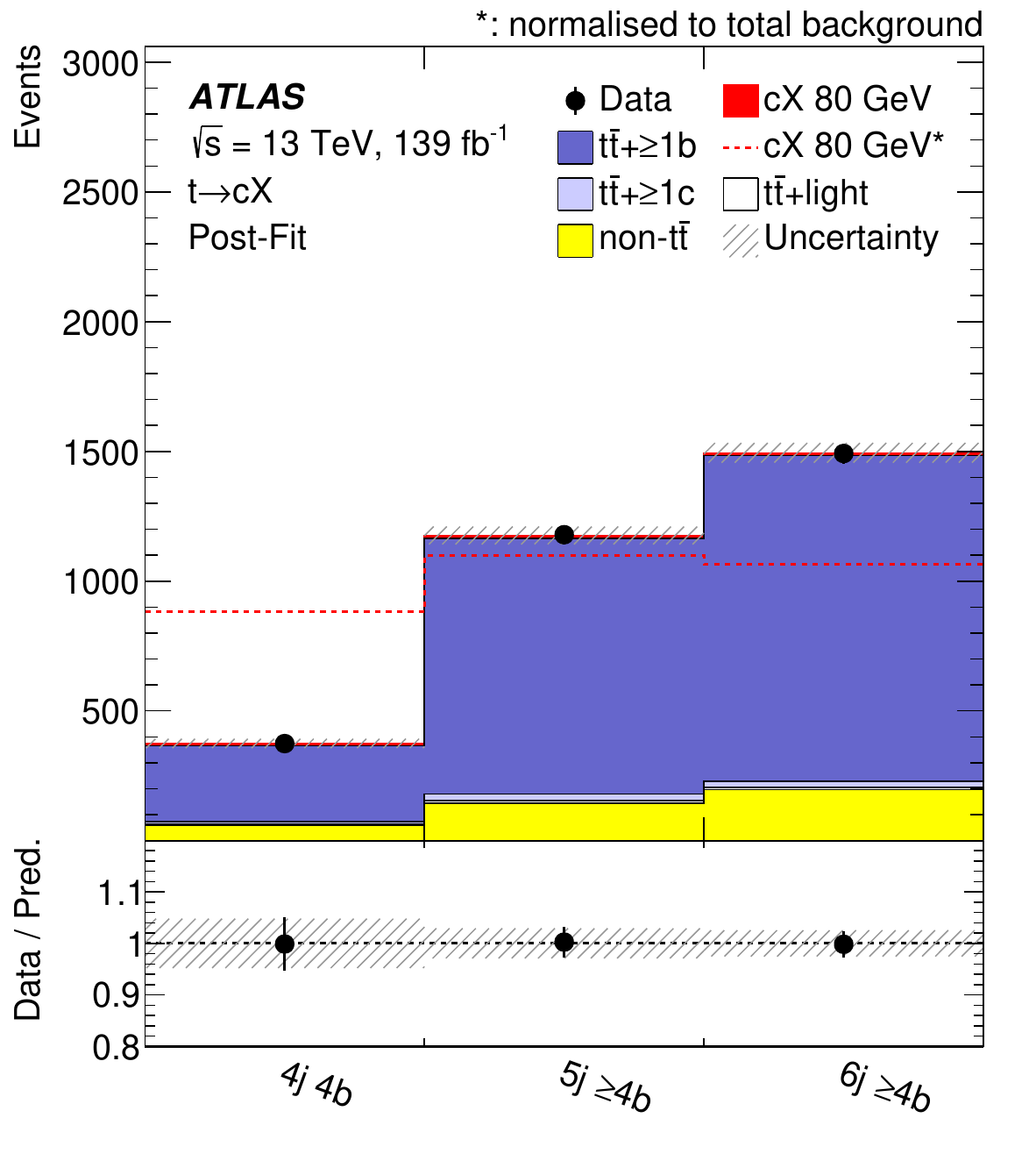}}
\caption{Comparison between the data and prediction for the NN output in the 3b regions for the \tuX\ ((a) to (c))
and the \tcX\ ((d) to (f)) processes,
and the yields in the $\ge 4$b regions for the \tuX\ (g) and the \tcX\ (h) processes
after the signal-plus-background fit to data for the 80~GeV $X$ scalar mass hypothesis.
The dashed red line corresponds to a $m_X = 80$~GeV signal normalised to the total background.
The uncertainty bands show the total uncertainty after the fit.}
\label{fig:NNfit80}
\end{figure}
 
\begin{figure}[htb]
\centering
\subfloat[\tuX, 4j 3b]{\includegraphics[width = 0.30\textwidth]{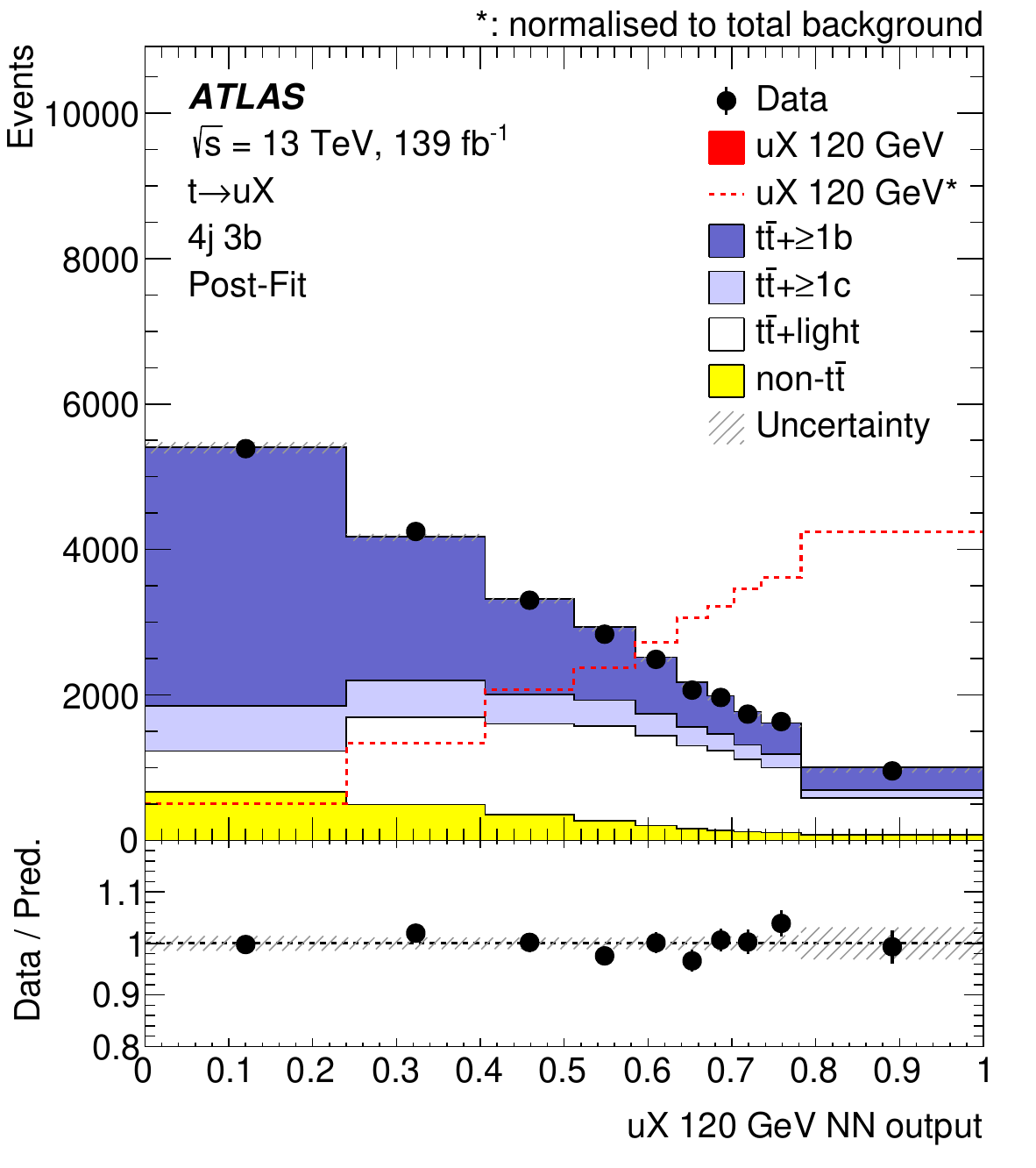}}
\subfloat[\tuX, 5j 3b]{\includegraphics[width = 0.30\textwidth]{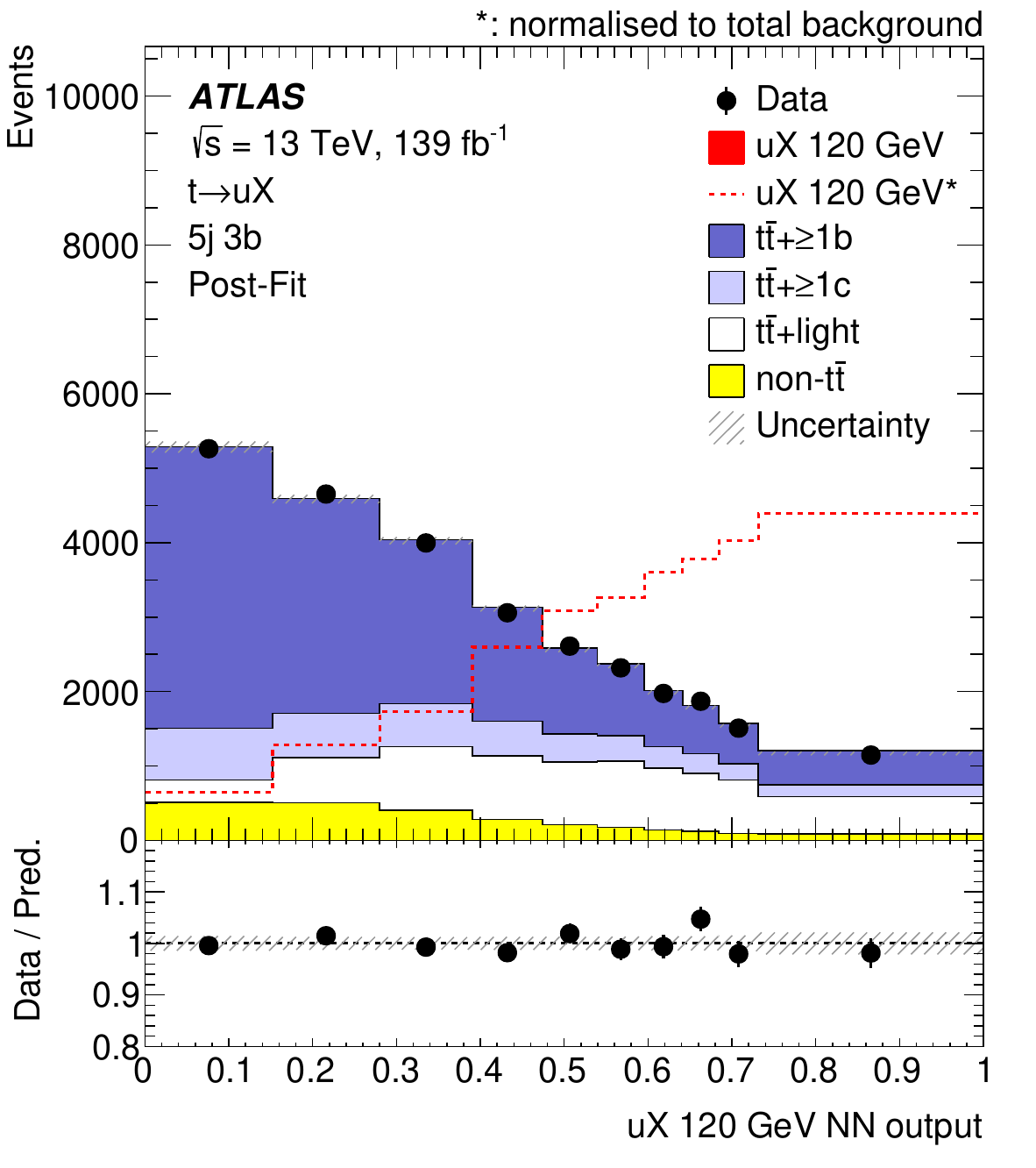}}
\subfloat[\tuX, 6j 3b]{\includegraphics[width = 0.30\textwidth]{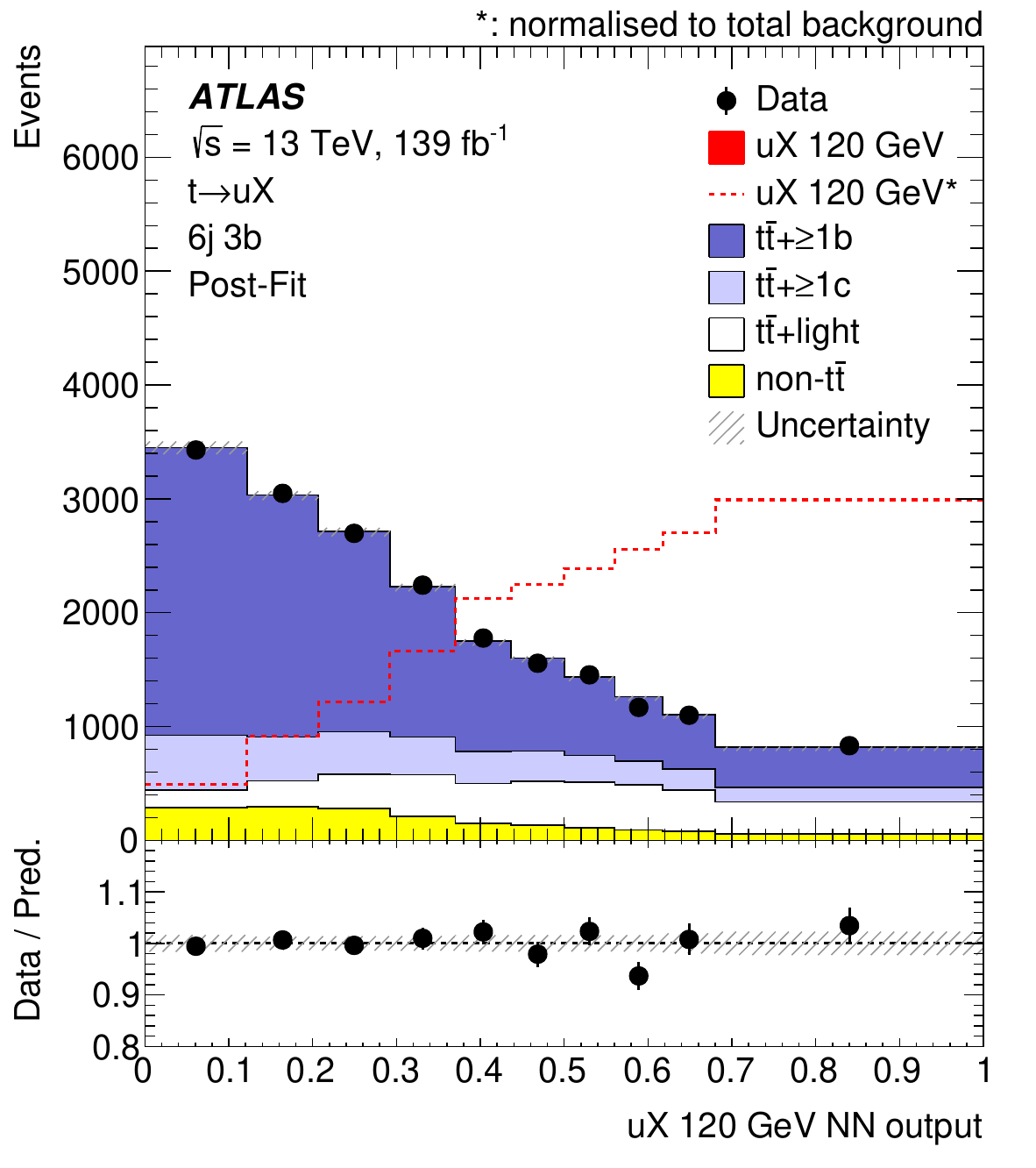}} \\
\subfloat[\tcX, 4j 3b]{\includegraphics[width = 0.30\textwidth]{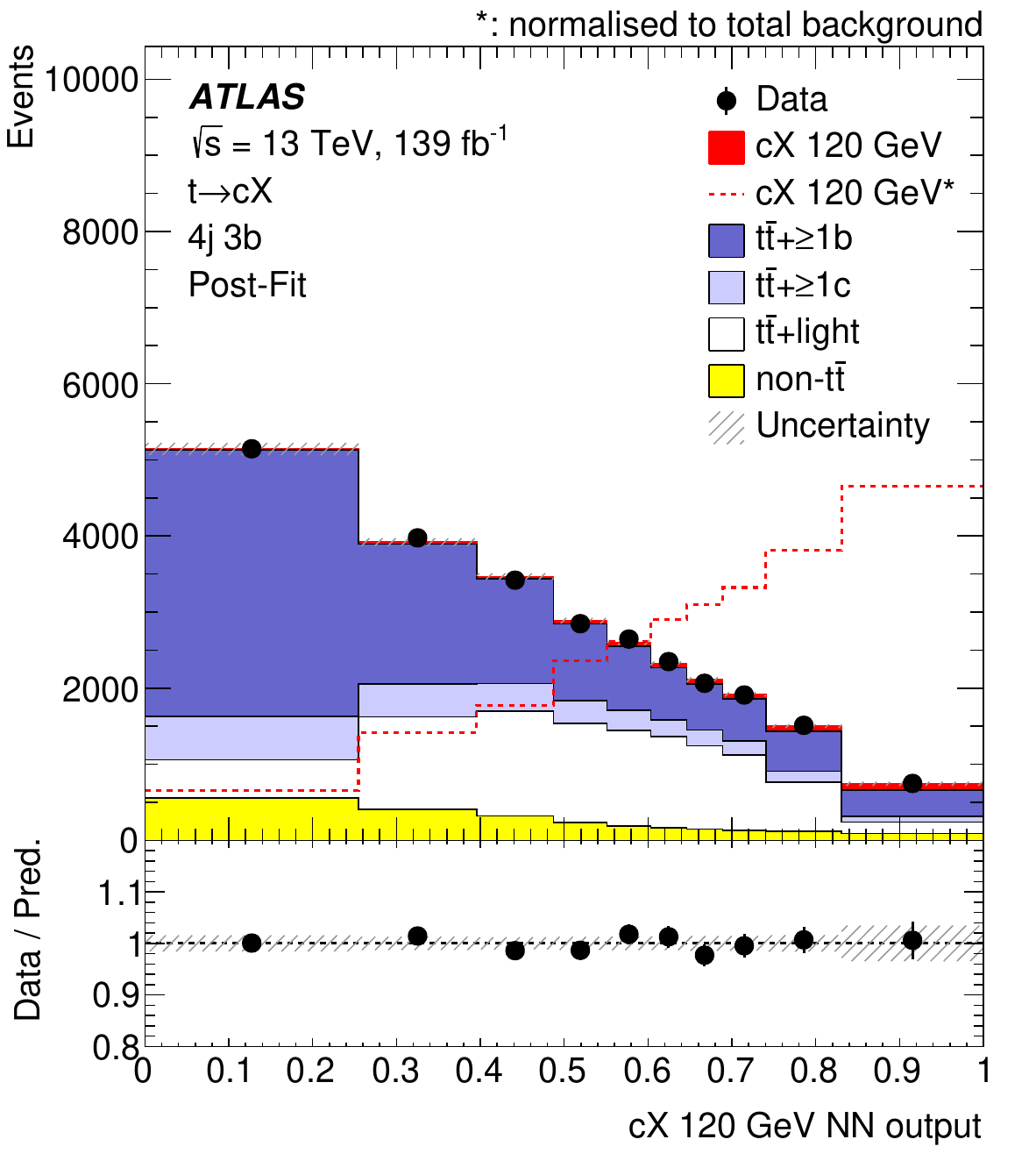}}
\subfloat[\tcX, 5j 3b]{\includegraphics[width = 0.30\textwidth]{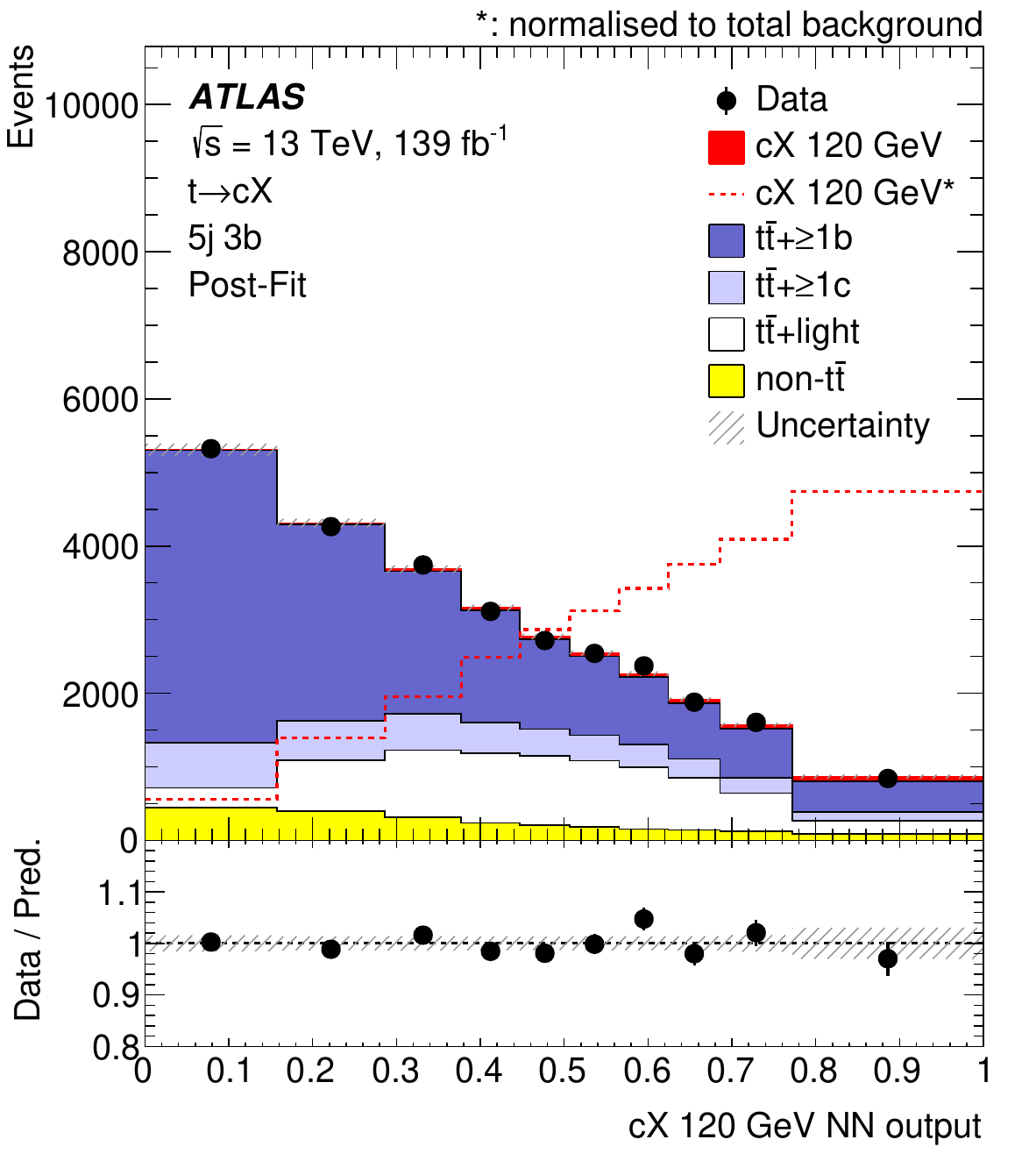}}
\subfloat[\tcX, 6j 3b]{\includegraphics[width = 0.30\textwidth]{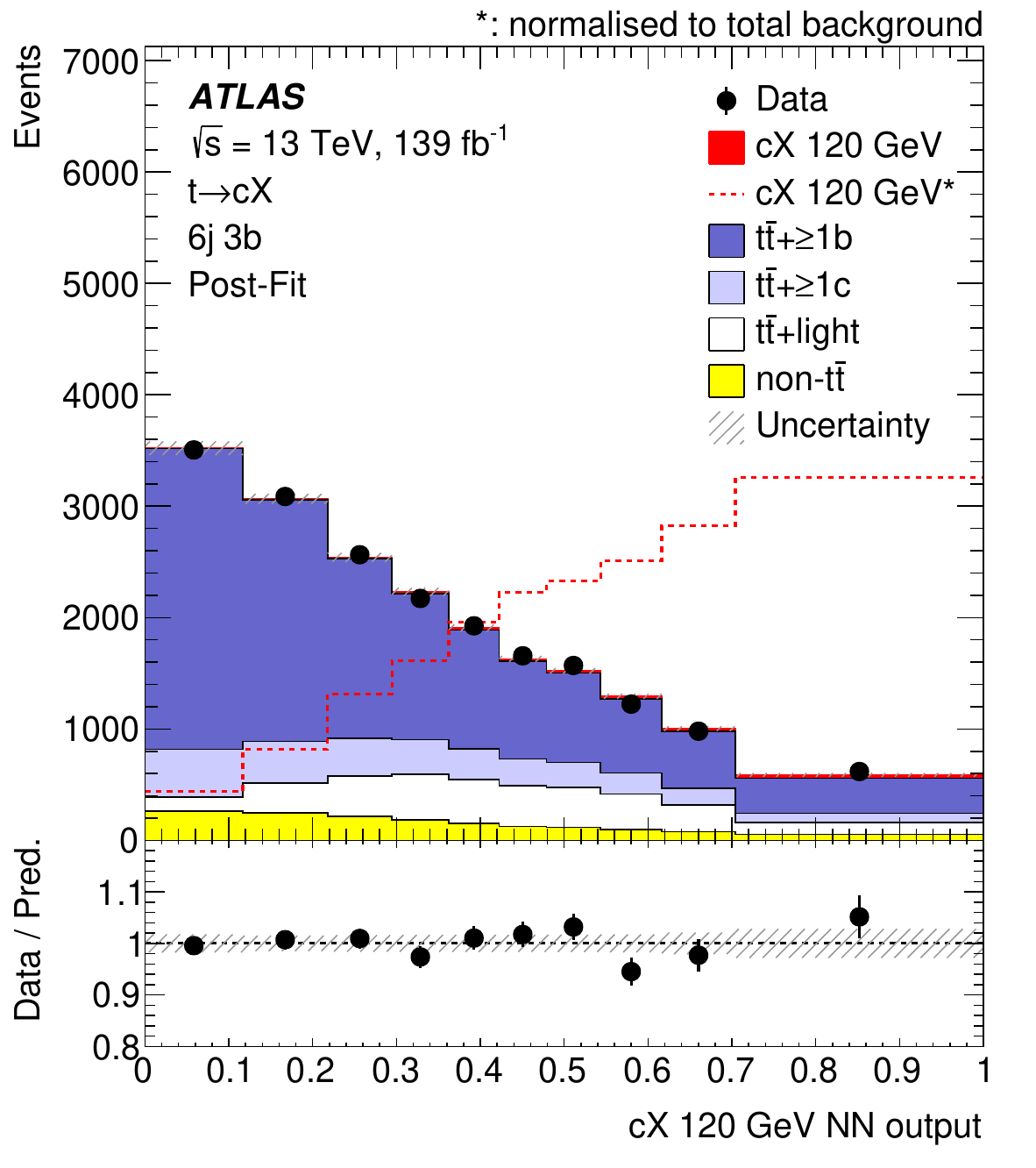}} \\
\subfloat[\tuX, $\ge 4$b]{\includegraphics[width = 0.30\textwidth]{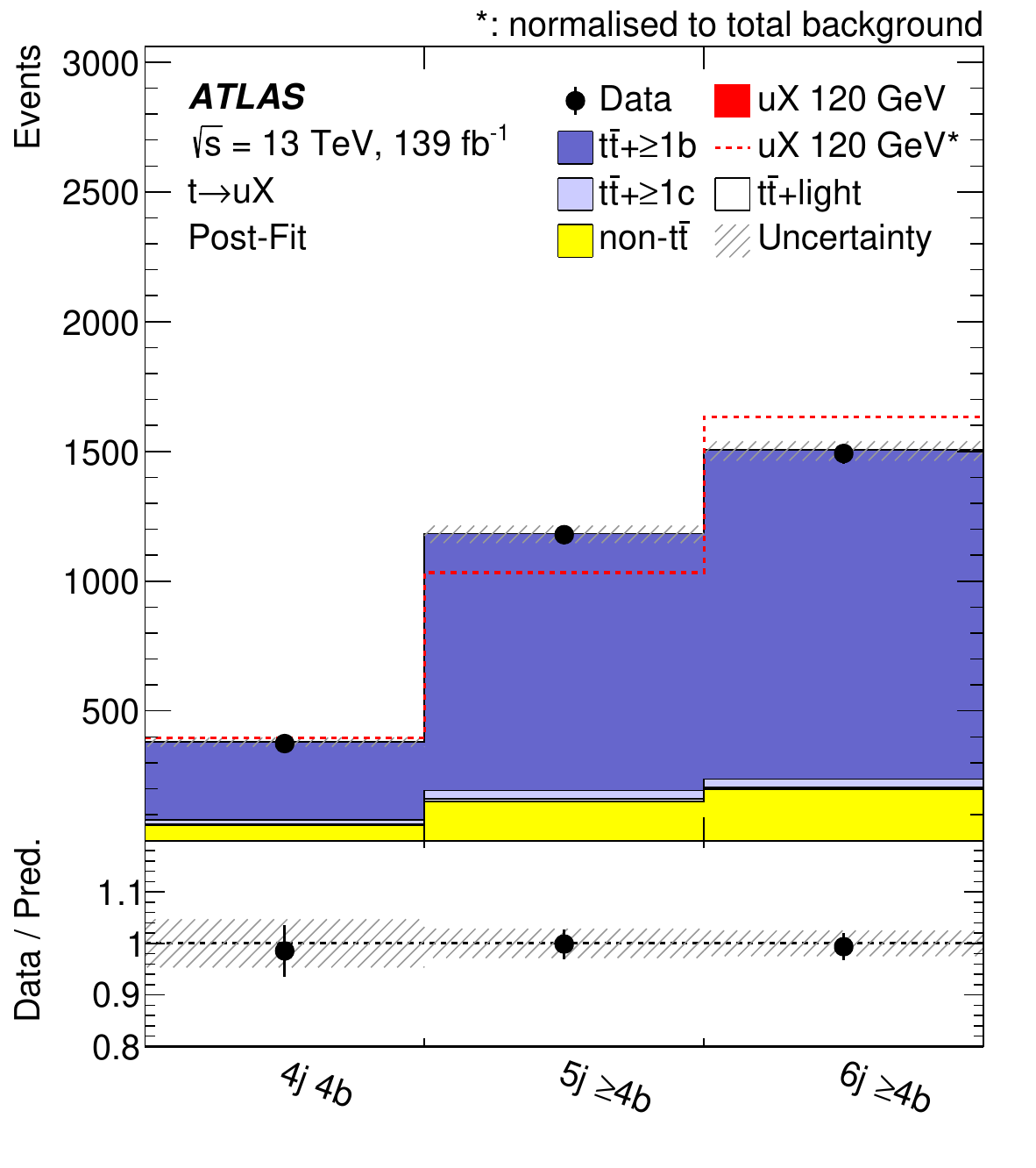}}
\subfloat[\tcX, $\ge 4$b]{\includegraphics[width = 0.30\textwidth]{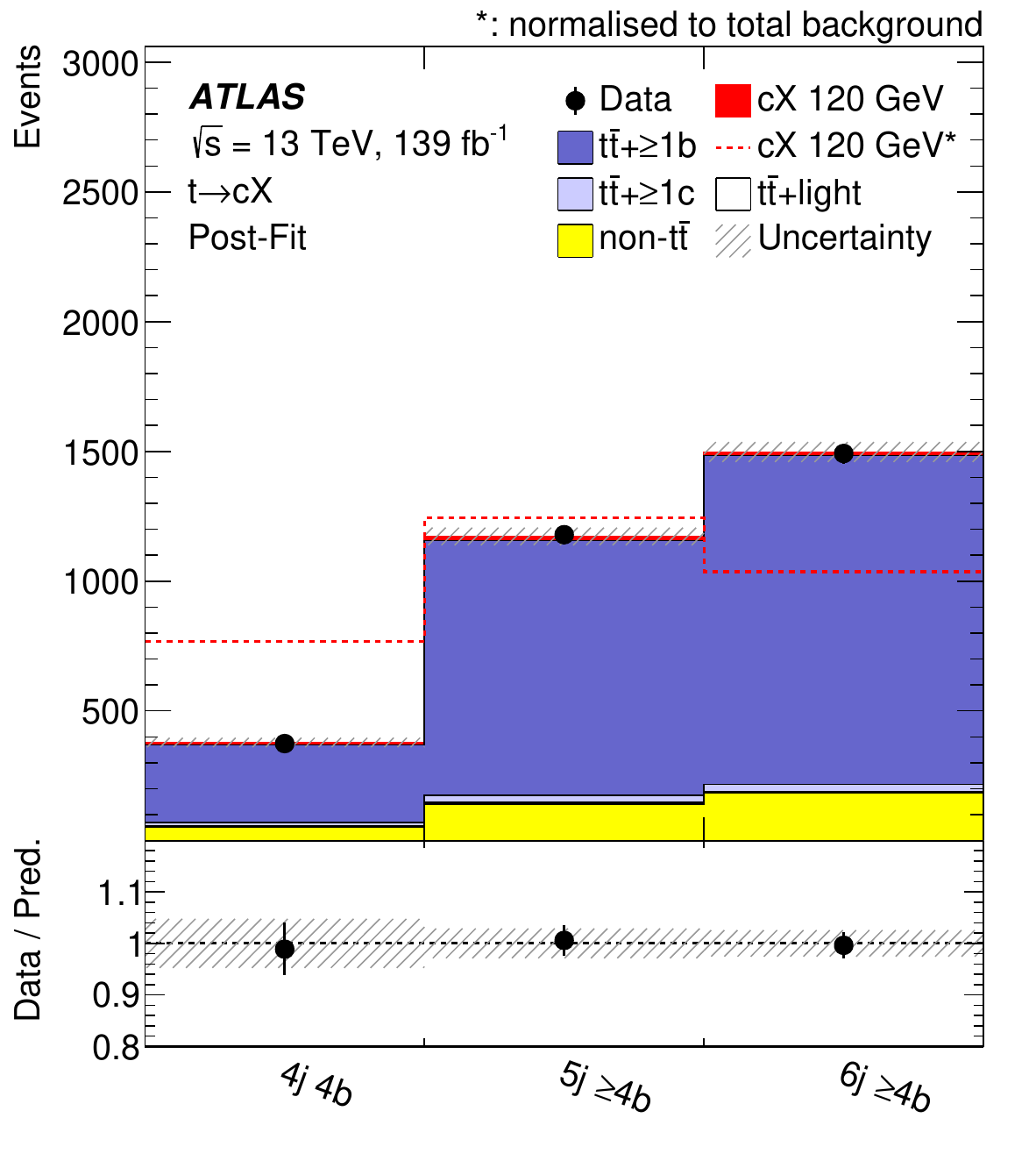}}
\caption{Comparison between the data and prediction for the NN output in the 3b regions for the \tuX\ ((a) to (c))
and the \tcX\ ((d) to (f)) processes,
and the yields in the $\ge 4$b regions for the \tuX\ (g) and the \tcX\ (h) processes
after the signal-plus-background fit to data for the 120~GeV $X$ scalar mass hypothesis.
The dashed red line corresponds to a $m_X = 120$~GeV signal normalised to the total background.
The uncertainty bands show the total uncertainty after the fit.}
\label{fig:NNfit120}
\end{figure}
 
The 95\% CL upper limits on $\BR(\tuX) \times \BR(X\rightarrow b\bar b)$ and
$\BR(\tcX) \times \BR(X\rightarrow b\bar b)$ obtained using the
$\mathrm{CL_s}$ method are presented in Figure~\ref{fig:limit}.
A local excess of 1.8 standard deviations is seen in the \tuX\ channel at $\mX = 40$~GeV.
Also,
a roughly two-standard deviation excess can be seen in the \tcX\ observed limit over almost the entire range of $\mX$.
This excess,
slightly visible in Figures~\ref{fig:NNfit80} and~\ref{fig:NNfit120} (e) to (h),
is not compatible with the presence of a scalar particle $X$,
which would show up as a narrower,
resonance-like, excess in the limit plot.
 
The observed (expected) limits range from 0.019\% (0.017\%) to 0.062\% (0.056\%) for
$\BR(\tuX) \times \BR(X\rightarrow b\bar b)$
and from 0.018\% (0.015\%) to 0.078\% (0.056\%) for $\BR(\tcX) \times \BR(X\rightarrow b\bar b)$.
The $p_0$ values range from 0.033 (under the $\mX = 40$~GeV hypothesis) to 0.908 for the \tuX\ fits and
from 0.016 (under the $\mX = 80$~GeV hypothesis) to 0.332 for the \tcX\ fits.
Although the kinematics of \tuX\ and \tcX\ events are very similar,
they slightly differ in the fourth jet due to its different flavour.
Given the use of $b$-tagging information in the NN training,
the discrimination achieved between background and \tuX\ or \tcX\ signals
slightly differs too and depends on the mass of the scalar.
Similarly,
the observed (expected) 95\% CL upper limit on \BR(\tuH) is 0.077\% (0.088\%) and
that on \BR(\tcH) is 0.12\% (0.076\%),
where the single top-quark contribution to the process has not been included because it is negligible
and SM decay branching fractions have been assumed for the Higgs boson.
 
The limits can be compared with previous results on FCNC \tqH\ searches,
where $H$ is the SM Higgs boson,
obtained by ATLAS with 36~\ifb of data,
0.52\% (0.49\%) for the observed (expected) limit on \BR(\tuH($b\bar b$))
and 0.42\% (0.40\%) for the observed (expected) limit on \BR(\tcH($b\bar b$))~\cite{TOPQ-2017-07}.
The observed (expected) limits obtained by CMS with 137~\ifb\ of data are
0.079\% (0.11\%) for \BR(\tuH($b\bar b$)) and
0.094\% (0.086\%) for \BR(\tcH($b\bar b$))~\cite{CMS:2021gfa}.
The expected limits presented in this paper are on average
a factor of three better than the previous ATLAS results scaled to the same integrated luminosity,
and slightly better than the CMS results.
The use of a NN instead of a likelihood discriminant,
a better $b$-tagging algorithm,
and better \ttbar\ background modelling are the main improvements over the previous ATLAS analysis.
In these comparisons the fact that this analysis assumes a $\BR(\Xbb)=100$\%,
whereas the \tqH\ results are affected by a smaller $\BR(\Hbb)$,
has been taken into account.
More stringent limits were obtained by CMS using the $H \rightarrow \gamma\gamma$ decay mode:
0.019\% (0.031\%) for the observed (expected) limit on \BR(\tuH($\gamma\gamma$)) and
0.073\% (0.051\%) for the observed (expected) limit on \BR(\tcH($\gamma\gamma$))~\cite{CMS-TOP-20-007}.
Finally,
the observed (expected)
ATLAS limits on \BR(\tqH) with $H\rightarrow\tau\bar\tau$,
obtained with 139~\ifb\ of data,
are
0.072\% (0.036\%) for \tuH\ and
0.099\% (0.050\%) for \tcH~\cite{TOPQ-2019-17},
which are comparable with the results presented in this paper.

\begin{figure}[htb]
\centering
\subfloat[\tuX]{\includegraphics[width =0.6\textwidth]{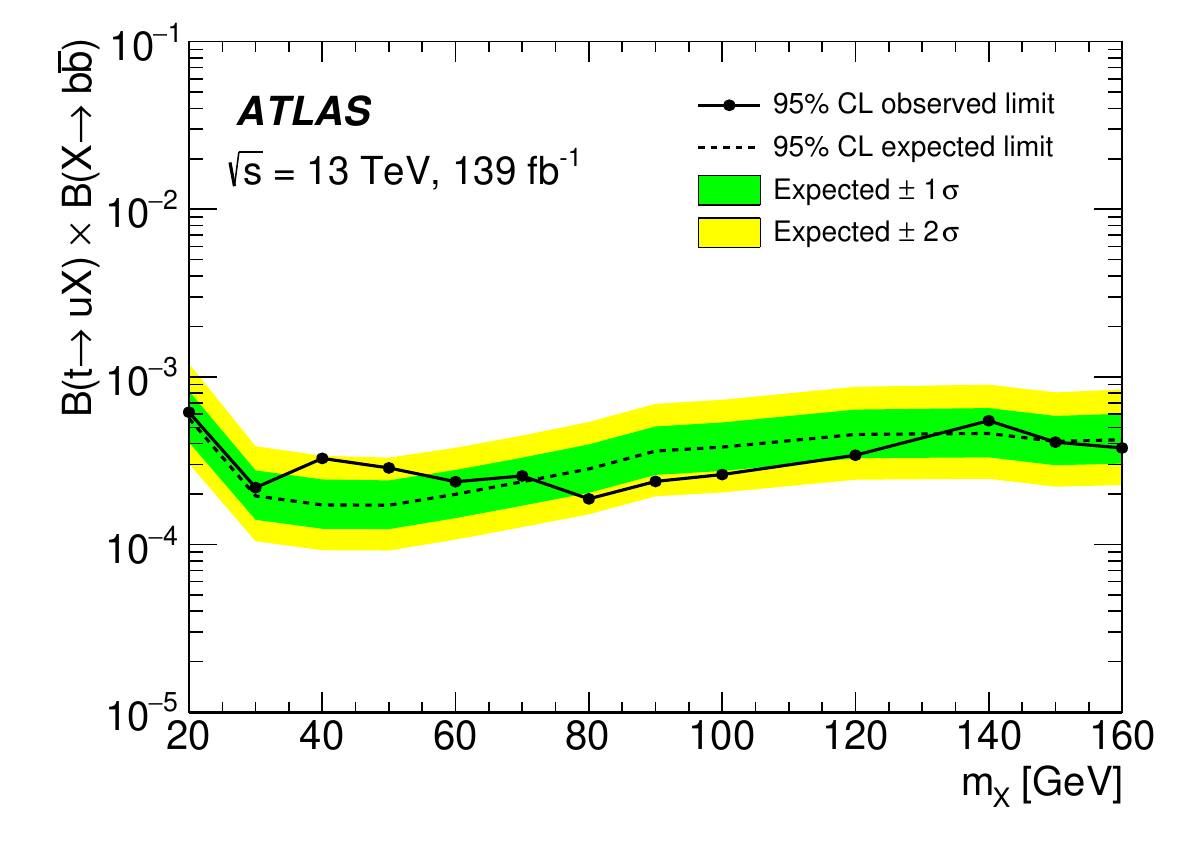}} \\  
\subfloat[\tcX]{\includegraphics[width =0.6\textwidth]{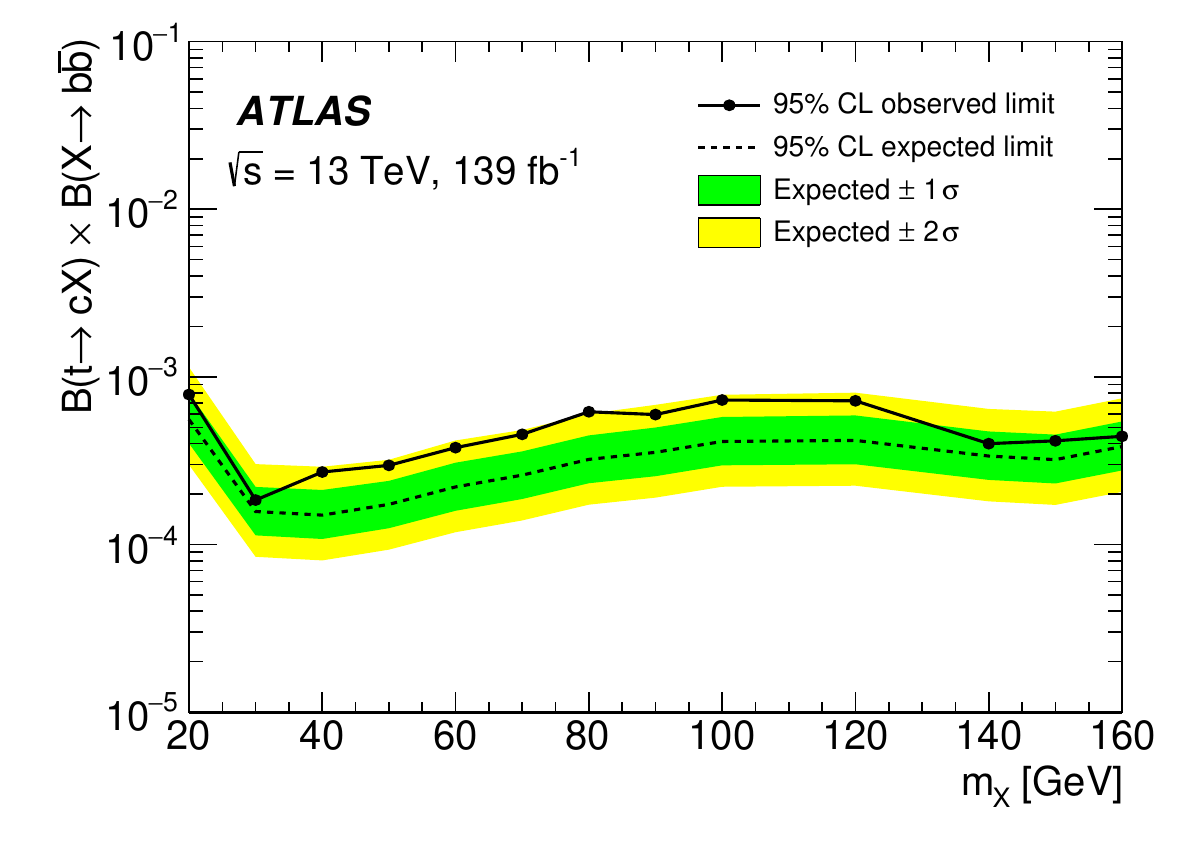}}
\caption{Expected and observed 95\% CL upper limits for \BR(\tuX) $\times$ \BR($X\rightarrow b\bar{b}$) (a) and \BR(\tcX) $\times$ \BR($X\rightarrow b\bar{b}$) (b).
The bands surrounding the expected limits show the 68\% and 95\% confidence intervals,
respectively.}
\label{fig:limit}
\end{figure}


\FloatBarrier
 
\section{Conclusion}
\label{sec:conclusion}

A search for a neutral scalar particle $X$ produced in flavour-changing neutral-current top-quark decays is presented,
based on a data sample corresponding to an integrated luminosity of 139~fb$^{-1}$ from proton-proton collisions at $\sqrt{s}$=13~TeV,
recorded with the ATLAS detector at the LHC.
The search for \tqX,
with $X \rightarrow b\bar{b}$,
produced in \ttbar\ events is performed in the $X$ mass range from 20 to 160~GeV in single-lepton final states.
A discriminant neural network has been trained for each \tuX\ or \tcX\ process to distinguish between signal and background.
The output of the neural network depends on the $X$ mass,
and a fit to the data is performed simultaneously on these output distributions
in the signal regions and the total yields in the control regions,
separately for the various mass hypotheses.
 
No significant excess above the expected Standard Model background is found and
observed (expected) 95\% confidence-level upper limits between 0.019\% (0.017\%) and 0.062\% (0.056\%)
are derived for the branching fraction $\BR(\tuX) \times \BR(X\rightarrow b\bar b)$
and between 0.018\% (0.015\%) and 0.078\% (0.056\%)
for the branching fraction $\BR(\tcX)\times \BR(X\rightarrow b\bar b)$ in the explored mass range.
The expected limits are on average a factor of three better than the previous ATLAS
results scaled to the same integrated luminosity.
The same neural network is used to derive limits for the branching fraction of
a top quark into the Standard Model Higgs boson and a quark,
resulting in 95\% confidence level upper limits of 0.077\% (0.088\%) for the observed (expected) $\BR(\tuH)$ and
0.12\% (0.076\%) for the observed (expected) $\BR(\tcH)$.


\section*{Acknowledgements}


We thank CERN for the very successful operation of the LHC, as well as the
support staff from our institutions without whom ATLAS could not be
operated efficiently.
 
We acknowledge the support of
ANPCyT, Argentina;
YerPhI, Armenia;
ARC, Australia;
BMWFW and FWF, Austria;
ANAS, Azerbaijan;
SSTC, Belarus;
CNPq and FAPESP, Brazil;
NSERC, NRC and CFI, Canada;
CERN;
ANID, Chile;
CAS, MOST and NSFC, China;
Minciencias, Colombia;
MEYS CR, Czech Republic;
DNRF and DNSRC, Denmark;
IN2P3-CNRS and CEA-DRF/IRFU, France;
SRNSFG, Georgia;
BMBF, HGF and MPG, Germany;
GSRI, Greece;
RGC and Hong Kong SAR, China;
ISF and Benoziyo Center, Israel;
INFN, Italy;
MEXT and JSPS, Japan;
CNRST, Morocco;
NWO, Netherlands;
RCN, Norway;
MEiN, Poland;
FCT, Portugal;
MNE/IFA, Romania;
JINR;
MES of Russia and NRC KI, Russian Federation;
MESTD, Serbia;
MSSR, Slovakia;
ARRS and MIZ\v{S}, Slovenia;
DSI/NRF, South Africa;
MICINN, Spain;
SRC and Wallenberg Foundation, Sweden;
SERI, SNSF and Cantons of Bern and Geneva, Switzerland;
MOST, Taiwan;
TENMAK, T\"urkiye;
STFC, United Kingdom;
DOE and NSF, United States of America.
In addition, individual groups and members have received support from
BCKDF, CANARIE, Compute Canada and CRC, Canada;
COST, ERC, ERDF, Horizon 2020 and Marie Sk{\l}odowska-Curie Actions, European Union;
Investissements d'Avenir Labex, Investissements d'Avenir Idex and ANR, France;
DFG and AvH Foundation, Germany;
Herakleitos, Thales and Aristeia programmes co-financed by EU-ESF and the Greek NSRF, Greece;
BSF-NSF and GIF, Israel;
Norwegian Financial Mechanism 2014-2021, Norway;
NCN and NAWA, Poland;
La Caixa Banking Foundation, CERCA Programme Generalitat de Catalunya and PROMETEO and GenT Programmes Generalitat Valenciana, Spain;
G\"{o}ran Gustafssons Stiftelse, Sweden;
The Royal Society and Leverhulme Trust, United Kingdom.
 
The crucial computing support from all WLCG partners is acknowledged gratefully, in particular from CERN, the ATLAS Tier-1 facilities at TRIUMF (Canada), NDGF (Denmark, Norway, Sweden), CC-IN2P3 (France), KIT/GridKA (Germany), INFN-CNAF (Italy), NL-T1 (Netherlands), PIC (Spain), ASGC (Taiwan), RAL (UK) and BNL (USA), the Tier-2 facilities worldwide and large non-WLCG resource providers. Major contributors of computing resources are listed in Ref.~\cite{ATL-SOFT-PUB-2021-003}.


\printbibliography
 
\clearpage
\input{atlas_authlist}

\end{document}

%% file: atlas_authlist.tex
 
\begin{flushleft}
\hypersetup{urlcolor=black}
{\Large The ATLAS Collaboration}

\bigskip

\AtlasOrcid[0000-0002-6665-4934]{G.~Aad}$^\textrm{\scriptsize 101}$,
\AtlasOrcid[0000-0002-5888-2734]{B.~Abbott}$^\textrm{\scriptsize 119}$,
\AtlasOrcid[0000-0002-7248-3203]{D.C.~Abbott}$^\textrm{\scriptsize 102}$,
\AtlasOrcid[0000-0002-1002-1652]{K.~Abeling}$^\textrm{\scriptsize 55}$,
\AtlasOrcid[0000-0002-8496-9294]{S.H.~Abidi}$^\textrm{\scriptsize 29}$,
\AtlasOrcid[0000-0002-9987-2292]{A.~Aboulhorma}$^\textrm{\scriptsize 35e}$,
\AtlasOrcid[0000-0001-5329-6640]{H.~Abramowicz}$^\textrm{\scriptsize 150}$,
\AtlasOrcid[0000-0002-1599-2896]{H.~Abreu}$^\textrm{\scriptsize 149}$,
\AtlasOrcid[0000-0003-0403-3697]{Y.~Abulaiti}$^\textrm{\scriptsize 116}$,
\AtlasOrcid[0000-0003-0762-7204]{A.C.~Abusleme~Hoffman}$^\textrm{\scriptsize 136a}$
% ...
\bigskip
\\

% ...
$^{29}$Physics Department, Brookhaven National Laboratory, Upton NY; United States of America.\\
% ...
$^{35}$$^{(a)}$Facult\'e des Sciences Ain Chock, R\'eseau Universitaire de Physique des Hautes Energies - Universit\'e Hassan II, Casablanca;$^{(b)}$Facult\'{e} des Sciences, Universit\'{e} Ibn-Tofail, K\'{e}nitra;$^{(c)}$Facult\'e des Sciences Semlalia, Universit\'e Cadi Ayyad, LPHEA-Marrakech;$^{(d)}$LPMR, Facult\'e des Sciences, Universit\'e Mohamed Premier, Oujda;$^{(e)}$Facult\'e des sciences, Universit\'e Mohammed V, Rabat;$^{(f)}$Institute of Applied Physics, Mohammed VI Polytechnic University, Ben Guerir; Morocco.\\
% ...
$^{55}$II. Physikalisches Institut, Georg-August-Universit\"{a}t G\"ottingen, G\"ottingen; Germany.\\
% ...
$^{101}$CPPM, Aix-Marseille Universit\'e, CNRS/IN2P3, Marseille; France.\\
$^{102}$Department of Physics, University of Massachusetts, Amherst MA; United States of America.\\
% ...
$^{116}$Department of Physics, New York University, New York NY; United States of America.\\
% ...
$^{119}$Homer L. Dodge Department of Physics and Astronomy, University of Oklahoma, Norman OK; United States of America.\\
% ...
$^{136}$$^{(a)}$Departamento de F\'isica, Pontificia Universidad Cat\'olica de Chile, Santiago;$^{(b)}$Millennium Institute for Subatomic physics at high energy frontier (SAPHIR), Santiago;$^{(c)}$Instituto de Investigaci\'on Multidisciplinario en Ciencia y Tecnolog\'ia, y Departamento de F\'isica, Universidad de La Serena;$^{(d)}$Universidad Andres Bello, Department of Physics, Santiago;$^{(e)}$Instituto de Alta Investigaci\'on, Universidad de Tarapac\'a, Arica;$^{(f)}$Departamento de F\'isica, Universidad T\'ecnica Federico Santa Mar\'ia, Valpara\'iso; Chile.\\
% ...
$^{149}$Department of Physics, Technion, Israel Institute of Technology, Haifa; Israel.\\
$^{150}$Raymond and Beverly Sackler School of Physics and Astronomy, Tel Aviv University, Tel Aviv; Israel.\\
% ...
$^{a}$ Also Affiliated with an institute covered by a cooperation agreement with CERN.\\
$^{b}$ Also at An-Najah National University, Nablus; Palestine.\\
$^{c}$ Also at Borough of Manhattan Community College, City University of New York, New York NY; United States of America.\\
$^{d}$ Also at Bruno Kessler Foundation, Trento; Italy.\\
$^{e}$ Also at Center for High Energy Physics, Peking University; China.\\
$^{f}$ Also at Centro Studi e Ricerche Enrico Fermi; Italy.\\
% ...

\end{flushleft}
